\documentclass[a4paper,11pt]{article}
\pdfoutput=1 

\usepackage{jcappub} 
\usepackage[T1]{fontenc}

\usepackage{cancel}
\usepackage{enumitem}    
\usepackage{amsthm}         
\usepackage{amsmath} 	   
\usepackage{amssymb}       
\usepackage{amsfonts}
\usepackage{graphicx} 	    
\usepackage{verbatim}
\usepackage{epsfig,multicol,bbm}
\usepackage{color}
\usepackage{colortbl}
\usepackage{float}
\usepackage{multirow}

\usepackage{subfig}

\allowdisplaybreaks

\title{Stability and the Gauge Problem in Non-Perturbative Cosmology}

\author[a,b,1]{Anna Ijjas,}
\note{Corresponding Author}
\author[b,c]{Frans Pretorius,}
\author[b,c]{and Paul J. Steinhardt}

\affiliation[a]{Institute for Theory and Computation, Harvard University\\Cambridge, MA, 02138, USA}
\affiliation[b]{Department of Physics, Princeton University, Princeton, NJ 08544, USA}
\affiliation[c]{Princeton Center for Theoretical Science, Princeton University, Princeton, NJ 08544, USA}

\emailAdd{anna.ijjas@cfa.harvard.edu}
\emailAdd{fpretori@princeton.edu}
\emailAdd{steinh@princeton.edu}

\abstract{
In this paper, we describe the first steps towards fully non-perturbative cosmology.   We explain why the conventional methods used by cosmologists based on the ADM formulation are generally inadequate for this purpose and why it is advantageous instead to adapt the harmonic formulation pioneered and utilized in mathematical and numerical relativity.  Here we focus on using this approach to evaluating the linear mode stability in homogeneous and nearly homogeneous backgrounds and devising a valid scheme and diagnostics for numerical computation.  We also briefly touch on the relevance of these methods for extracting cosmological observables from non-perturbative simulations.
}


\begin{document}
\maketitle 
\flushbottom

\section{Introduction}

The study of gauge-invariant perturbations has been the cornerstone of both cosmological model building and astrophysical data analysis since its introduction in the early 80's \cite{Gerlach:1979rw,Bardeen:1980kt,Bardeen:1983qw,Sasaki:1983kd,Mukhanov:1988jd}. Theoretical proposals are characterized and judged by specifying their predictions for the spectra of these variables. Accordingly, most codes ({\it e.g.,} {\sc camb} \cite{Lewis:1999bs}, {\sc cosmomc} \cite{Lewis:2002ah}, or {\sc class} \cite{Lesgourgues:2011re}, to name a few) are written in the language of these variables to extract constraints from observational data, such as the microwave background and other surveys. 
The reason for this wide-spread use of gauge-invariant variables is that
they provide an accurate and efficient way to extract predictions from a gravitational field theory applied to the universe whenever the large-scale evolution is linear. Instead of solving a highly complex, coupled system of non-linear partial differential equations (PDEs) and then extracting the observable predictions, gauge-invariant variables reduce the analysis to the study of only three decoupled second-order ordinary differential equations (ODEs). Moreover, by representing physical quantities they immediately connect to observations.

However, cosmological perturbation theory has its limitations when taken out of its original context. In this paper we show how the conventional method based on scalar-vector-decomposition (SVT) and the Arnowitt-Deser-Misner (ADM) form is ill-suited to address important questions concerning non-linear dynamics, or to evaluate the viability of scenarios based on classical modifications of Einstein gravity.  
We will instead introduce a new formulation along with a gauge fixing protocol that enables the study of these issues in a wide range of cosmological scenarios.
Our scheme is the implementation of a method that has been successful in analyzing dynamical systems in mathematical and numerical general relativity: we will linearize cosmological scalar field theories based on the {\it generalized harmonic formulation} of the Einstein equations \cite{Garfinkle:2001ni,Pretorius:2004jg}.

The generalized harmonic formulation promotes each spacetime coordinate $x^{\mu}$ to a dynamical variable obeying a scalar wave equation with source function $J^{\mu}$ that itself is a function of the coordinates,
\begin{equation}
\label{def-harm-coo}
\Box x^{\mu} = J^{\mu}(x^{\alpha})\,.
\end{equation}
Throughout, we will use Greek indices to refer to spacetime coordinates and Latin indices to refer to purely spatial coordinates.
Notably, Eq.~\eqref{def-harm-coo} does not directly define a particular gauge or coordinate choice.
Rather, any known metric $g_{\mu\nu}$ in any coordinate system has a harmonic representation in terms of the corresponding source functions. As we will emphasize, the essence of any generalized harmonic formulation of the field equations is to treat the source functions $J^{\mu}$ as new degrees of freedom, with Eq.~\eqref{def-harm-coo} then becoming a set of constraint equations. To close the system, we have to specify additional equations for the source functions $J^{\mu}$ that  define the gauge.

In order to identify observables and extract cosmological predictions within our scheme, we specify a way of finding harmonic source functions that correspond to the commonly used cosmological gauges. This provides a `dictionary' between conventional cosmological perturbation theory and our scheme.
The method we propose is closely related to the fact that gauge invariance is in a certain sense equivalent to gauge fixing \cite{Bardeen:1980kt} -- that is, each gauge-invariant variable or observable of the linear theory can be associated with a particular space-time slicing. One example is that the Bardeen potentials correspond to the lapse and the scalar part of the spatial metric perturbations in zero-shear gauge. Another example is in the case of a canonical scalar field, where the co-moving curvature mode is the scalar part of the spatial metric perturbation in unitary gauge.

As a working example, we will analyze the conformally-coupled ${\cal L}_4$-Horndeski theory given by the action
\begin{equation}
\label{L4-action}
{\cal S} = \int d^4 x \sqrt{-g}\left( \frac12 \big(1+G_4(\phi) \big) R + G_2(\phi, X) +G_3(\phi, X) \Box\phi \right) + {\cal S}_m
\,.
\end{equation}
Here $g$ is the metric determinant, $R$ is the Ricci scalar, $X\equiv -(1/2)g^{\alpha\beta}\nabla_{\alpha}\phi\nabla_{\beta}\phi$ the canonical kinetic term associated with the scalar $\phi$, $G_i(X,\phi) \;(i=1,2,3)$ is the $i$-th Horndeski interaction, and ${\cal S}_m$ is the matter action. We denote the covariant derivative with respect to the coordinate $x_{\mu}$ by $\nabla_{\mu}$, and the partial derivative with respect to the coordinate $x_{\mu}$ by $\partial_{\mu}$. Throughout, we work in reduced Planck units $M_{\rm Pl}^2=8\pi G_{\rm N}=1$, where $G_{\rm N}$ is Newton's constant. Obviously, this theory is particularly well-suited as a representative working example because it admits second-order equations of motion and at the same time encompasses all the most commonly used gravitational field theories in cosmology: setting $G_3 = 0$ recovers Brans-Dicke theory; and setting both $G_3, G_4 \equiv 0$ recovers all $P(X,\phi)$ theories. 

However, our motivation is entirely physical. What makes conformally-coupled ${\cal L}_4$-Horndeski theories particularly interesting is their cosmological application. Most recently, it has been found that the {\it null convergence condition} (NCC) can be violated in these theories at energies well below the Planck scale without encountering pathologies at linear order in perturbation theory \cite{Pirtskhalava:2014esa,Ijjas:2016tpn,Ijjas:2016vtq,Ijjas:2017pei}. The NCC requires that for all null vectors $n^{\alpha}$,
\begin{equation}
R_{\alpha\beta}n^{\alpha}n^{\beta} \geq 0\,,
\end{equation}
where $R_{\alpha\beta}$ is the Ricci tensor. It is apparent that the NCC is a statement about geometry. Assuming Einstein gravity, the NCC coincides with the null energy condition (NEC), $T_{\alpha\beta}n^{\alpha}n^{\beta} \geq 0$ for all $n^{\alpha}$, though this is in general not the case in modified gravity theories, such as Horndeski. Indeed, Horndeski modifications of Einstein gravity mix the metric with scalar fields in a way that makes the NEC ill-defined as a condition on the properties of matter. At the same time the NCC remains perfectly well-defined. It is also the very condition that has to be violated to describe a so-called non-singular cosmological bounce -- the transition from a contracting to an expanding phase at energies well below the Planck scale accurately described by classical equations of motion -- because the notion of contraction and expansion refer precisely to a geometrical condition described by the physical metric. More concretely, on a Friedmann-Robertson-Walker (FRW) cosmological background with scale factor $a(t)$ and time coordinate $t$, violating the NCC means having a period with $\dot{H}>0$ (where $H\equiv \dot{a}/a$ is the Hubble parameter and dot denotes differentiation with respect to $t$), and this is exactly the condition required for a cosmological bounce. 

The main goal of our study is to stress test the `stability' claim about the linearized conformally-coupled ${\cal L}_4$-Horndeski theory and lay the foundations for its non-perturbative, numerical study that will ultimately establish the validity and robustness of non-singular bounce solutions and enable observational tests.

There have been earlier studies of cosmological scenarios using tools of numerical relativity, such as \cite{Garfinkle:2008ei,Xue:2013bva,East:2015ggf,Giblin:2016mjp,Clough:2016ymm,Clough:2017efm}. However, all the currently existing work involves scalar field theories with canonical kinetic terms and minimally-coupled to Einstein gravity. It is known that these theories are locally well-posed, {\it e.g.}, in harmonic formulations. Our goal is to go beyond these theories and study the linear well-posedness of scalar field theories with non-canonical kinetic terms and/or ${\cal L}_3$-Horndeski modifications of Einstein gravity and beyond. 

In addition, previous analyses were not designed to extract precise information from the non-perturbative analysis suitable for observational tests.
Our goal is to fill this gap and develop a scheme  that provides a complete set of non-perturbative  quantities that correspond to physical observables

The paper is organized as follows: We begin with a brief review of cosmological perturbation theory in Sec.~\ref{sec:review}. Then, in Sec.~\ref{sec:harmonic} we derive the linearized Einstein and scalar field equations. After performing a characteristic analysis of the system, we define necessary conditions for dynamical mode stability. 
In Sec.~\ref{sec:rel-to-old}, we clarify why in certain special cases cosmological perturbation theory yields reliable conclusions about mode stability while generically the conventional treatment does not fully characterize the dynamical system and hence cannot be used to decide about mode stability and well-posedness of the linearized theory. Finally, in Sec.~\ref{sec:dictionary} we describe a scheme for choosing harmonic source functions that drive the coordinates towards familiar gauges in cosmological perturbation theory. We conclude with some general remarks and point towards possible future directions in Sec.~\ref{sec:discussion}.

This paper is intended for both cosmologists as well as the mathematical and numerical general relativity communities. Those familiar with conventional cosmological treatment of perturbations can skip Sec.~\ref{sec:review}. The key results are presented in Secs.~\ref{sec:char}~and~\ref{sec:rel-to-old}.

\section{Cosmological perturbation theory}
\label{sec:review}

Simply put, cosmological perturbation theory is a particular application of the ADM formalism \cite{Arnowitt:1959ah}, fully exploiting the symmetry-properties of a homogeneous and isotropic spacetime to extract observables from the linear theory in an accurate and effective way. 
Its development has a long history, starting with Lifshitz's decomposition theorem of the linearized metric in the 40s \cite{Lifshitz:1945du} and culminating in Bardeen's solution of the gauge problem in 1980 \cite{Bardeen:1980kt}.
The significance of Bardeen's approach for cosmology first became clear when Bardeen, Steinhardt, and Turner applied it to describing super-horizon perturbations of quantum origin in the inflationary universe \cite{Bardeen:1983qw}, even though there were various other approaches to the problem \cite{Mukhanov:1981xt,Guth:1982ec,Hawking:1982cz,Starobinsky:1982ee}. This allowed them to reliably extract observational predictions from the theory. The formalism as further developed by Mukhanov \cite{Mukhanov:1988jd} and Sasaki \cite{Sasaki:1983kd} serves as the basis of all theoretical, computational, and observational cosmology until today, without any significant change. 

Without diminishing its virtues, the goal of this section is to show why cosmological perturbation theory has its limitations when taken out of its original context, namely to link theory with observations. First, we will give a compact review of the basic underlying principles and methods. Then we will present a worked example by applying the scheme to the conformally-coupled ${\cal L}_4$-Horndeski theory defined in Eq.~\eqref{L4-action}. We close the section by pointing out the shortcomings and open issues whose resolution is the subject of this paper.

\subsection{Basics}

In linearizing gravitational field theories, the principle of general covariance translates into a twofold freedom, namely to choose coordinates that describe the background and to separately choose the coordinates that describe the perturbed spacetime. Note that, while both choices are manifestations of the gauge freedom of general relativity, the term `gauge'  in cosmological perturbation theory usually refers to a particular slicing of the perturbed spacetime. Obviously, any pair of such choices introduces a correspondence between the coordinates of the background and the perturbed spacetime. So any change in the slicing conditions of the background and/or perturbed spacetime can be described by the following two coordinate transformations or  a combination thereof: 
\begin{itemize}
\item[-] a transformation of the background coordinates that leaves the correspondence unchanged (fixed gauge) and thus induces a coordinate transformation in the perturbed spacetime; and 
\item[-] a coordinate transformation of the perturbed spacetime  under fixed background slicing (gauge transformation) that induces a change in the correspondence.
\end{itemize}

Cosmological perturbation theory is a particular way of handling both of these aspects of coordinate freedom to characterize physical perturbations on a homogeneous and isotropic   FRW spacetime, given by the line element
\begin{equation}
{\rm d}s^2 = \bar{g}_{00}(t){\rm d}t^2 + a^2(t)\delta_{ij}{\rm d}x^i{\rm d}x^j \,, 
\end{equation}
where $\bar{g}_{00}(t)$ is the homogeneous part of the $00$ metric component and $a(t)$ is the scale factor.
(Throughout, we denote by bar if a quantity is evaluated for the homogeneous FRW background and hence is a function of the time coordinate only.) 
Typically $\bar{g}_{00}$ is set to $-1$ (physical time) or $- a^2$ (conformal time). Unless otherwise noted, we do not fix $\bar{g}_{00}$ for reasons that we explain below in Sec.~\ref{sec:harmonic}.

Cosmological perturbation theory rests on two pillars: 
\begin{enumerate}
\item[-] the scalar-vector-tensor (SVT) decomposition of the linearized metric to reflect the behavior of the linearized metric under coordinate transformations of the FRW background; and
\item[-] the use of gauge-invariant perturbation variables to distinguish physical from unphysical fluctuations and to connect with observations.
\end{enumerate}
Next we will explain how the combination of these two elements enables the study of cosmological scenarios.

\subsubsection{Scalar-vector-tensor decomposition}
\label{sec:SVT}

On generic backgrounds, to describe the evolution of small perturbations, we have to solve the coupled system of Einstein and scalar field partial differential equations. In most cases, this is only possible using computer simulations. 
The SVT decomposition of the linearized metric,\begin{equation}
h_{\mu\nu}(t,{\bf x}) = g_{\mu\nu}(t,{\bf x}) - \bar{g}_{\mu\nu}(t)\,, 
\end{equation}
with $\bar{g}_{\mu\nu}(t)$ being the homogeneous background metric,
takes advantage of the symmetry properties of an FRW space-time, allowing for a particularly economical treatment of the perturbations and making the dynamics in representative cases analytically computable.

In the ADM formulation, the geometrical symmetries of an FRW spacetime reduce to the rotational symmetries of the spatial metric of constant-time hyper-surfaces. These are the symmetries of the 3-d Euclidean group $SO(3)$.  As a Lie group, $SO(3)$ is irreducibly represented by spin-0 scalar, spin-1 vector, and spin-2 tensor harmonics, $Q^{(0)}(x^m)$, $Q_i^{(1)}(x^m)$, and $Q_{ij}^{(2)}(x^m)$, respectively, where $m=1,2,3$. 
The spatial harmonics are solutions of generalized Helmholtz equations,
\begin{equation}
\nabla^2 Q^{(0)} = k^2 Q^{(0)}
,\quad \nabla^2 Q_i^{(1)} = k^2 Q_i^{(1)},\quad \nabla^2 Q_{ij}^{(2)} = k^2 Q_{ij}^{(2)}
,
\end{equation}
where $\nabla^2 \equiv \nabla_m\nabla^m$ is the covariant Laplacian operator and $k$ the wavenumber.
As a consequence, each metric component $h_{\mu\nu}$ can be written as a linear combination of the spatial harmonics with time-dependent coefficients,
\begin{eqnarray}
\label{svt00}
h_{00} &=& 2 \bar{g}_{00}(t) \alpha
\,,\\
h_{0i} &=& \sqrt{-\bar{g}_{00}(t)}\,a(t)\big( \beta_{,i} + B_i \big) 
\,,\\
\label{svtij}
h_{ij} &=& 2\, a^2(t)\Big( -\psi \delta_{ij} + \epsilon_{,ij} + 2S_{(i,j)} + u_{ij} \Big) \,,
\end{eqnarray}
where
\begin{align}
&\alpha \equiv  \alpha(t) Q^{(0)}(x^m) 
\,,\quad
\beta \equiv  \beta(t) Q^{(0)}(x^m) \,, \quad 
\psi \equiv  \psi(t) Q^{(0)}(x^m) \,, \quad 
\epsilon \equiv  \epsilon(t) Q^{(0)}(x^m) 
\,,\\
&B_i \equiv B^{(1)}(t)Q_i^{(1)}(x^m)\,, \quad S_i \equiv S^{(1)}(t)Q_i^{(1)}(x^m)
\,,\\
&u_{ij} \equiv u^{(2)}(t) Q_{ij}^{(2)}(x^m)
\,;
\end{align}
and
\begin{equation}
\label{SVT-constraints}
\partial^i B_i = 0\,; \quad \partial^i S_i = 0\,; \quad u_{ij} = u_{ji}\,; \quad \partial^i u_{ij} = 0\,; \quad u^i_i = 0\,.
\end{equation}

Similarly, the components of the perturbed stress-energy tensor $\delta T_{\mu\nu}$ can be decomposed into scalar, tensor, and vector components: In an FRW universe, at zeroth order, any type of stress-energy takes the form of a `perfect fluid' due to the symmetry properties of the background spacetime. That means, we can characterize the background stress-energy through its energy density $\bar{\rho}(t)$ and pressure $\bar{p}(t)$,
\begin{equation}
\bar{T}_{\mu\nu} = \bar{p}\bar{g}_{\mu\nu} + (\bar{\rho}+\bar{p})\bar{u}_{\mu}\bar{u}_{\nu}\,,
\end{equation}
where $\bar{u}_{\mu}$ is the co-moving velocity (in particular, $u^{\mu}u_{\nu}=1$ to all orders) and in the rest frame of the `fluid' $\bar{u}_{\mu}\equiv (\bar{u}_0, 0,0,0)$.
Accordingly, the perturbed stress-energy tensor is given by
\begin{eqnarray}
\delta T_{00} &=& - \bar{\rho} h_{00} + \delta \rho
\,,\\
\delta T_{0i} &=& \bar{p} h_{0i} - \Big(\bar{\rho}+\bar{p} \Big) \Big( \partial_i\delta u + \delta U_i \Big)
\,,\\
\delta T_{ij} &=& \bar{p} h_{ij} + a^2(t)\Big( \delta p\delta_{ij} + \partial_i\partial_j \pi + 2 \partial_{(i}P_{j)} + \Pi_{ij} \Big)
\,.
\end{eqnarray}
Here, $\delta \rho$ is the linearized energy density, $\delta p$ is the linearized pressure, $\delta u$ is the linearized velocity potential, $\delta U_i$ is the linearized divergenceless velocity vector ($\partial^i \delta U_i = 0$), and $\pi, P_i$, and $\Pi_{ij}$ are the scalar, divergenceless vector ($\partial^i \delta P_i = 0$),  and transverse, traceless tensor components ($\partial^i \Pi_{ij} = 0, \Pi_i^i = 0$) of the anisotropic stress, respectively. As the linearized metric, the components of $\delta T_{\mu\nu}$ can each be separated as a product of spatial harmonics and time-dependent amplitudes.

By construction, the scalar, vector and the  tensor components decouple at linear order. In addition, transforming to Fourier space, modes corresponding to different co-moving wave-numbers $k$ evolve independently such that the Einstein and scalar field equations become ordinary differential equations in time for the time-dependent coefficients of the spatial harmonics.

This is the so-called {\it decomposition theorem} that Lifshitz found in 1945. More precisely, Lifshitz showed in Ref.~\cite{Lifshitz:1945du} that in synchronous gauge ($h_{00}, h_{0i}\equiv 0$) linear perturbations of the FRW metric can be fully characterized according to their transformation properties under spatial rotations: $h_{00}$ transforms as a scalar, $h_{0i}$ transforms as a 3-vector, and $h_{ij}$ transforms as a  3-tensor. Then, he used Helmholtz's theorem to decompose both the vectors and tensors into curl-free and divergence-free parts. In Ref.~\cite{Bardeen:1980kt}, Bardeen generalized the decomposition theorem to arbitrary gauges using the ADM formulation.

\subsubsection{Gauge invariance and algebraic gauge fixing}

The major advantage of the SVT decomposition is to greatly reduce the complexity of the Einstein-scalar PDE system to decoupled ODEs. However, the scheme has to be supplemented with a method that solves the gauge problem, {\it i.e.}, identifies a complete set of  variables that characterizes the linearized system and connects it to observations.
The gauge problem arises due to the freedom to choose the coordinates of the perturbed spacetime, while keeping the background coordinates fixed. 
The result is an ambiguity in the correspondence between the coordinates of the background and the perturbed spacetime; in particular, it is not immediately clear how to  connect with observations since, in an arbitrary gauge, perturbations can reflect physical quantities as well as fictitious modes that are artifacts of the particular slicing. Note, however, that, at linear order in perturbation theory, tensors are invariant under gauge transformations such that the gauge problem only affects the scalar and vector sectors.

The gauge problem was solved in the early 80s by Gerlach/Sengupta \cite{Gerlach:1979rw}, though it was Bardeen who first applied the scheme to cosmology in Ref.~\cite{Bardeen:1980kt}. The proposal was to resolve the ambiguity that arises due to the gauge freedom using {\it gauge-invariant variables}, {i.e.}, linear combinations of the perturbation variables that remain unchanged under infinitesimal coordinate transformations of the perturbed spacetime,
\begin{equation}
t \rightarrow t + \chi^0\,, \qquad x^i \rightarrow x^i + \chi^{,i}\,.
\end{equation}
 The original set of gauge-invariant variables applied to cosmology~\cite{Bardeen:1980kt} are given by
 \begin{equation}
 \Phi = \alpha - a^{-1}\partial_t \big( a \dot{\epsilon} - \beta \big), \quad 
 \Psi = \psi + a^2 H \big( \dot{\epsilon} - a^{-1}\beta \big) \,, \quad
\Sigma_i = \dot{S}_i - a B_i .
 \end{equation}
For simplicity we set $\bar{g}_{00} =-1$ for the remainder of the current section. However, as noted above, it will be essential to release this constraint in Sec.~\ref{sec:harmonic} and below. 
Today, $\Phi$ and $\Psi$ are called the {\it Bardeen variables}. Notably, $\Phi$ is the gauge-invariant Newtonian potential. 
Of course, once the principle is known, one can construct infinitely many gauge invariant variables, for example as a linear combination of just a few simple invariants. 

In addition to the Bardeen variables, there are only a few observationally relevant quantities used in practice, such as the {\it co-moving curvature perturbation}
\begin{equation}
{\cal R} =  \psi - H \delta u \,;
\end{equation}
the {\it energy density perturbation}
\begin{equation}
- \zeta = \psi + H \frac{\delta \rho}{\dot{\bar{\rho}}} \,;
\end{equation}
or, in a gravitational scalar-field theory with scalar components $\phi^I$ $(I=1,...,N)$, the gauge-invariant field perturbation
\begin{equation}
\label{def-Q}
{\cal Q^I} = \delta \phi^I + \frac{\dot \phi^I}{H} \psi \,.
\end{equation}
For a compendium of several gauge invariant quantities and associated slicing conditions see  Ref.~\cite{Kodama:1985bj}.

Obviously, the use of gauge invariant variables allows for putting the field equations into a particularly  simple form. But to follow the linearized dynamics, particular initial conditions can usually only be set in terms of the variables of a given slicing. 
Hence, what happens in practice is that one identifies the relevant scalar (or vector) observable -- a gauge invariant quantity -- and {\it algebraically} fixes the gauge by introducing two algebraic conditions on both the  scalar and vector gauge variables.
For example, in co-moving gauge ($\delta q \equiv 0$), the scalar variable $\psi$ is the co-moving curvature perturbation; in Newtonian gauge ($\beta, \epsilon \equiv 0$), the Bardeen variables coincide with the metric variables $\alpha = \Phi$ and $\psi = \Psi$; in constant density gauge ($\delta \rho \equiv 0$), $\psi$ is the gauge-invariant perturbation of the normalized energy density; and in spatially-flat gauge ($\psi =0$), scalar-field perturbations are gauge-invariant. 

Once a preferred slicing is identified, it is straightforward to reduce the dynamics of the scalar sector to the evolution equation of the relevant gauge-invariant variables;
this is the advantage of using the ADM formulation combined with the SVT decomposition.  
It is well-known that in the ADM formulation two of the Einstein equations are constraints: the $00$-component is the Hamiltonian constraint and can be obtained by varying the action with respect to the lapse $\alpha$; the $0i$-component is the momentum constraint and can be obtained by varying the action with respect to the shift $\beta$. Since any gauge invariant scalar can be expressed as a linear combination of the remaining two scalar metric variables and the three scalar variables of the stress-energy tensor, the gauge freedom can now be used to eliminate two more gauge degrees of freedom and express the system in terms of just three scalar variables.

Throughout this paper, we consider gravitational field theories with a single scalar, though the results can be straightforwardly generalized to multi-component stress-energy. In the case of a single scalar field, algebraic gauge fixing combined with the SVT decomposition to separate the scalar, vector, and tensor sectors reduces the analysis of the linearized theory to the study of only three decoupled ODEs, each  describing the scalar, vector and tensor degrees of freedom by a  single gauge invariant variable, respectively, as we will show in the next subsection.

\subsection{Worked example}

Next we will illustrate the use of cosmological perturbation theory by applying it to characterize scalar perturbations of the linearized conformally-coupled ${\cal L}_4$-Horndeski theory  as given in Eq.~\eqref{L4-action}, first linearizing the equations of motion without gauge fixing and then evaluating them for three of the most common gauges, Newtonian, spatially-flat and unitary.

\subsubsection{Covariant equations of motion}

Varying the action~\eqref{L4-action} with respect to the metric yields the Einstein equations 
\begin{equation}
G_{\mu\nu} = T_{\mu\nu}
\end{equation}
with the stress-energy tensor taking the form
\begin{eqnarray}
\label{Horndeski-Tmunu}
T_{ \mu \nu}  & = & \Big(G_2(X,\phi) + b(\phi)  \nabla _{\mu} \phi \nabla ^{\mu} X - 2 b_{, \phi}(\phi) X ^2 + 2G_{4,\phi\phi}(\phi)X - G_{4,\phi}\Box\phi \Big) g _{\mu\nu} 
\nonumber\\
&+& \Big( G_{2,X}(X,\phi) -b( \phi) \Box \phi - 2 b_{,\phi}(\phi) X + G_{4,\phi\phi}(\phi) \Big)\nabla _{\mu} \phi\, \nabla _{\nu} \phi 
\nonumber\\
 &-&  b( \phi) \Big(\nabla _{ \mu} \phi\, \nabla _{ \nu} X +  \nabla _{ \nu} \phi\, \nabla _{ \mu} X\Big) + G_{4,\phi} \nabla _{\mu} \nabla_{\nu} \phi  - G_4(\phi) G_{\mu\nu}
 \,.
\end{eqnarray}
and variation of the action with respect to the scalar yields the evolution equation for $\phi$,
\begin{eqnarray}
\label{cov-sf}
-G_{2,X}  \Box\phi &=&    \left( G_{2,XX} - 2b_{,\phi} \right) \nabla_{\mu}X\nabla^{\mu}\phi 
- 2X\left( G_{2,X\phi} -b_{,\phi\phi}X \right) + G_{2,\phi}  + \frac12 G_{4,\phi} R 
\nonumber\\
& -& b(\phi) \left( (\Box \phi)^2 - (\nabla_{\mu}\nabla_{\nu}\phi)^2 - R_{\mu\nu}\nabla^{\mu}\phi\nabla^{\nu}\phi \right)
\,.
\end{eqnarray}
For the purposes of this study, we fixed the ${\cal L}_3$-Horndeski interaction as $G_3\equiv - b(\phi)X$ without loss of generality. Notably, Horndeski theories admit second-order equations of motion that protects them from Ostrogradsky ghost instabilities.

\subsubsection{Linearized equations of motion without gauge fixing}

Linearizing Eq.~(\ref{Horndeski-Tmunu}) around space-time and using the SVT decomposition defined in Eqs.~(\ref{svt00}-\ref{svtij}), the scalar part of the linearized Einstein equations for a single Fourier mode with co-moving wavenumber $k$ takes the following form:
\begin{eqnarray} 
\label{0-0-u-simple0}
& & \Big( 6 H\gamma(t) - 3A _h(t)H^2(t) - \rho_K(t)  \Big) \alpha + \gamma(t)\left(3\dot{\psi} + \frac{k^2}{a^2}\sigma \right) +  \frac{k^2}{a^2} A_h(t) \psi  
\qquad\\
&+&  \Big( 3H(t)\gamma(t) -3A_h(t)H^2(t)  - \rho_K(t) \Big)  \delta\dot{u} 
 -   \left( 3\dot{H}(t) \gamma(t) + \frac{ k^2}{a^2}\Big(A_h(t)H(t) - \gamma(t) \Big)  \right) \delta u 
= 0
\,,\nonumber\\
\label{t-i-u-simple0} 
& &  A_h(t) \dot{\psi} 
-  \Big( A_h(t)H(t) - \gamma(t)  \Big) \delta \dot{u} 
= -\gamma(t)\alpha + A_h(t) \dot{H}(t)  \delta u 
\,,\\
\label{offdiag-u-simple0}
&& A_h(t)  (\alpha - \psi - \dot{\sigma} - H\sigma)  =  \dot{A}_h(t) \left(\delta u + \sigma\right)
\,, 
\\
\label{diag-u-simple00}
& & \big(\gamma(t)\alpha\big)^{\cdot} +   3H \big(\gamma(t) \alpha \big) +
A_h(t)\ddot{\psi}  +  \left(3A_h(t)H + \dot{A}_h(t) \right) \dot{\psi} 
\\
&-&  \left( A_h(t)H(t) - \gamma(t)  \right)  \delta\ddot{u}  
+  \left(  \dot{\gamma}(t) - 3H\left( A_h(t)H(t) - \gamma(t)  \right) - 2A_h(t)\dot{H}(t) - \dot{A}_h(t)H(t)
 \right) \delta\dot{u} 
\nonumber\\
&-&  \left( A_h\ddot{H} + \dot{A}_h\dot{H}   + 3A_hH\dot{H} \right)\delta u
 = 0 \,,
 \nonumber
\end{eqnarray}
where Eq.~\eqref{0-0-u-simple0} is the linearized Hamiltonian constraint; Eq.~\eqref{t-i-u-simple0} is the linearized momentum constraint; Eq.~\eqref{offdiag-u-simple0} is the linearized anisotropy equation; and Eq.~\eqref{diag-u-simple00} is the linearized pressure equation;  $\delta u  \equiv - \delta \phi/\dot{\phi} $; and the scalar shear perturbation is defined through
\begin{equation}
\label{shear}
\frac{\sigma(t,{\bf x})}{a(t)}\equiv  a(t)\dot{\epsilon}(t,{\bf x}) - \beta(t,{\bf x}) \,;
\end{equation} 
where $\sigma$ is the scalar component of the linearized shear tensor
\begin{equation}
\sigma_{\mu\nu} = \frac13 K\gamma_{\mu\nu} - K_{\mu\nu}
\,.
\end{equation}

Note that, using the SVT decomposition, the scalar part of the $0i$ and $ij$ field equations are higher than second order and take the form $(...)_{,i}=0$ and $(...)_{,ij}=0$, respectively. The momentum constraint~\eqref{t-i-u-simple0} as well as the anisotropy and pressure equations~(\ref{offdiag-u-simple0}-\ref{diag-u-simple00}) were obtained  by partial integration to eliminate the overall spatial derivatives and by setting the time-dependent integration constant to zero.

The background quantities
\begin{eqnarray}
\label{A_h-def}
A_h(t) &=& 1+\bar{G}_4(\phi)
\,,\\
\label{gamma}
\gamma(t) &=& A_h(t) H(t) - \frac12\left( \bar{b}( \phi)\dot{\phi}^3(t) - \dot{A}_h(t) \right) 
\,,\\
\label{rho_K}
\rho_K(t) &=&  \frac12\bar{G}_{2,X} \dot{\phi}^2 + \frac12 \big(\bar{G}_{2,X X} - 2\bar{b}_{,\phi}  \big) \dot{\phi}^4  + 3H\bar{b}(\phi)\dot{\phi}^3
\end{eqnarray}
are functions of the homogeneous background solution. 
If there is no mixing between the kinetic energy of the scalar and the metric  ($G_3, G_4 \equiv 0$), the function $\rho_K(t)$ that measures the kinetic energy of the field is independent of the metric and the function $\gamma(t)$ that measures the kinetic energy of the metric is field independent, $\gamma(t) = H(t)$.
In ${\cal L}_3$-Horndeski theories and beyond, this is not anymore the case: both $\rho_K$ and $\gamma$ involve metric and scalar kinetic terms.  The direct mixing between the kinetic energy of the scalar and the metric -- a feature also called `braiding' \cite{Deffayet:2010qz} -- can be characterized by the deviation of $\gamma$ from $H$. 
In addition, the conformal quartic Horndeski interaction leads to a mixing between the scalar field and the four-Ricci scalar as measured by the deviation between $A_h$ and unity.

The Hubble parameter and its time derivative are related to the Horndeski interactions through the FRW background equations 
\begin{eqnarray}
\label{FRW1}
3 H^2  &=&  - \bar{G}_2(X,\phi) + \bar{G}_{2,X}(X,\phi) \dot{\phi}^2 - \frac{1}{2} \bar{b}_{,\phi}(\phi) \dot{\phi}^4 + 3 H \bar{b}(\phi) \dot{\phi}^3 \\
&-& 3 \bar{G}_{4,\phi}H\dot{\phi} - 3\bar{G}_4(\phi)H^2 + \bar{\rho}_{\rm matter}
\nonumber
\,, \\
\label{FRW2}
-2 \dot{H}   &=&  \bar{G}_{2,X}(X,\phi) \dot{\phi}^2 - \bar{b}_{,\phi}(\phi) \dot{\phi}^4 + 3H\bar{b}(\phi)\dot{\phi}^3 - \bar{G}_{4,\phi}\dot{\phi}H + \bar{G}_{4,\phi\phi}\dot{\phi}^2
\\
&+& \left( \bar{G}_{4,\phi} - \bar{b}(\phi)\dot{\phi}^2 \right)\ddot{\phi} + 2 \bar{G}_4(\phi)\dot{H} + \left(\bar{\rho}_{\rm matter} + \bar{p}_{\rm matter}\right)
\,;\nonumber
\end{eqnarray}
and, finally, the linearized scalar-field equation is given by 
\begin{eqnarray}
\label{scalar-eq-u0}
&&  \left( \rho_K + 3H \big(A_hH - \gamma\big) \right) \dot{\alpha} -\big( A_hH - \gamma \big)\frac{k^2}{a^2}\alpha
\\
&+& \left( \dot{\rho}_K + 3H\rho_K + ( 6\dot{H} + 9H^2) \big(A_hH - \gamma \big)  
+  3H \big( \dot{A}_hH -\dot{\gamma} \big) \right) \alpha
\nonumber\\
& + & \Big(A_h(t)H(t) - \gamma(t)\Big) \frac{k^2}{a^2}(\dot{\sigma} + H(t)\sigma) + \frac{k^2}{a^2}\left( \dot{A}_hH -\dot{\gamma}\right)\sigma
\nonumber\\
& + &  3\big(A_h(t)H(t) - \gamma(t)\big) \ddot{\psi}   - \dot{A}_h(t) \frac{k^2}{a^2}  \psi 
+ 3 \left(  3H(t) \big(A_h(t)H(t) - \gamma(t)\big) + \dot{A}(t)H(t) -\dot{\gamma}(t) \right) \dot{\psi}
\nonumber\\
&+&   \rho_K(t) \delta\ddot{u} + \left( \dot{\rho}_K(t)  + 3H \rho_K(t)  \right) \delta\dot{u} 
+  \Big( H\big(A_h(t)H(t) - \gamma(t)\big) + 2\dot{A}_hH  -\dot{\gamma}  \Big)  \frac{k^2}{a^2}\delta u
\nonumber\\
&-&  3\left( \big(\dot{A}_hH - \dot{\gamma}\big)\dot{H} + \big(A_hH - \gamma\big) \big(\ddot{H} + 3 H\dot{H}\big)\right)  \delta{u}  
= 0
\,.
\nonumber
\end{eqnarray}
These equations were first obtained in Ref.~\cite{Ijjas:2017pei}; for the derivation see the Appendix of the same paper. Of course, due to the gauge freedom, only three of the five equations are independent. In the following we will utilize exactly this freedom to illustrate the scheme of cosmological perturbation theory.

\subsubsection{Newtonian gauge: $\beta, \epsilon \equiv 0$}
\label{def-newt-gauge}

In the first example, we derive the Newtonian (or zero-shear) gauge equations for the scalar sector.

The Newtonian gauge is defined through the two constraints $\beta, \epsilon \equiv 0$, eliminating two of the five scalar gauge variables $\alpha, \beta, \psi, \epsilon$, and $\delta u$. We use the linearized Einstein equations to eliminate further two of the remaining three scalar gauge variables - the scalar velocity  potential $\delta u$ using the anisotropy equation
\begin{equation}
\label{lin-anis}
  (\dot{A}_h/A_h) \delta u = \Phi - \Psi
\,,
\end{equation}
and the Newtonian potential $\Phi \;(\equiv \alpha)$ using the momentum constraint,
\begin{align}   
 \left(\dot{H} - \frac{\dot{A}_h}{A_h} H - \frac{ k^2}{a^2} \frac{\left( A_hH - \gamma \right)^2}{\det(P)} \right)\left(\Phi - \Psi  \right) = \frac{\dot{A}_h}{A_h}\left(\dot{\Psi} + H \Psi - \frac{k^2}{a^2} A_h  \frac{A_h H - \gamma }{\det(P) }   \Psi \right)
 \,,
\end{align}
where   
\begin{eqnarray}
\det(P) &=& A_h\rho_K +  3\left(A_hH - \gamma  \right)^2  
\end{eqnarray}
is the determinant of the kinetic matrix associated with the ODE system describing the evolution of  $(\Psi, \delta u)$.

As a result, the system of linearized Einstein equations reduces to a single dynamical equation for the Bardeen potential $\Psi \;(\equiv \psi)$: 
\begin{equation}
\label{psi-final-eq}
\ddot{\Psi} + F(t, k) \dot{\Psi} + \left( m_0^2(t, k) + c_S^2(t, k)\frac{k^2}{a^2} + u_H^2(t, k)\frac{k^4}{a^4} \right) \Psi=0
\,.
\end{equation}
The coefficient of the friction term $\propto \dot{\Psi}$ is given by
\begin{align}
\label{friction-final}
F(t, k) \equiv &  \Bigg( \det(P) \left( 
\left(H + \frac{\dot{A}_h}{A_h}\right) \left(-\dot{H} + \frac{\dot{A}_h}{A_h} H \right) - \frac{d}{dt}\left(-\dot{H} + \frac{\dot{A}_h}{A_h} H \right) \right) 
\\
& 
+ \left(   \frac{d}{dt} \ln \frac{a^3\, A_h\, \det(P)}{ \left( A_hH - \gamma \right)^2 }  \right) \left( A_hH - \gamma \right)^2 \frac{ k^2}{a^2} \Bigg) \frac{1}{d(t,k)}
\,.\nonumber
\end{align}
In the limit of large and small $k$, $F(t, k)$ is a function of $t$ only and so, generally, does not affect the characteristic behavior. 

The coefficient of the term $\propto \Psi$ is given by
\begin{align}
\label{m_0-final}
m_0^2(t, k) \equiv &  \Bigg(   2\dot{H} - H\frac{d}{dt}\ln \left(-\dot{H} + \frac{\dot{A}_h}{A_h} H \right)
\Bigg) \left(-\dot{H} + \frac{\dot{A}_h}{A_h} H\right)  \frac{ \det(P)}{d(t,k)}
\, ,\\
\label{c_S-final}
c_S^2(t, k) \equiv& \Bigg(  \left(-\dot{H} + \frac{\dot{A}_h}{A_h} H \right)\left(  \det(P)c_{\infty}^2(t) + 2A_h\left( \dot{\gamma} + (A_hH - \gamma)H  - \frac{d}{dt}(A_hH) \right) \right) 
\\
&+ 2(\dot{H}+H^2)(A_hH - \gamma)^2 + A_h(A_hH - \gamma) \frac{d}{dt} \left(-\dot{H} + \frac{\dot{A}_h}{A_h} H \right) - H \frac{d}{dt} \left( A_hH - \gamma \right)^2
\nonumber\\
&+  \left(A_h(A_hH - \gamma) (-\dot{H} + \frac{\dot{A}_h}{A_h} H) + H(A_hH - \gamma)^2\right) \frac{d}{dt}\ln \det(P) 
\Bigg)\frac{1}{d(t,k)} 
\,,\nonumber\\
\label{u_H-final}
u_H^2(t, k) \equiv&   \frac{1}{d(t,k)} \left( A_hH - \gamma \right)^2 c_{\infty}^2(t) 
\,.
\end{align}
Notice that all coefficients share the common denominator 
\begin{equation}
\label{def-d}
d(t, k) = 
\det(P)\left(-\dot{H} + \frac{\dot{A}_h}{A_h} H \right)  +  \left( A_hH - \gamma \right)^2\frac{ k^2}{a^2}\,.
\end{equation}

Finally, the quantity
\begin{equation}
\label{c_s^2}
c_{\infty}^2(t) \equiv \frac{2\dot{A}_h\gamma  + (A_hH-\gamma)\gamma -A_h\dot{\gamma} }{\det(P)}
\end{equation}
is the square of the propagation speed in the limit of $k\to \infty$.

\subsubsection{Unitary gauge: $\delta u, \epsilon \equiv 0$ / Spatially-flat gauge: $\psi, \epsilon \equiv 0$}

While the Newtonian gauge analysis is already significantly less complex than solving the full coupled system of linearized Einstein equations, choosing the unitary ($\delta u, \epsilon \equiv 0$) and/or spatially-flat ($\psi, \epsilon \equiv 0$) gauges reduces the complexity by another level.

In both of these cases, (since the gauge constraints do not apply to either the lapse $\alpha$ or the shift $\beta$) it is straightforward to eliminate the linearized lapse and the gradient of the linearized shift by using the Hamiltonian and momentum constraints Eqs.~(\ref{0-0-u-simple0}-\ref{t-i-u-simple0}),
\begin{eqnarray}
\label{lapse-eq}
 \alpha &=& \frac{A_h(t)}{\gamma(t)} \left( -\dot{\psi}  + H(t)  \delta \dot{u} + \dot{H}(t)  \delta u \right) - \delta \dot{u} 
 \,,\\
\label{shear-eq}
\frac{k^2}{a^2}\sigma  & = & 
 \frac{r(t) }{\gamma^2(t)} \left( -\dot{\psi}  + H(t)  \delta \dot{u} + \dot{H}(t)  \delta u \right)
+  \frac{k^2}{a^2} \frac{A_h(t)}{\gamma(t)}\left(- \psi + H(t)\delta u \right) - \frac{ k^2}{a^2}\delta u\,,
\end{eqnarray}
where we define
\begin{equation}
\label{scriptA}
r(t) \equiv A_h(t)\rho_K(t) + 3\Big(A_h(t)H(t) -\gamma(t) \Big)^2
\,.
\end{equation}

Substituting the expressions for $\alpha$ and $\sigma$ into the anisotropy equation, 
we obtain a simple second-order differential equation for the gauge-invariant quantity $-\psi + H\delta u$:
\begin{eqnarray}
\label{v-eq}
&&  \frac{{\rm d}^2}{{\rm d}t^2}\Big( -\psi  + H(t)  \delta u  \Big)  
+\frac{d}{dt}\ln \left( a^3(t)A_h(t) \frac{r(t) }{\gamma^2(t)}\right) \frac{{\rm d}}{{\rm d}t}\Big( -\psi  + H(t)  \delta u  \Big)
\qquad\\
&+& \frac{2\dot{A}_h(t)\gamma(t) - A_h(t) \dot{\gamma}(t) - \big(A_h(t)H(t)-\gamma(t)\big)\gamma(t)}{r(t)} \frac{k^2}{a^2} \Big(- \psi + H(t)\delta u \Big)
 = 0\,.
\nonumber
\end{eqnarray}
The sound speed of the modes  is given by 
\begin{equation}
\label{c-zeta}
c_{\zeta}^2(t) = \frac{2\dot{A}_h(t)\gamma(t) - A_h(t) \dot{\gamma}(t) - \big(A_h(t)H(t)-\gamma(t)\big)\gamma(t)}{r(t)}
\,.
\end{equation}
It is immediately apparent that in unitary gauge Eq.~\eqref{v-eq} is the evolution equation for the gauge invariant scalar variable $v\equiv \psi$; and in spatially-flat gauge it is the evolution equation for the gauge invariant scalar variable $v \equiv H \delta u$ (or equivalently, $\delta \phi \equiv - (\dot{\phi}/H)\, v $). 

\subsection{Open issues}

The case of linearized conformally-coupled ${\cal L}_4$-Horndeski makes clear why cosmological perturbation theory has been increasingly popular since its introduction in the early 80s: it connects seemingly complicated gravitational field theories to observations at linear order in an accurate and economical way, by following the evolution of only a few gauge invariant variables.

However, we also chose this example because it enables us to point out some shortcomings of the conventional scheme that will be the focus of the remainder of this paper.
Note that in the case of a canonical scalar or any minimally-coupled $P(X)$-theory where $\gamma \equiv H$, both the Newtonian and unitary/spatially-flat gauges yield the same type of dynamical behavior for the associated gauge variables:
In Newtonian gauge, Eq.~\eqref{psi-final-eq} takes the simple form
\begin{equation}
\label{newt-sec2}
\ddot{\Psi} = \frac{\dot{H}}{\rho_K}\frac{k^2}{a^2}\Psi + \left( \frac{\ddot{H}}{\dot{H}} - H\right)\dot{\Psi} +\left( \frac{\ddot{H} H}{\dot{H}} - 2\dot{H} \right)\Psi
\,;
\end{equation}
and in the spatially-flat/unitary gauges the evolution equation~\eqref{v-eq} reduces to 
\begin{eqnarray}
\label{v-eq-min}
 \frac{d^2}{ d t^2}\Big( -\psi  + H(t)  \delta u  \Big)  
&=& \frac{ \dot{H}(t) }{\rho_K(t)} \frac{k^2}{a^2} \Big(- \psi + H(t)\delta u \Big)
\\
&-&\frac{d}{dt}\ln \left( a^3(t) \frac{\rho_K(t) }{H^2(t)}\right) \frac{d}{d t}\Big( -\psi  + H(t)  \delta u  \Big)
\,.\nonumber\\
\nonumber
\end{eqnarray}

Obviously, in both cases the evolution of each Fourier mode is governed by an ordinary second-order differential equation characterized with a single sound speed $\propto -\dot{H}/\rho_K$ for all wavenumbers. 
On the other hand, introducing braiding ($\gamma \neq H$), the evolution equations in different gauges have different characteristics. (Note that, throughout this paper, we use the expressions `characteristics', `characteristic feature,' `dynamical character,' etc. exclusively as they are being used in mathematics, {\it i.e.}, to describe the dynamical structure of ODEs and PDEs.)

The ambiguity immediately raises the two questions:
\begin{itemize}
\item[i.] what is a reliable approach (and in particular, a proper {\it formulation} of the field equations) to study the characteristics of the PDEs describing gravitational field theories that include modifications of Einstein gravity?
\item[ii.] what is the set of variables that fully characterizes the system in this new formulation and accurately connects it to observations? 
\end{itemize}

The first question concerns the issues of well-posedness and mode stability of gravitational field theories. 
In Sec.~\ref{sec:harmonic}, we will show that addressing these issues requires the implementation of techniques used in non-perturbative mathematical and numerical general relativity. As much as cosmological perturbation theory was ahead of its time in the 80s, the scheme does not capture more recent developments of mathematical and numerical relativity that are essential to reliably determine the dynamical behavior of the system.
By the early 2000s it was widely recognized that the ADM formulation, the basis of the SVT decomposition, is ill-posed in its traditional implementation combined with algebraic gauge fixing.  Hence, we must turn to a different formulation and/or make different gauge choices to correctly analyze the dynamical character of the Einstein-scalar PDE system.

The second question is related to the gauge choice in the context of a new, `well-posed' formulation that replaces the ADM decomposition. As we stressed above, the great advantage of cosmological perturbation theory was to identify physical quantities of the linearized theory. 
But the conventional scheme heavily relied on the ADM form of the field equations  combined with the SVT decomposition of the linearized metric and algebraic gauge fixing. In Sec.~\ref{sec:dictionary}, we will use the insights of the old scheme. By employing the harmonic formulation, we will introduce a protocol to identify dynamical gauge source functions that correspond to common cosmological gauges. Then, using the fact that each of these gauges can be associated with gauge-invariant variables, we can identify harmonic gauge variables with observables of the linear theory. 
In forthcoming work we will show how it is straightforward to generalize our approach non-perturbatively and to lift observables of the linear theory to the fully covariant theory. With the conventional method of cosmological perturbation theory this would not be possible since non-perturbative analyses require the use of well-posed gauges.

\section{Linear perturbation theory in the generalized harmonic formulation}
\label{sec:harmonic}

The first test any classical (or effective) theory must pass is {\it well-posedness} {\it i.e.}, for given initial data, there must exist unique solutions of the linearized theory that depend continuously  on the initial conditions. Short of meeting this criterion, arbitrarily small wavelength mode fluctuations can grow to large amplitudes on arbitrarily small timescales such that it is not meaningful to talk about a predictive theory.

A common strategy for proving linear well-posedness around a given background is to show that the linearized system of PDEs  is  {\it strongly hyperbolic}. Perhaps the best-known example is the wave equation. In the case of covariant PDEs describing gravitational systems, such as the Einstein equations, well-posedness is typically shown by finding a {\it formulation} of the theory, that is strongly hyperbolic. Note, though, that `well-posedness' is a property of differential equations, and not a theory {\it per se}. For example, the Einstein equations are well-posed in generalized harmonic form but generically ill-posed in ADM form. In particular, since cosmological perturbation theory conventionally employs the ADM decomposition with algebraic gauge fixing, it cannot be used to decide well-posedness.

While there is a plethora of modified gravity proposals, to date, we know of only a few well-posed theories; most prominently, Einstein gravity \cite{FouresBruhat:1952zz} and classical supergravity \cite{ChoquetBruhat:1984hn}. Theories that involve higher than second derivatives and cannot be reduced to a second-order system are ill-posed  because they suffer from the so-called {\it Ostrogradsky instability} \cite{Ostrogradsky,Woodard:2015zca}. 
 
The local well-posedness of all Horndeski theories  is an open question. In Ref.~\cite{Papallo:2017qvl}, Papallo and Reall claim that conformally-coupled ${\cal L}_4$-Horndeski theories are linearly well-posed on generic weak-field backgrounds but only a subclass, including Brans-Dicke theories, are non-linearly well-posed for arbitrary initial data. They find that, in their formulation,  the specific gauge condition required for well-posedness of the linearized theory  cannot in general be covariantly lifted and hence well-posedness of the linearized theory does not generalize to the non-linear theory. Ref.~\cite{Papallo:2017qvl} leaves open whether some other scheme can establish non-linear well-posedness.

In this section, we will re-visit this claim. We stress, though, that our motivation is to study the non-linear structure of these theories in cosmological contexts, such as bouncing scenarios, as described in the Introduction. In particular, it is not our goal to provide another local well-posedness argument of the covariant theory on generic backgrounds in the full rigor of a proper mathematical proof. Rather, we will study the well-posedness of the linearized theory on homogeneous backgrounds,  formulate all necessary conditions for strong hyperbolicity, and provide a scheme of gauge fixing readily applicable for non-perturbative, numerical studies.
In the Appendix~\ref{sec:appD}, we will discuss under what conditions our conclusions extend to backgrounds without symmetry assumptions and the precise relationship between our analysis and~Ref.~\cite{Papallo:2017qvl}. 

\subsection{Covariant equations of motion in the generalized harmonic formulation}

A key to our analysis is the application of the generalized harmonic formulation. We will show that, in this decomposition, the initial value problem of the linearized conformally coupled ${\cal L}_4$-Horndeski theory is strongly hyperbolic around homogeneous backgrounds. 

For a PDE system with constant coefficients defined through a complex $n\times n$ matrix ${\cal M}$, the initial value problem 
\begin{equation}
\label{def-PDE}
\partial_t {\bf v}(t,x) = {\cal M} \partial_x {\bf v}(t,x)\, \quad {\bf v}(0,x) = f(x) \,,
\end{equation}
where ${\bf v}(t,x)$ is an $n$-dimensional vector function of time $t$ and space $x$ and $f \in {\cal C}^{\infty}(x)$, is {\it strongly hyperbolic} if all eigenvalues of the matrix ${\cal M}$ are real and there is a complete set of eigenvectors, {\it i.e.}, ${\cal M}$ is diagonalizable; see {\it e.g.}, \cite{Gustafsson:2122877}. (For simplicity of the definition, we only assumed a single spatial dimension.)
The initial value problem  is  {\it weakly hyperbolic} if all eigenvalues of the matrix ${\cal M}$ are real but ${\cal M}$ is not diagonalizable. Note that, for a strongly hyperbolic initial value problem, there are constants $K, \alpha$ such that the solution satisfies the {\it energy estimate}  
\begin{equation}
| {\bf v} (t, x) | \leq K e^{\alpha t} |f(x)| \,.
\end{equation}

More exactly, if the matrix ${\cal S}$ transforms ${\cal M}$ to a diagonal form, any solution of the Fourier transformed system,  
\begin{equation}
\partial_t \tilde{{\bf v}} = i k {\cal M} \tilde{{\bf v}} \,,
\end{equation}
obeys the inequality 
\begin{equation}
| \tilde{{\bf v}} (t, k) | \leq |{\cal S}|\, |{\cal S}^{-1}|\, |\tilde{{\bf v}}(0, k)| \,.
\end{equation}
It is this `mode stability' of the system that ensures robustness against arbitrarily small wavelength perturbations, which is the relevant necessary condition for the non-perturbative, numerical applications in which we are interested. For this reason, and also to distinguish our more pragmatic approach from a strict mathematical proof, we will henceforth use `mode stability'  to characterize our study.

As noted above in the Introduction,  the defining feature of the generalized harmonic formulation is that each of the spacetime coordinates $x^{\mu}$ obeys a scalar wave equation with source function $J^{\mu}$ that itself is a function of the coordinates,
\begin{equation}
\label{harm-coo-def2}
\Box x^{\mu} = J^{\mu}(x^{\alpha})\,.
\end{equation}
This equation can be viewed as a set of four scalar equations, or a single vector-like equation. Note that this relation is not immediately a coordinate choice. Rather, {\it any} known metric $g_{\mu\nu}$ in {\it any} coordinate system can be expressed in generalized harmonic form. The corresponding source functions are then given by evaluating Eq.~\eqref{harm-coo-def2}. 

The generalized harmonic formulation treats the source functions $J^\mu$ as additional degrees of freedom, with Eq.~\eqref{harm-coo-def2} then becoming a set of constraint equations. The reason for doing so is that, as with the harmonic coordinates (where $J^\mu=0$), substituting Eq.~\eqref{harm-coo-def2} into the Einstein equations transforms their principal part into a strongly hyperbolic system of equations for the metric tensor $g_{\mu\nu}$. To close the system, one must specify additional equations for the source functions, which can be viewed as gauge equations. For more details, see, {\it e.g.}, \cite{Garfinkle:2001ni,Pretorius:2004jg,Lindblom:2005qh}. 

In generalized harmonic decomposition, the trace-reversed Einstein equations,
\begin{equation}
R_{\mu\nu} = T_{\mu\nu} - \frac12 g_{\mu\nu}T^{\lambda}_{\lambda}\,,
\end{equation}
take the following form: 
\begin{eqnarray}
\label{trace-rev-EEq-J}
&& \Big(1+ G_4(\phi) \Big)\left(-\frac12 g^{\alpha \beta}g_{\mu \nu}{}_{,\beta\alpha}
-J_{(\nu, \mu)}
- \frac12  g^{\alpha \beta}_{,\mu} g_{\beta \nu}{}_{,\alpha}
- \frac12 g^{\alpha \beta}_{,\nu}  g_{\beta \mu}{}_{,\alpha}
-\Gamma^{\beta}_{\alpha\mu}\Gamma^{\alpha}_{\beta\nu} 
 + \Gamma^{\alpha}_{\mu\nu}J_{\alpha}
\right) 
\qquad\\
&+&   \left(  b( \phi) \left( \phi _{,\mu}  \phi_{,\nu} -\frac12 \phi _{,\alpha}  \phi_{,\beta} g ^{\alpha\beta} g _{\mu\nu} \right) - \frac12 G_{4,\phi}(\phi) g _{\mu\nu}  \right)   \Big( g^{\alpha \beta} \phi_{,\beta\alpha} + J^{\alpha}\phi_{,\alpha} \Big)
\nonumber\\
 &-&  b( \phi) g^{\rho\sigma}  \phi_{,\rho}  \Big( \phi_{,\mu}  \phi_{,\sigma\nu} +  \phi_{,\nu} \phi_{,\sigma\mu}
 - \phi_{,\mu}\Gamma_{\nu\sigma}^{\lambda}\phi_{,\lambda} - \phi_{,\nu}  \Gamma_{\mu\sigma}^{\lambda}\phi_{,\lambda}  \Big)   
 - G_{4,\phi} (\phi)  \left(  \phi_{,\nu\mu} - \Gamma_{\mu \nu}^{\sigma}\phi_{,\sigma} \right) 
\nonumber\\
&-&\left(-G_2(X,\phi) - \frac12 \Big( G_{2,X}(X,\phi)   - G_{4,\phi\phi}(\phi)\Big) g^{\alpha \beta} \phi_{,\alpha} \phi_{,\beta}   \right) g _{\mu\nu} 
\nonumber\\
&-& \Big( G_{2,X}(X,\phi)  + b_{,\phi}(\phi) g^{\alpha \beta} \phi_{,\alpha} \phi_{,\beta}  
+ G_{4,\phi\phi}(\phi) \Big) \phi_{,\mu} \phi_{, \nu} 
= 0
 \,;
 \nonumber
 \end{eqnarray}
and the scalar field equation is given by
\begin{eqnarray}
\label{scalar-field-eq-J-3}
&-&G_{2,X}  \left( g^{\alpha \beta} \phi_{,\beta\alpha} + J^{\alpha}\phi_{,\alpha} \right)  
+    \Big( G_{2,XX} - 2b_{,\phi} \Big) g^{\mu \alpha}g^{\nu \beta}
\phi_{,\alpha} \phi_{,\beta} \Big(\phi_{,\mu\nu} - \Gamma_{\mu\nu}^{\lambda}\phi_{,\lambda}\Big)
\\
& +& b(\phi) \left( \left( g^{\alpha \beta} \phi_{,\beta\alpha} + J^{\alpha}\phi_{,\alpha} \right)^2 - g^{\mu \alpha}g^{\nu \beta} \Big(\phi_{,\mu\nu} -  \Gamma_{\mu\nu}^{\lambda}\phi_{,\lambda}\Big) \Big(\phi_{,\alpha\beta} -  \Gamma_{\alpha\beta}^{\sigma}\phi_{,\sigma}\Big) \right)  
\nonumber
\\
&-& b(\phi) g^{\mu \alpha}g^{\nu \beta} \phi_{,\alpha} \phi_{,\beta} R_{\mu\nu}
- \frac12 G_{4,\phi} R 
- g^{\alpha \beta} \phi_{,\alpha} \phi_{,\beta} \left( G_{2,X\phi} + \frac12 b_{,\phi\phi} g^{\alpha \beta} \phi_{,\alpha} \phi_{,\beta} \right) - G_{2,\phi} 
=0
\,;
\nonumber
\end{eqnarray}
for the derivation see the Appendix~\ref{app:A}.
Throughout, we shall assume that the coupling to the 4-Ricci scalar, $1+G_4(\phi)$, is positive definite for all values of the field $\phi$.

\subsection{Linearized equations of motion in the generalized harmonic formulation}

Next, we shall linearize the system. Keeping in mind that our goal is to evaluate Horndeski theories for cosmological applications, we will perform the linearization around homogeneous backgrounds. We extend our conclusions and comment on applications to generic backgrounds in the Appendix~\ref{sec:appD}.

In the generalized harmonic formulation, components of the linearized trace-reversed Einstein equations, 
\begin{equation}
\label{lin-Eeq.-gen-harm}
\delta R_{\mu\nu} = \delta T_{\mu\nu} - \frac12h_{\mu\nu}\bar{T}^{\alpha}_{\alpha} - \frac12\bar{g}_{\mu\nu}\delta{T}^{\alpha}_{\alpha}
\,,
\end{equation}
take the following form. Note that, for simplicity, we amend the bar convention when referring to the background scalar field, {\it i.e.}, $\bar{\phi} \equiv \phi$ and denote the scalar field perturbation by $\pi$.
\begin{eqnarray}
\label{trace-rev-EEq-J-lin-00}
&& \Big(1+ \bar{G}_4(\phi) \Big)\delta R_{00} 
\quad\\
&-& \frac32 \left(  \bar{b}( \phi) \bar{g}^{00} \dot{\phi}^2  + \bar{G}_{4,\phi}(\phi) \right) \ddot{\pi} 
+ \frac12  \left( \bar{b}( \phi)  \dot{\phi}^2  - \bar{G}_{4,\phi}(\phi) \bar{g} _{00}  \right)  \Big(  \bar{g}^{mn} \pi_{,mn} + \dot{\phi}\delta J^0 \Big)
  \nonumber\\
 &+&  \Big( 2 \bar{b}( \phi) \bar{g}^{00} \dot{\phi}^2 + \bar{G}_{4,\phi} \Big) \dot{\phi} \delta \Gamma_{00}^0 
\nonumber\\
&+& \left( \bar{G}_2(X,\phi) + \frac12 \bar{G}_{2,X} \bar{g}^{00}\dot{\phi}^2 
-  \left( \frac14 \bar{G}_{2,XX}  - \bar{b}_{,\phi} \right) \dot{\phi}^4  \bar{g}^{00} \bar{g}^{00}
-\frac12 \bar{G}_{4,\phi} \bar{J}^0\dot{\phi}  \right) h _{00} 
\nonumber\\
&+&  \frac12 \bar{b}( \phi) \left( 3 \ddot{\phi} -  4\bar{\Gamma}^0_{00} \dot{\phi}  \right)  \dot{\phi}^2 \bar{g}^{00} \bar{g}^{00} h_{00} 
\nonumber\\
&-&  \left( 2 \bar{G}_{2,X} - \left(\frac12 \bar{G}_{2,XX} - 4\bar{b}_{,\phi} \right) \bar{g}^{00}\dot{\phi}^2 + 3 \bar{G}_{4,\phi\phi} 
+ 3 \bar{b}( \phi) \Big( \bar{g}^{00} \ddot{\phi} - \frac12 \bar{J}^0\dot{\phi} - 2 \bar{g}^{00} \bar{\Gamma}^0_{00}\dot{\phi}  \Big)
\right) \dot{\phi} \dot{\pi}
\nonumber\\
&+& \bar{G}_{4,\phi} \left( \bar{\Gamma}_{00}^0 - \frac12 \bar{g} _{00}  \bar{J}^0  \right) \dot{\pi} 
\nonumber\\
&+& \left( \bar{G}_{2,\phi} \bar{g} _{00} 
- \frac12 \left(  \bar{G}_{2,X\phi}    
+ \bar{b}_{,\phi} \left( 3 \bar{g}^{00} \ddot{\phi} 
- 4\bar{g}^{00}  \bar{\Gamma}^0_{00} \dot{\phi} -  \bar{J}^0\dot{\phi}    \right) 
+ 2 \bar{b}_{,\phi\phi} \bar{g}^{00}\dot{\phi}^2 \right)  \dot{\phi}^2 \right) \pi  
\nonumber\\
&+& \bar{G}_{4,\phi}  \left(-\frac12 \bar{g}^{00} \ddot{\bar{g}}_{00}
- \dot{\bar{J}}_0
- \dot{\bar{g}}^{00} \dot{\bar{g}}_{0 0}{}
- \bar{\Gamma}^0_{0 0} \bar{\Gamma}^0_{00} 
- \bar{\Gamma}^{n}_{m 0} \bar{\Gamma}^{m}_{n 0} 
 + \bar{\Gamma}^0_{00}\bar{J}_0
\right) \pi 
\nonumber\\
&-&  \frac12 \left( \bar{G}_{4,\phi \phi} \left( 3 \ddot{\phi} +    \bar{J}_0\dot{\phi} - 2 \bar{\Gamma}_{00}^0 \dot{\phi} \right) + 3 \bar{G}_{4,\phi\phi\phi}  \dot{\phi}^2\right) \pi  
= 0
 \,;
 \nonumber\\
\label{trace-rev-EEq-J-lin-0i}
&& \Big(1+ \bar{G}_4(\phi) \Big) \delta R_{0i} 
\quad\\
&-& \Big( - \bar{b}( \phi) (-\bar{g}^{00})  \dot{\phi}^2 + \bar{G}_{4,\phi}(\phi) \Big) \Big( \dot{\pi}_{,i} - \dot{\phi}\delta \Gamma_{i 0}^0 \Big)
\nonumber\\
 &-&  \bar{b}( \phi)\dot{\phi}^2 \left( - \dot{\phi} \bar{\Gamma}_{i l}^0 (-\bar{g}^{00}) \bar{g}^{lk}h_{0 k}  - \bar{\Gamma}_{i m}^0 \bar{g}^{mn}\pi_{,n} + (-\bar{g}^{00}) \bar{\Gamma}_{i 0}^k \pi_{,k}    \right)
\nonumber\\
&-& \left( \bar{G}_{2,X}(X,\phi) + \bar{b}( \phi) \Big( (-\bar{g}^{00}) \bar{\Gamma}_{00}^0 - \bar{J}^0 \Big)\dot{\phi} + \bar{b}_{,\phi}(\phi)\bar{g}^{00}\dot{\phi}^2 + \bar{G}_{4,\phi\phi}(\phi) \right) \dot{\phi}\, \pi_{, i} 
- \bar{G}_{4,\phi} (\phi) \bar{\Gamma}_{0i}^j\pi_{,j}
\nonumber\\
&+& \left( \bar{G}_2(X,\phi) + \frac12 \Big( \bar{G}_{2,X}    -  \bar{G}_{4,\phi\phi} \Big) \bar{g}^{00}\dot{\phi}^2   
- \frac12   \left(  \bar{b}( \phi)\bar{g}^{00}\dot{\phi}^2 +\bar{G}_{4,\phi} \right) \Big( \bar{g}^{00} \ddot{\phi} + \bar{J}^0\dot{\phi} \Big) \right) h_{0i}  
= 0
 \,;
 \nonumber\\
\label{trace-rev-EEq-J-lin-ij}
&& \Big(1+ \bar{G}_4(\phi) \Big)\delta R_{ij} 
\quad\\
&-&  \frac12 \left(  \bar{b}( \phi) \bar{g}^{00}\dot{\phi}^2 + \bar{G}_{4,\phi}(\phi) \right) \bar{g} _{i j}   \Big( \bar{g}^{\alpha \beta} \pi_{,\beta\alpha} +  \ddot{\phi}(-\bar{g}^{00})\bar{g}^{00}h_{00} + \bar{g}^{00} \bar{J}_0 \dot{\pi} +  \dot{\phi}\delta J^0 \Big)
\nonumber\\
 &-& \bar{G}_{4,\phi} (\phi)  \left(  \pi_{,j i} - \bar{\Gamma}_{i j}^0 \dot{\pi} - \dot{\phi} \delta \Gamma_{i j}^0 \right) 
 \nonumber\\
&-&  \left( - \bar{G}_2(X,\phi) + \frac12 \Big( \bar{G}_{2,X}  - \bar{G}_{4,\phi\phi} \Big)(- \bar{g}^{00})\dot{\phi}^2   
- \frac12 \left(  \bar{b}( \phi)  \bar{g}^{00}\dot{\phi}^2 + \bar{G}_{4,\phi} \right) (-\bar{g}^{00})  \left( \ddot{\phi} + \bar{J}_0\dot{\phi} \right)  \right) h _{i j} 
\nonumber\\
&-& (-\bar{g}^{00}) \bar{g} _{i j} \left( \frac12 \bar{G}_{2,XX}(-\bar{g}^{00})\dot{\phi}^2  
+  \bar{b}( \phi)  (-g ^{00})  \Big(  \ddot{\phi} + \bar{J}_0\dot{\phi} \Big)  
- G_{4,\phi\phi}  \right)  \left(  \frac12 \dot{\phi}^2 (-\bar{g}^{00}) h_{00} + \dot{\phi} \dot{\pi} \right) 
\nonumber\\
&-& \bar{g} _{i j}  \left(-G_{2,\phi} + \frac12 \Big( G_{2,X\phi} -  G_{4,\phi\phi\phi} \Big)  (-\bar{g}^{00})\dot{\phi}^2 
+ \frac12 (-\bar{g}^{00}) \left(  \bar{b}_{, \phi}  (-g^{00})\dot{\phi}^2  - G_{4,\phi \phi}  \right)  \left(  \ddot{\phi} + \bar{J}_0\dot{\phi} \right) \right) \pi
\nonumber\\
&+& \bar{G}_{4,\phi}  \left(-\frac12 \bar{g}^{00} \ddot{\bar{g}}_{i j}
- \bar{\Gamma}^{0}_{m i} \bar{\Gamma}^{m}_{0 j} 
- \bar{\Gamma}^{n}_{0 i} \bar{\Gamma}^{0}_{n j} 
 + \bar{\Gamma}^0_{i j}\bar{J}_0
\right) \pi + \bar{G}_{4,\phi \phi}   \bar{\Gamma}_{i j}^0 \dot{\phi}\, \pi
= 0
 \,.
 \nonumber
\end{eqnarray}
Here bar denotes background quantities that only depend on time $t$ in the homogeneous case. The unperturbed (background) metric,
\begin{equation}
\bar{g}_{\mu\nu} = \left(\begin{array}{cc} \bar{g}_{00} &  0 \\0 &  \bar{g}_{ij}\end{array}\right)\,,\quad  \bar{g}_{\mu\nu, i} =0 \quad {\rm for\; all}\; i=1,2,3\,;
\end{equation}
 and 
 \begin{equation}
 h_{\mu\nu} \equiv g_{\mu\nu} - \bar{g}_{\mu\nu}
 \end{equation}
 is the linear perturbation to $\bar{g}_{\mu\nu}$. Note that we do not impose any constraints on the spatial part of the background metric $\bar{g}_{ij}$.  
In Appendix~\ref{sec:appB}, we provide the expressions for the linearized Ricci tensor $\delta R_{\mu\nu}$ and the linearized connection coefficients $\delta \Gamma_{\alpha\beta}^{\gamma}$.

The linearized scalar field equation takes the form:
\begin{eqnarray}
\label{cov-sf-lin00}
&-&  \left( \bar{G}_{2,X} - 2 \bar{b}(\phi) \bar{g}^{00} \left( \ddot{\phi} +  \bar{J}_0  \dot{\phi} \right) \right) \bar{g}^{00} \left(  \ddot{\pi} + \bar{g}_{00} \bar{g}^{ij} \pi_{,ji}   + \bar{J}_0 \dot{\pi}  
+  \dot{\phi} \delta J_0 - \bar{g}^{00} \left(\ddot{\phi} + \bar{J}_0\dot{\phi} \right) h_{00}  \right)
\;\qquad\\
&+& \left( \left( \bar{G}_{2,XX} - 2\bar{b}_{,\phi} \right)\dot{\phi}^2 - 2\bar{b}(\phi) \left(  \ddot{\phi} - \bar{\Gamma}_ {00}^0 \dot{\phi}  \right) \right) \bar{g}^{00}\bar{g}^{00} \left( \ddot{\pi}  - \bar{\Gamma}_ {00}^0 \dot{\pi} - \dot{\phi} \delta \Gamma_ {00}^0  \right) 
\nonumber\\
&+& 2\bar{b}(\phi)  \bar{g}^{ij}\bar{g}^{kl}  \bar{\Gamma}_ {jl}^0 \dot{\phi} \left(\pi_{,ik}  - \bar{\Gamma}_ {ik}^0 \dot{\pi} - \dot{\phi} \delta \Gamma_ {ik}^0  \right) 
\nonumber\\
&+& 2\left( \bar{G}_{2,XX} - 2\bar{b}_{,\phi} \right) \left( \ddot{\phi}  - \bar{\Gamma}_ {00}^0 \dot{\phi} \right)  \bar{g}^{00}\bar{g}^{00} \left( - \bar{g}^{00}\dot{\phi}^2 h_{00}  
+  \dot{\phi}\dot{\pi} \right) 
- 2 \bar{b}(\phi) \bar{g}^{00} \bar{g}^{00} \bar{R}_{00} \dot{\phi} \dot{\pi}
\nonumber\\
&-& \bar{G}_{2,XX} \bar{g}^{00} \left( \ddot{\phi} +  \bar{J}_0  \dot{\phi} \right) (-\bar{g}^{00})\left(  \frac12 (-\bar{g}^{00})\dot{\phi}^2 h_{00}  +  \dot{\phi}\dot{\pi} \right) 
\nonumber\\
&-&  \left(   \bar{G}_{2,X\phi} - \left( \bar{G}_{2,XX\phi}   - 2\bar{b}_{,\phi\phi}  \right) \bar{g}^{00} \dot{\phi}^2
+ \bar{G}_{2,XXX}   \bar{g}^{00}\bar{g}^{00} \dot{\phi}^2  \left( \ddot{\phi}  - \bar{\Gamma}_ {00}^0 \dot{\phi} \right) \right) \bar{g}^{00}\left( \frac12 (-\bar{g}^{00})\dot{\phi}^2 h_{00}  +  \dot{\phi}\dot{\pi} \right) 
\nonumber\\
&-& \left( \bar{G}_{2,\phi\phi}    
+ \left( \bar{G}_{2,X\phi\phi} + \frac12 \bar{b}_{,\phi\phi\phi}\bar{g}^{00} \dot{\phi}^2  \right) \bar{g}^{00} \dot{\phi}^2
- \left( \bar{G}_{2,XX\phi} - 2 \bar{b}_{,\phi\phi} \right)  \bar{g}^{00}\bar{g}^{00}   \left( \ddot{\phi}  - \bar{\Gamma}_ {00}^0 \dot{\phi} \right)\dot{\phi}^2  \right) \pi
\nonumber\\
&-& \bar{G}_{2,X\phi} \bar{g}^{00} \left( \ddot{\phi} +  \bar{J}_0  \dot{\phi} \right) \pi 
\nonumber\\
& +& \bar{b}_{,\phi}  \bar{g}^{00}\bar{g}^{00} \left( 2\Big( \bar{J}_0 + \bar{\Gamma}^0_{00}\Big) \dot{\phi} \ddot{\phi} +  \Big(\bar{J}_0\bar{J}_0 - \bar{\Gamma}^0_{00}\bar{\Gamma}^0_{00}  \Big)  \dot{\phi}^2 \right) \pi
-  \bar{b}_{,\phi} \bar{g}^{ik}\bar{g}^{jl} \bar{\Gamma}_ {ij}^0\bar{\Gamma}_ {kl}^0 \dot{\phi}^2  \pi
\nonumber\\
&-& \left( \bar{b}_{,\phi} \bar{g}^{00} \bar{g}^{00}  \dot{\phi}^2 \bar{R}_{00} 
+ \frac12 \bar{G}_{4,\phi\phi} \bar{R} \right)\pi 
\nonumber\\
& +& 2 \bar{b}(\phi) \bar{g}^{00}\bar{g}^{00} \left(  \dot{\phi}^2 \bar{R}_{00} 
+  \left(  \ddot{\phi} - \bar{\Gamma}_ {00}^0 \dot{\phi}  \right)^2 \right) \bar{g}^{00}h_{00}
\nonumber\\
&+& 2\bar{b}(\phi) \bar{g}^{kl} \bar{g}^{im}\bar{g}^{jn} \bar{\Gamma}_ {ik}^0\bar{\Gamma}_ {jl}^0 \dot{\phi}^2 h_{mn}
\nonumber\\
&-& \bar{b}(\phi)  \bar{g}^{00} \bar{g}^{00}  \dot{\phi}^2 \delta R_{00}
- \frac12 \bar{G}_{4,\phi} \delta R 
= 0 \,.
\nonumber
\end{eqnarray}
We note that, evaluating in the SVT decomposition, we checked that all Einstein and scalar field equations agree with the results obtained in Ref.~\cite{Ijjas:2017pei} and given above in Eqs.~(\ref{0-0-u-simple0}-\ref{diag-u-simple00}); for details see the Appendix~\ref{sec:appB}~and~\ref{sec:appE}.

\subsection{Mode stability analysis} 
\label{sec:char}

In order to decide whether the linearized system of Einstein and scalar field partial differential equations is stable under mode fluctuations for given initial data, we will perform a characteristic analysis and ask if Eqs.~(\ref{trace-rev-EEq-J-lin-00}-\ref{trace-rev-EEq-J-lin-ij}, \ref{cov-sf-lin00}) satisfy necessary conditions to form a strongly hyperbolic system. Here, we will adapt the canonical analysis as presented, {\it e.g.}, in Refs.~\cite{Sarbach:2012pr,Gustafsson:2122877}.

First, we re-express the linearized system of PDEs~(\ref{trace-rev-EEq-J-lin-00}-\ref{trace-rev-EEq-J-lin-ij}, \ref{cov-sf-lin00}) in matrix form,
\begin{eqnarray}
\label{lin-eq}
A(t) \ddot{\bf v} (t,x^m) &=& 
\sum_{m,n=1}^3 B^{mn}(t)\frac{\partial^2 {\bf v}}{\partial x^m\partial x^n}(t,x^m) 
+ \sum_{m=1}^3 D^{m}(t)\frac{\partial \dot{{\bf v}}}{\partial x^m}(t,x^m)
\\
&+& \sum_{m=1}^3 E^{m}(t)\frac{\partial {\bf v}}{\partial x^m}(t,x^m)
+ F(t) \dot{\bf v} (t,x^m)
+ M(t) {\bf v }(t,x^m)
\,, \nonumber
\end{eqnarray}
where the vector ${\bf v}$ is defined by
\begin{equation}
\mathbf{v} \equiv (h_{00}, h_{0x},h_{0y},h_{0z},h_{xx},h_{xy},h_{xz},h_{yy}, h_{yz}, h_{zz}, \pi)^{\rm T} \in  \mathbb{R}^{11}
\,,
\end{equation} 
and $A, B^{mn}, D^{m}, E^{m}, F$, and $M \in \mathbb{R}^{(11\times11)}$ are each real $11\times11$ matrices. 
The system is second-order in both space and time derivatives and has variable coefficients but we can reduce it to a system that is first-order in time derivatives. Further, we will use the so-called {\it frozen coefficient approximation}, {\it i.e.}, we will treat the system as one with constant coefficients for each fixed coordinate $x_{\mu}$.

 Transforming into Fourier space, 
\begin{eqnarray}
 \ddot{\bf v} &=& 
i \left( |k| \sum_{m=1}^3  \hat{D}^{m}\tilde{k}_m  - i \hat{F} \right) \dot{\bf v} 
\\
&+& \left( - |k| \sum_{m,n=1}^3 \hat{B}^{mn} \tilde{k}_m\tilde{k}_n  + i \sum_{m=1}^3 \hat{E}^{m}  \tilde{k}_m  + \frac{1}{|k|}\hat{M} \right) |k| {\bf v }
\,, \nonumber
\end{eqnarray}
where $\tilde{k}_m = k_m/|k|$, and the hat denotes that a matrix is multiplied by $A^{-1}$, {\it i.e.}, $\hat{B}^{mn} = A^{-1}\times B^{mn}, \hat{D}^{m} = A^{-1}\times D^{m},$ etc., assuming that $A$ is invertible ($\det(A)\neq 0$).

Now, introducing the new variable $u = ( |k| v, -i \dot{v})^{\rm T} \in  \mathbb{R}^{22}$, the PDE can be rewritten as a first-order system,
\begin{equation}
\label{init-valp-def}
\partial_t u = i {\cal P}(ik_m) u
\,,
\end{equation}
where the matrix ${\cal P}$ is given by 
 \begin{equation}
 \label{P-def-full}
{\cal P}(ik_m)  = \left(
\begin{array}{cc}
0& |k| {\mathbb I}_{11}   \\
 |k|  \hat{B}^{mn} \tilde{k}_m\tilde{k}_n  - i  \hat{E}^{m}  \tilde{k}_m  - \frac{1}{|k|}\hat{M} & \quad  |k|  \hat{D}^{m}\tilde{k}_m  - i \hat{F} 
\end{array}
\right)\,.
\end{equation}
Here, ${\mathbb I}_{11}$ is the ${11\times11}$ identity matrix.
The principal part of ${\cal P}$ is defined by
 \begin{equation}
 \label{principal-symbol}
{\cal P}^0 = |k| \left(
\begin{array}{cc}
0&  {\mathbb I}_{11}   \\
 \hat{B}^{mn} \tilde{k}_m\tilde{k}_n   & \;   \hat{D}^{m}\tilde{k}_m  
\end{array}
\right)\,,
\end{equation}
The principal symbol determines mode stability, by definition, {\it i.e.}, for the system Eq.~\eqref{init-valp-def} to be strongly hyperbolic for given initial data it is necessary that ${\cal P}^0$ is diagonalizable.

To show that this is indeed the case here, we next compute the eigenvalues corresponding to the linearized conformally-coupled ${\cal L}_4$-Horndeski in Eqs.~(\ref{trace-rev-EEq-J-lin-00}-\ref{trace-rev-EEq-J-lin-ij}, \ref{cov-sf-lin00}) and show that there is a complete set of eigenvectors. For simplicity, we use the trace-reversed Einstein equations~(\ref{trace-rev-EEq-J-lin-00}-\ref{trace-rev-EEq-J-lin-ij}) to eliminate second derivatives of the metric from the scalar field equation~\eqref{cov-sf-lin00}. Note that reformulating the scalar field equation this way amounts to a simple re-ordering of the PDE system and, hence, does not alter the characteristic structure of the theory; for a proof see the Appendix~\ref{sec:appC}.

Performing the re-ordering and keeping only terms with second derivatives,  Eqs.~(\ref{trace-rev-EEq-J-lin-00}-\ref{trace-rev-EEq-J-lin-ij}, \ref{cov-sf-lin00}) reduce to
\begin{eqnarray}
\label{trace-rev-EEq-00-J-char}
&-& \frac12 \Big(1+ \bar{G}_4(\phi) \Big)  \left( \bar{g}^{00}  \ddot{h}_{00} + \bar{g}^{mn}  h_{00}{}_{,nm} \right)
\quad\\
&-& \frac32 \left(  \bar{b}( \phi) \bar{g}^{00} \dot{\phi}^2  + \bar{G}_{4,\phi}(\phi) \right) \ddot{\pi} 
- \frac12  \left( -\bar{b}( \phi)  \dot{\phi}^2  + \bar{g} _{00} \bar{G}_{4,\phi}(\phi)   \right)   \bar{g}^{mn} \pi_{,mn} 
+ ...  
= 0
 \,;
 \nonumber\\
\label{trace-rev-EEq-0i-J-char}
&-& \frac12 \Big(1+ \bar{G}_4(\phi) \Big) \left( \bar{g}^{00}  \ddot{h}_{0i} + \bar{g}^{mn}  h_{0i,nm} \right) 
- \Big( - \bar{b}( \phi) (-\bar{g}^{00})  \dot{\phi}^2 + \bar{G}_{4,\phi}(\phi) \Big) \dot{\pi}_{,i} 
+ ... = 0
 \,;
\\
\label{trace-rev-EEq-ij-J-char}
&-& \frac12 \Big(1+ \bar{G}_4(\phi) \Big)  \left( \bar{g}^{00}  \ddot{h}_{ij} + \bar{g}^{mn}  h_{ij,nm} \right)
\quad\\
&-&  \frac12 \left(  \bar{b}( \phi) \bar{g}^{00}\dot{\phi}^2 + \bar{G}_{4,\phi}(\phi) \right) \bar{g} _{i j}   \Big( \bar{g}^{00} \ddot{\pi} + \bar{g}^{mn} \pi_{,nm}  \Big)
- \bar{G}_{4,\phi} (\phi)   \pi_{,j i}  
+ ... = 0
 \,;
  \nonumber\\
 \label{scalar-field-eq-J-2-char}
&& \Big( \bar{G}_{2,X} + 2 \bar{b}(\phi) \Big(  \bar{J}_0  + \bar{\Gamma}_ {00}^0 \Big) (-\bar{g}^{00}) \dot{\phi}
+ \Big( \bar{G}_{2,XX} - 2\bar{b}_{,\phi} \Big)(-\bar{g}^{00}) \dot{\phi}^2  
\Big)  (-\bar{g}^{00}) \ddot{\pi}  
\qquad \\
&+& \frac32 \Big(1+ \bar{G}_4(\phi)\Big)^{-1} \left( \bar{b}(\phi) (-\bar{g}^{00})\dot{\phi}^2 - \bar{G}_{4,\phi} \right)^2  (-\bar{g}^{00}) \ddot{\pi}
\nonumber\\
&-& \left( \bar{G}_{2,X} + 2 \bar{b}(\phi) (-\bar{g}^{00}) \left( \ddot{\phi} +  \bar{J}_0  \dot{\phi} \right) \right) \bar{g}^{mn} \pi_{,nm}  
+   2\bar{b}(\phi)  \bar{g}^{mk}\bar{g}^{nl}  \bar{\Gamma}_ {kl}^0 \dot{\phi} \pi_{,mn}  
\nonumber\\
&-&  \frac12 \Big(1+ \bar{G}_4(\phi)\Big)^{-1} \left( \bar{b}(\phi) \bar{g}^{00}  \dot{\phi}^2 
  \left( -\bar{b}( \phi) \bar{g}^{00}\dot{\phi}^2  + \bar{G}_{4,\phi}  \right)  
  + \bar{G}_{4,\phi}  \left( \bar{b}(\phi) \bar{g}^{00}\dot{\phi}^2 + 3\bar{G}_{4,\phi} \right)
  \right) \bar{g}^{mn} \pi_{,mn} 
\nonumber\\
&+&  ... 
= 0 \,;
\nonumber
 \end{eqnarray}
where $...$ stands for lower than second order terms that do not contribute to the principal symbol. The coefficient matrices corresponding to second-derivative terms and hence relevant for the principal symbol take an upper-triangular form,
 \begin{equation}
 \label{coeff-matrices}
A(t) = \left(
\begin{array}{cc}
 A_{h}  & \; A_{h \pi}  \\
0 & \; A_{\pi}
\end{array}
\right), \quad B^{mn}(t) = \left(
\begin{array}{cc}
B_{h}^{mn} & \; B_{h \pi}^{mn}  \\
0 & \; B_{\pi}^{mn}
\end{array}
\right),\;  
D^m(t) = \left(
\begin{array}{cc}
0 & \; D _{h \pi}^m  \\
0 & \; 0
\end{array}
\right)\,.
\end{equation}
Here 
$A_{h}$ and $B_{h}^{mn}$ are $10\times10$ real diagonal matrices;  $A_{h \pi}, B^{mn}_{h \pi},$ and $D _{h \pi}^m$ are 10-dim real vectors; and $A_{\pi},$ and $B_{\pi}^{mn} \in \mathbb{R}$ are real scalars. For $A$ to be invertible, we must require that $1+G_4\neq0, A_{\pi} \neq0$.

It is immediately apparent from the matrix representation in what way the PDE structure of Horndeski theories deviates from Einstein gravity with a minimally-coupled scalar field: in the Horndeski case, $A_{h \pi}, B^{mn}_{h \pi}$, and $D _{h \pi}^m$ are non-zero; the `braiding' effect manifests itself through the non-zero off-diagonal terms.  In the case of Einstein gravity, the same quantities are all zero such that the matrices $A, B^{mn}$ are diagonal and the matrix $D^m\equiv0$.

The characteristic polynomial $\chi(\lambda)$ corresponding to ${\cal P}^0$ as given in Eq.~\eqref{principal-symbol} takes the simple form
\begin{equation}
\chi(\lambda) \equiv \det | \lambda\, {\mathbb I}_{22} - {\cal P}^0 | = \Big( \lambda^2 - (-\bar{g}_{00}) \bar{g}^{mn} \tilde{k}_m \tilde{k}_n \Big)^{10} \left( \lambda^2 - \frac{B^{mn}_{\pi}}{A_{\pi}} \tilde{k}_m \tilde{k}_n \right)\,.
\end{equation}
That means, the eigenvalues of ${\cal P}^0$, 
\begin{eqnarray}
\lambda^{\pm} &=& \pm \sqrt{(-\bar{g}_{00}) \bar{g}^{mn} \tilde{k}_m \tilde{k}_n} \,,\\
c_S^{\pm} &=& \pm \sqrt{A_{\pi}^{-1}B^{mn}_{\pi} \tilde{k}_m \tilde{k}_n }\,,
\end{eqnarray}
are all real if and only if 
\begin{equation}
\label{weak-hp}
A_{\pi}^{-1} B^{mn}_{\pi}\tilde{k}_m \tilde{k}_n > 0.
\end{equation}
It follows that the system is weakly hyperbolic when Eq.~\eqref{weak-hp} is satisfied.

The eigenvectors corresponding to $\lambda^{\pm}$ are inherited from Einstein gravity and, in accordance with the well-posedness of the Einstein equations, the associated twenty eigenvectors are linearly independent and bounded, given reasonable assumptions on the background metric; for details see the Appendix~\ref{sec:appD}.

The eigenvectors corresponding to the remaining two eigenvalues $c_S^{\pm}$ take the form
\begin{equation}
\label{pi-ev-hom}
{\bf s}^{\pm} = \left(v_{tt}^{\pm}, ...\,, v_{zz}^{\pm}, 1/c_S^{\pm} , w_{tt}^{\pm}, ...\,, w_{zz}^{\pm},1 \right)\,,
\end{equation}
where
\begin{eqnarray}
\label{v-pm-h}
v_{\mu\nu}^{\pm} &=& \frac{   c_S^{\pm} A_{h \pi}^{\mu\nu} + (1/c_S^{\mp} )  \, B_{h \pi}^{\mu\nu} - D_{h \pi}^{\mu\nu}
}{  c_S^2 \, \bar{g}^{00} +  \bar{g}^{mn} \tilde{k}_m \tilde{k}_n
   }\,,\quad \\
\label{w-pm-h}
w_{\mu\nu}^{\pm} &=& \frac{ c_S^2 A_{h \pi}^{\mu\nu}  - B_{h \pi}^{\mu\nu} -  c_S^{\mp} D_{h \pi}^{\mu\nu} 
}{
 c_S^2\, \bar{g}^{00} +  \bar{g}^{mn} \tilde{k}_m \tilde{k}_n  
   }\,;\quad 
\end{eqnarray}
and the coefficients $A_{h \pi}^{\mu\nu}, B_{h \pi}^{\mu\nu}, D_{h \pi}^{\mu\nu}$ can be read off from the perturbed Einstein equations~(\ref{trace-rev-EEq-J-lin-00}-\ref{trace-rev-EEq-J-lin-ij}).
Both eigenvectors ${\bf s}^{\pm}$ are linearly independent and finite if $c_S^{\pm} \neq 0$ and the denominator of $v_{\mu\nu}^{\pm}$ and $w_{\mu\nu}^{\pm}$ is non-zero, {\it i.e.},
\begin{equation}
\label{denom-cond}
c_S^2 \neq (-\bar{g}_{00})  \bar{g}^{mn} \tilde{k}_m \tilde{k}_n \,.
\end{equation}
This condition is straightforward for cosmological backgrounds of interest. For example, for an FRW background, it is equivalent to choosing a time coordinate such that Eq.~\eqref{denom-cond} is satisfied.

Together with the 20 eigenvectors corresponding to $\lambda^{\pm}$, ${\bf s}^{\pm}$ form a complete set, for details see the Appendix~\ref{sec:appD}.
In particular, the principal symbol  ${\cal P}^0$ is diagonalizable and hence the initial value problem for the linearized Horndeski theory around homogeneous backgrounds as in Eqs.~ (\ref{trace-rev-EEq-J-lin-00}-\ref{trace-rev-EEq-J-lin-ij}, \ref{cov-sf-lin00}) is strongly hyperbolic in the frozen coefficient approximation and choosing any generalized harmonic source function.

This result is significant because it implies that arbitrarily small wavelength mode fluctuations do not grow to large amplitudes on arbitrarily small timescales, which is the {\it sine qua non} of any non-perturbative, numerical application.
Notably, strong hyperbolicity holds for more general backgrounds and for any generalized harmonic source function; for details see the Appendix~\ref{sec:appD}.

\section{Relation to cosmological perturbation theory}
\label{sec:rel-to-old}

We close our analysis by connecting our results obtained using the generalized harmonic formulation to cosmological perturbation theory. 

First, we explain why mode stability of the coupled Einstein-scalar field PDE system cannot be studied in the SVT decomposition and discuss what the notion of `gradient instability' in earlier studies  (see, {\it e.g.}, \cite{Hsu:2004vr,Creminelli:2006xe,deRham:2017aoj}) actually describes.
In particular, we contrast the implications for theories with scalar fields minimally coupled to Einstein gravity and scalar field theories that involve ${\cal L}_3$-Horndeski modifications to Einstein gravity and beyond.

Second, we briefly outline how the cosmological gauges can be defined using generalized harmonic source functions. We will provide a detailed analysis including worked examples in forthcoming publications \cite{Ijjas-to-appear}. 

\subsection{Mode stability and SVT decomposition}

As presented above in Sec.~\ref{sec:SVT}, the underlying idea of the SVT decomposition was to reduce the study of the coupled linearized Einstein-scalar field PDE system to the study of decoupled ODEs that all describe the time evolution of  amplitudes corresponding to different co-moving wavenumbers for scalar, vector, and tensor perturbations. Such a simplification is perfectly reasonable and fruitful when it comes to extracting observables from linearized Einstein gravity with minimally coupled scalars for FRW spacetimes, as was the original purpose of the scheme. But, when it comes to  analyses of mode stability for a PDE system, {\it i.e.}, verifying that arbitrarily small wavelength mode fluctuations do not grow to large amplitudes on arbitrarily small timescales, the SVT decomposition has a threefold shortcoming:
\begin{itemize}[leftmargin=*]
\item[-]First, and most obviously, the SVT decomposition combined with algebraic gauge fixing {\it cannot} be used to study the characteristic structure or mode stability of the associated Einstein-scalar PDE system, by construction. 
\item[-]
Second, such a simplification is only possible due to the symmetry properties of FRW spacetimes. More generic spacetimes without special symmetry properties do not admit a corresponding basis. Hence, unlike the harmonic formulation, the SVT decomposition is limited to linearizing around FRW spacetimes, by construction. 
\item[-]
Third, due to replacing  linearized metric components by spatial derivatives of scalars,  ODEs describing the amplitudes  of scalar and vector modes are generically higher than second order; for details see the Appendix~\ref{sec:appE}. In practice, the higher derivatives are removed by a combination of algebraic gauge fixing ({\it e.g.}, setting $\epsilon = 0$), elimination of gauge variables using the Hamiltonian and momentum constraints, and integration of the $0i$- and $ij$-components that each take the form $(...)_{,i}=0$ and $(...)_{,ij}=0$, until there remain only three decoupled ODEs each describing the evolution of the scalar, vector, and tensor amplitudes, respectively.
\end{itemize}

At the same time, having completed a proper mode stability analysis of the linearized Einstein-scalar PDE system using the generalized harmonic formulation, we can relate earlier `stability' analyses of the decoupled ODEs using cosmological perturbation theory to mode stability of the PDE system. Below, we will show the implications both for Einstein gravity ($G_i \equiv 0, i\geq3$) and ${\cal L}_3$-Horndeski modifications to Einstein gravity and beyond ($G_3 \neq 0$).

\subsubsection{Einstein gravity with minimally coupled scalars ($G_i(X,\phi) \equiv 0, i\geq3$)}

Around a homogeneous FRW background, the linearized trace-reversed Einstein equations take the form,
\begin{eqnarray}
\label{trace-rev-EEq-h-min}
 \delta R_{\mu\nu} &=&
 \bar{G}_{2,X}  \big( \pi_{,\mu} \phi_{, \nu} + \phi_{,\mu} \pi_{, \nu} \big)		
+ \frac12 G_{2,XX}  \bar{g}^{00} \dot{\phi}^3 \left( \frac12  \bar{g}^{00}  \bar{g} _{\mu\nu} 
- \delta_{0}^{\mu}\delta_{0}^{\nu}\right)  \Big ( 2\dot{\pi} +  \dot{\phi} (- \bar{g}^{00})h_{00}   \Big)
\qquad 
\\
&-& \left(  
\left( \bar{G}_{2,\phi} + \frac12 G_{2,X\phi}  \bar{g}^{00} \dot{\phi}^2  \right) \bar{g} _{\mu\nu}  
- \bar{G}_{2,X\phi} \dot{\phi}^2 \delta_{0}^{\mu}\delta_{0}^{\nu} \right)  \pi
- \left( \bar{G}_2 + \frac12 \bar{G}_{2,X}  \bar{g}^{00} \dot{\phi}^2  \right) h _{\mu\nu} 
 \,;
 \nonumber 
 \end{eqnarray}
 and the linearized scalar-field equation is given by
\begin{eqnarray}
\label{cov-sf-lin00-min}
&-&  \bar{G}_{2,X}  \bar{g}^{00} \left(  \ddot{\pi} + \bar{g}_{00} \bar{g}^{ij} \pi_{,ji}   + \bar{J}_0 \dot{\pi}  +  \dot{\phi} \delta J_0 - \bar{g}^{00} \left(\ddot{\phi} + \bar{J}_0\dot{\phi} \right) h_{00}  \right)
\;\qquad\\
&+&  \bar{G}_{2,XX} \dot{\phi}^2  \bar{g}^{00}\bar{g}^{00} \left( \ddot{\pi}  - \bar{\Gamma}_ {00}^0 \dot{\pi} - \dot{\phi} \delta \Gamma_ {00}^0  \right) 
\nonumber\\
&+& 2 \bar{G}_{2,XX}\left( \ddot{\phi}  - \bar{\Gamma}_ {00}^0 \dot{\phi} \right)  \bar{g}^{00}\bar{g}^{00} \left( - \bar{g}^{00}\dot{\phi}^2 h_{00}  
+  \dot{\phi}\dot{\pi} \right) 
\nonumber\\
&-& \bar{G}_{2,XX} \bar{g}^{00} \left( \ddot{\phi} +  \bar{J}_0  \dot{\phi} \right) (-\bar{g}^{00})\left( - \frac12 \bar{g}^{00}\dot{\phi}^2 h_{00}  +  \dot{\phi}\dot{\pi} \right) 
\nonumber\\
&-&  \left(   \bar{G}_{2,X\phi} -  \bar{G}_{2,XX\phi} \bar{g}^{00} \dot{\phi}^2
+ \bar{G}_{2,XXX}   \bar{g}^{00}\bar{g}^{00} \dot{\phi}^2  \left( \ddot{\phi}  - \bar{\Gamma}_ {00}^0 \dot{\phi} \right) \right) \bar{g}^{00}\left( \frac12\dot{\phi}^2 (-\bar{g}^{00}) h_{00}  +  \dot{\phi}\dot{\pi} \right) 
\nonumber\\
&-& \left( \bar{G}_{2,\phi\phi}    
+ \bar{G}_{2,X\phi} \bar{g}^{00} \left( \ddot{\phi} +  \bar{J}_0  \dot{\phi} \right)
+ \bar{G}_{2,X\phi\phi} \bar{g}^{00} \dot{\phi}^2
- \bar{G}_{2,XX\phi} \bar{g}^{00}\bar{g}^{00}   \left( \ddot{\phi}  - \bar{\Gamma}_ {00}^0 \dot{\phi} \right)\dot{\phi}^2  \right) \pi
= 0 \,.
\nonumber
\end{eqnarray}
It is apparent that, in matrix representation, both coefficient matrices $A, B^{mn}$ entering the principal symbol as introduced in Eq.~\eqref{lin-eq} are diagonal and $D^m \equiv 0$.
The condition for weak hyperbolicity~\eqref{weak-hp} reduces to $c_S^2 = A_{\pi}^{-1}B_{\pi}^{mn}\tilde{k}_m\tilde{k}_n \geq 0$. Since $1/c_S^{\pm}$ is the only non-trivial entry in the $\pi$-eigenvectors given above in Eq.~\eqref{pi-ev-hom}, weak hyperbolicity implies strong hyperbolicity if we {\it additionally} demand $c_S^{\pm} \neq 0$, as otherwise the eigenvectors would blow up. 

In cosmological perturbation theory, from the three ODEs characterizing the scalar, vector, and tensor amplitudes of the linearized Einstein-scalar field system in SVT decomposition, the only non-trivial evolution equation is the Mukhanov-Sasaki equation~\eqref{v-eq-min},
\begin{eqnarray}
\label{Muk-Sas-eq}
 \frac{d^2}{ d t^2}\Big( -\psi  + H(t)  \delta u  \Big)  
&=& \frac{ \dot{H}(t) }{\rho_K(t)} \frac{k^2}{a^2} \Big(- \psi + H(t)\delta u \Big)
\\
&-&\frac{d}{dt}\ln \left( a^3(t) \frac{\rho_K(t) }{H^2(t)}\right) \frac{d}{d t}\Big( -\psi  + H(t)  \delta u  \Big)
\,;
\nonumber
\end{eqnarray}
where the Mukhanov variable $v \equiv -\psi  + H(t)  \delta u$ (with $\delta u \equiv - \pi/\dot{\phi}$) is invariant under infinitesimal coordinate transformations.
It is straightforward to see that the Mukhanov-Sasaki equation has the same characteristic structure as the $\pi$-equation, either by gauge fixing ($\psi = 0$) or by direct comparison of the coefficients  of second-derivative terms while setting $\bar{g}_{00}=-1, \bar{g}_{ij}= a^2(t) \delta_{ij}$ in Eq.~\eqref{cov-sf-lin00-min},
\begin{equation}
A_{\pi}^{-1}B_{\pi}^{mn}\tilde{k}_m\tilde{k}_n = \frac{ \bar{G}_{2,X}}{\bar{G}_{2,X} + \bar{G}_{2,XX}\dot{\phi^2}} = - \frac{ \dot{H}(t) }{\rho_K(t)} 
\,.
\end{equation}
Hence, the Mukhanov-Sasaki equation in the case of Einstein gravity with minimally-coupled scalar fields can indeed be used to analyze the characteristic structure of the PDE: Requiring positivity of the term $\propto k^2$ (called `no gradient instability' in the cosmology literature) in this ODE is sufficient to ensure mode stability of the linearized PDE system. Conversely, negativity of the  term $\propto k^2$ (called `imaginary sound speed') implies that both the Mukhanov-Sasaki equation and the linearized Einstein-scalar field PDE system turn {\it elliptic}. In this latter case, since we are interested in solving an initial value problem, uniqueness of the solution is immediately lost and arbitrarily small perturbations can carry away the system from the background solution. This fact invalidates a common claim made in the literature that an imaginary sound speed in Eq.~\eqref{Muk-Sas-eq} can lead to healthy cosmological scenarios provided the sound speed remains imaginary for  a sufficiently short time \cite{Creminelli:2006xe,deRham:2017aoj}. Quite the opposite, as soon as the sound speed turns imaginary, the theory instantaneously becomes ill-posed.

Note that this relation between the Mukhanov-Sasaki ODEs and the mode stability of the PDE system was not obvious without  actually doing the full principal symbol analysis, as first done here. Nor is it generally the case, as will be shown next.
Furthermore, the mode stability analysis provides a deeper understanding of the role of $c_S^{\pm}$. In fact, while cosmologists conventionally cite $c_S^{\pm}\neq0$ as a necessary condition for canonical quantization of scalar amplitudes, the truth is that non-zero $c_S^{\pm}$ is already required for strong hyperbolicity of the purely classical system.

\subsubsection{${\cal L}_3$-Horndeski theories and beyond ($G_3(X, \phi) \neq 0$)}

Similar to the case of Einstein gravity with minimally coupled scalars, the Horndeski version of the Mukhanov-Sasaki equation~\eqref{v-eq} reflects the characteristic structure of the linearized scalar field ODE in ${\cal L}_3$-Horndeski theories and beyond when perturbing around an FRW background, as can be straightforwardly verified by direct comparison of the expressions. 

However, as we have seen above in Sec.~\ref{sec:harmonic}, one way the `braiding' effect manifests itself is by its altering the coefficient matrices $A, B^{mn}$, and $D^m$ from being strictly diagonal matrices to upper-triangular matrices with non-trivial off-diagonal components due to non-zero $A_{h\pi}, B^{mn}_{h\pi}, D^m_{h\pi}$. The presence of these terms  changes the structure of the principal symbol and the two $\pi$-eigenvectors, and introduces new constraints on their boundedness. For this reason, the condition that the two eigenvalues $c_S^{\pm}$ associated with the linearized scalar field equation be real and non-zero   only ensures weak hyperbolicity of the initial value problem when perturbing around FRW backgrounds. But it does {\it not} satisfy the necessary conditions for strong hyperbolicity. This is a crucial difference from the case of Einstein gravity with minimally coupled scalar fields.

Another way the `braiding' effect manifests itself is the explicit $\phi, \dot{\phi}$-dependence of the coefficient matrix $E^m$ describing the PDE system of Einstein and scalar field equations~\eqref{lin-eq}. It is immediately apparent from Eq.~\eqref{P-def-full} that the $E^m$ matrix governs the dynamics of long-wavelength modes. 
That means, while $E^m$ does not enter the principal symbol and is thus negligible when studying mode stability under arbitrarily small wavelength fluctuations, it is indispensable to include $E^m$ for physical applications since it is this matrix that determines the dynamics on long-wavelengths of cosmological interest and, depending on the model, possibly all scales larger than the Planck length.

A corollary is that the popular effective field theory (EFT) formulation of these theories, see {\it e.g.}, \cite{Creminelli:2016zwa}, using ADM slicing combined with unitary or spatially-flat gauge choice provides insufficient information to ensure linear well-posedness around FRW backgrounds. In particular, simply demanding that the coefficient of the gradient term (or the `sound speed') be positive in the $\zeta$ or $\pi$-action is not enough to prevent arbitrarily small wavelength fluctuations from carrying the system away from the background solution.
More than that, by reducing the analysis to the unitary (or spatially-flat) gauge variable $\zeta$ (or $\pi$), as is standard in EFT analyses, only the $|k|\to\infty$ limit of the theory is being studied while no proper account is taken of the dynamical behavior at macroscopic wave-lengths, {\it i.e.}, in the realm where the EFT is supposed to be valid.  

\subsection{Harmonic source functions for cosmological gauges}
\label{sec:dictionary}

The fact that cosmological perturbation theory is ill-suited for mode stability analyses is not surprising, as it was never developed for this purpose and especially not for evaluating the characteristic behavior of modified gravity theories. Rather, as we emphasized in Sec.~\ref{sec:review}, the real utility of the concept has been to extract observables of linearized scalar field theories minimally coupled to Einstein gravity in a particularly economical way. As we have shown, the harmonic formulation, on the other hand, provides a scheme to study mode stability and is thus well-suited for numerical implementation. But it is not immediately obvious how to extract observables within our harmonic scheme, whether from the linearized theory or non-perturbatively.

For completeness, we describe what the harmonic source functions are in terms of gauge invariant variables for several commonly used cosmological gauges on linearly perturbed FRW backgrounds. This will be helpful in devising dynamical gauge equations for the source functions that can evolve a spacetime along slicings similar to the corresponding cosmological gauges. A detailed study including fully worked examples will be given in forthcoming work; see  \cite{Ijjas-to-appear}.

\subsubsection{Basic strategy}

A generalized harmonic gauge is fixed by the four source functions $J_{\mu}$ $(\mu = 0, ..., 3)$ defined in Eq.~\eqref{def-harm-coo}. Conversely, each component of a harmonic source function can be expressed through elements of the metric; for details see the Appendix~\ref{sec:appB}.  We will exploit this latter feature to choose harmonic source functions for the linearized theory. 

As with other covariant quantities, $J_{\mu}$ can be defined perturbatively, order-by-order as 
\begin{equation}
J_{\mu} = \bar{J}_{\mu} + \delta J_{\mu} + \delta J_{\mu}^{(2)} + ...\,,
\end{equation}
where $\delta J_{\mu}$ is the linearized harmonic source function, $\delta J_{\mu}^{(2)}$ the second-order correction, etc. In particular, to fix the gauge at linear order in perturbation theory, we only need to define $\delta J_{\mu}$.

Using the SVT decomposition of the linearized metric, the linearized harmonic source functions can be expressed as follows,
\begin{eqnarray}
\label{SVT-J0-M}
\delta J_0 &=& 
-  \dot{\alpha}  - 3 \dot{\psi} +  \delta^{kl} \left(\dot{\epsilon} - a^{-1} \sqrt{-\bar{g}_{00}}\,\beta \right)_{,lk} 
\,,\\ 
\label{SVT-Ji-M}
\delta J_i &=&  \left(\alpha - \psi - \delta^{kl}\epsilon_{,lk}  \right)_{,i} 
  -   a\sqrt{-\bar{g}^{00}}\, \left(\dot{\beta}
-  \left(H  - \frac12 \dot{\bar{g}}_{00} \bar{g}^{00} \right) \beta  \right)_{,i} 
 \\ 
 &-&   a\sqrt{-\bar{g}^{00}}\, \left( \dot{B}_i 
-  \left(H  - \frac12 \dot{\bar{g}}_{00} \bar{g}^{00} \right)  B_i  \right)
- 2  \delta^{kl} S_{i,lk}
\,; \nonumber
\end{eqnarray}
for the derivation see the Appendix~\ref{sec:appE}. Now we are ready to use cosmological perturbation theory to find the expressions for the linearized harmonic source functions.

More precisely, to find the linearized harmonic source function for a given cosmological gauge and perturbation, perform following steps:
\begin{enumerate}
\item after fixing the gauge using cosmological perturbation theory, express each scalar and vector metric element in terms of a single scalar $\vartheta$ and vector $T_i$, respectively;
\item find the second-order ODEs for the dynamical gauge variables $\vartheta$ and $T_i$;
\item solve  the ODEs for $\vartheta$ and $T_i$; 
\item perform inverse Fourier transform to express $\vartheta$ and $T_i$ in terms of the coordinates;
\item substitute into Eqs.~(\ref{SVT-J0-M}-\ref{SVT-Ji-M}). 
\end{enumerate}
Note that it is essential for the harmonic formulation to actually solve the ODE for $\vartheta$ and $T_i$,  and to express $J_{\mu}$ directly as an {\it algebraic} function of the coordinates $x^{\mu}$.  

Obviously, the harmonic scheme is not as simple as the concept of cosmological perturbation theory. But, however simple the latter might be, it is ill-suited for applications such as mode stability analyses and non-perturbative, numerical implementation -- issues that our harmonic scheme is designed for and can readily handle.

\subsubsection{Example}

In the following, we give an example by showing how to fix the scalar part of $\delta J_{\mu}$ to represent Newtonian gauge in scalar field theories minimally-coupled to Einstein gravity.

The scalar part of the linearized harmonic gauge condition~(\ref{SVT-J0-M}-\ref{SVT-Ji-M}) takes the form
\begin{eqnarray}
\label{SVT-J0-FRW}
\delta J_0 &=& 
- \dot{\alpha}  - 3\dot{\psi}+ \delta^{kl} \left(\dot{\epsilon} - a^{-1} \beta \right)_{,lk} 
\,,\\ 
\label{SVT-Ji-FRW}
\delta J_i &=& \partial_i \delta J, \quad {\rm where}\quad 
\delta J = \alpha - \psi - \delta^{kl}\epsilon_{,lk}  +  a \left(\dot{\beta} -  H \beta  \right) \,. 
\end{eqnarray}
Here, for simplicity, we chose physical time ($\bar{g}_{00}= -1$) to fix the background time slicing, in particular, $\bar{J}_0= 3H, \bar{J}_i = 0$ ($i=1,2,3$); for details see Eq.~\eqref{harmonic-cond} in the Appendix~\ref{sec:appB}. 

The defining feature of Newtonian gauge, introduced in Sec.~\ref{def-newt-gauge}, is zero shear, {\it i.e.},  $\beta, \epsilon \equiv 0$. Evaluating the linearized anisotropy constraint~\eqref{lin-anis} for minimally-coupled scalar field theories in Newtonian gauge, it is well-known that 
\begin{equation}
\label{anis-sec4}
\Phi = \Psi\,.
\end{equation}
Substituting Eq.~\eqref{anis-sec4} into (\ref{SVT-J0-FRW}-\ref{SVT-Ji-FRW}), we obtain the expressions 
\begin{equation}
\delta J_0 = -4 \dot{\Psi} \quad {\rm and} \quad  \delta J \equiv 0. 
\end{equation}
Intriguingly, the generalized harmonic representation of Newtonian gauge is remarkably simple in that the spatial slicing coincides with harmonic gauge ($\Box \, x^{\mu}\equiv 0$) such that the two gauges differs only w.r.t. time slicing. 

Again, in practice it is essential to solve the ODE~\eqref{newt-sec2} for $\Psi$,
\begin{equation}
\ddot{\Psi} = \frac{\dot{H}}{\rho_K}\frac{k^2}{a^2}\Psi + \left( \frac{\ddot{H}}{\dot{H}} - H\right)\dot{\Psi} +\left( \frac{\ddot{H} H}{\dot{H}} - 2\dot{H} \right)\Psi
\,,
\end{equation}
then perform the inverse Fourier transform of the solution, and first then to substitute for $\delta J_0$ so that the source function is truly a function of the coordinates.

\section{Summary and Outlook}
\label{sec:discussion}

The goal of this study has been to identify and explain the first steps towards fully non-perturbative cosmology, a new avenue of theoretical analysis that introduces elements of mathematical and numerical general relativity into exploring the evolution of the universe.  The applications we have in mind range from analytically assessing the linear mode stability of cosmological scenarios, to setting a valid formulation for numerical relativity computations, to determining the proper method to extract cosmological observables from non-perturbative simulations.  

Until now, cosmologists have relied for the most part on conventional perturbation theory developed in the 1980s based on the SVT decomposition of linearized metric variables combined with the ADM formulation of the field equations and algebraic gauge fixing. This approach was adequate for analyzing cosmologies based on Einstein gravity and minimally-coupled scalar fields admitting FRW backgrounds solutions.   In some limited cases, effective field theory provides a convenient short cut. However, as we have emphasized, these techniques are not reliable or complete for analyzing more complex theories, a point that has been missed in numerous earlier papers.

The approach that we adopted is based on the harmonic formulation of the field equations pioneered in mathematical general relativity to show uniqueness and existence of the full non-linear Einstein equations, and later incorporated into the first successful numerical relativity codes used to analyze black hole inspiral, merger and ringdown.  
Here, we have discussed applying these techniques to cosmology, which typically focuses on homogeneous backgrounds described by the FRW metric.  
In exploring theories of the early universe or dark energy, cosmologists are interested in tracking the evolution over a short, finite period of time.   For example, in bouncing cosmologies, the application of non-perturbative cosmology is to studying the modifications of Einstein gravity that are only significant after a long period of cosmological smoothing and during a bounce phase that lasts perhaps 1000 or so Planck times.  

The use of the harmonic formulation as described in this paper is, for cosmological applications, not an endpoint but rather a first step.  Before proceeding towards numerical simulations, it is indispensable to show, as we have for the case of conformally coupled ${\cal L}_4$-Horndeski theories, that the theory is linearly well-posed around a typical cosmological background satisfying certain physically well-motivated conditions on the sound speed of scalar field perturbations. In addition, it is essential to determine if the same holds for small deviations from a homogeneous background  to be sure that a numerical simulation is not being set on a `knife-edge' of instability.  We have demonstrated how to perform these tests by computing and analyzing the eigenvalues and eigenvectors, as shown in Section~\ref{sec:harmonic} and Appendix~\ref{sec:appD}.

In cases like ${\cal L}_3$-Horndeski theories and beyond, this is currently the limit of purely mathematical analysis.  If the analytic tests are passed, the next stage is to develop a numerical relativity code that checks the non-perturbative behavior of the theory with initial data corresponding to cosmological background conditions.  Here, as in simulations of black hole mergers, the harmonic scheme is a powerful approach for defining a well-posed formulation. Furthermore, the explicit computation of the eigenvalues and eigenvectors of the principal symbol, as done in Section~\ref{sec:char}~and~Appendix~\ref{sec:appD}, can be used to identify the key diagnostics that need to be tracked in the simulation. 

Of course, black hole mergers are studied in asymptotically flat backgrounds and typically the only observable of interest is the spectrum of gravitational waves that propagate to the far field.  Cosmological backgrounds of interest are not asymptotically flat and there are different observables.  Here a refinement of the harmonic formulation, namely gauge fixing through the harmonic source function, plays a key role in extracting observable quantities from the non-perturbative numerical computation.  In this paper, we only briefly outlined the gauge fixing protocol. Precisely how this is done will be the subject of a companion publication~\cite{Ijjas-to-appear}.

\section*{Acknowledgements}

We thank David Garfinkle, Luis Lehner, Vasileios Paschalidis, and Harvey Reall for discussions and comments about issues related to this work. 
A.I. is supported by the Simons Foundation `Origins of the Universe Initiative' grant number 550202.  F.P. acknowledges support from NSF grant PHY1607449, the Simons Foundation, and the Canadian Institute For Advanced Research (CIFAR).
The work of P.J.S. is supported in by the DOE grant number DEFG02-91ER40671 and by the Simons Foundation grant number 548512. 

\newpage

\appendix

\section{Derivation of the Einstein and scalar field equations~\eqref{trace-rev-EEq-J} and~\eqref{scalar-field-eq-J-3}}
\label{app:A}

In this Appendix, we derive the covariant Einstein and scalar field equations in generalized harmonic formulation as given in Eqs.~\eqref{trace-rev-EEq-J} and~\eqref{scalar-field-eq-J-3} also including some formulae that we use throughout the paper.

Expanding all derivatives, the Einstein equations~\eqref{Horndeski-Tmunu} take the form:
\begin{eqnarray}
&-&  \frac14 \Big(1+ G_4(\phi) \Big) \delta^{\mu \alpha_1 \alpha_2}_{\nu  \beta_1 \beta_2}  R_{\alpha_1 \alpha_2}{}^{\beta_1 \beta_2} 
\\
&+& 2\Big( - b(\phi)X + \frac12 G_{4,\phi}\Big) \delta^{\mu \alpha}_{\nu \beta}\nabla _{ \alpha}\nabla^{ \beta} \phi
- b(\phi) \delta^{\mu \alpha_1 \alpha_2}_{ \nu \beta_1 \beta_2}  \Big(\nabla_{\alpha_1}\phi \nabla^{\beta_1}\phi \Big)  \nabla_{\alpha_2}\nabla^{\beta_2}\phi 
 \nonumber\\
&-& 
\Big(G_2(X,\phi)  + 2G_{4,\phi\phi}(\phi)X - 2 b_{, \phi}(\phi) X ^2   \Big) \delta^{\mu}_{\nu} 
\nonumber\\
&-& \Big( G_{2,X}(X,\phi)  - 2 b_{,\phi}(\phi) X + G_{4,\phi\phi}(\phi) \Big)\nabla ^{\mu} \phi\, \nabla _{\nu} \phi 
= 0
 \,;
 \nonumber
\end{eqnarray}
and the scalar field equation~\eqref{cov-sf} takes the form
\begin{eqnarray}
\label{sf-cov-1app}
&-& \Big(G_{2,X} + 2 \left(G_{2,XX} - 2b_{,\phi} \right) X \Big) \Box\phi 
\\
&-&  \left( G_{2,XX} - 2b_{,\phi} \right) \delta^{ \alpha_1 \alpha_2}_{ \beta_1 \beta_2}  \nabla_{\alpha_1}\phi \nabla^{\beta_1}\phi   \nabla_{\alpha_2}\nabla^{\beta_2}\phi 
+ b(\phi) \delta^{ \alpha_1 \alpha_2}_{ \beta_1 \beta_2} \nabla_{\alpha_1}\nabla^{\beta_1}\phi 
 \nabla_{\alpha_2}\nabla^{\beta_2}\phi 
 \nonumber \\
&-& \Big( - b(\phi)X + \frac12 G_{4,\phi} \Big)R  
+ \frac14 b(\phi) \delta^{\alpha_1 \alpha_2 \alpha_3}_{  \beta_1 \beta_2 \beta_3}   \Big(\nabla_{\alpha_1}\phi\nabla^{\beta_1}\phi \Big) R_{\alpha_2 \alpha_3}{}^{\beta_2 \beta_3}  
\nonumber\\
&+& 2X\left( G_{2,X\phi} -b_{,\phi\phi}X \right) - G_{2,\phi} 
= 0
\,.
\nonumber
\end{eqnarray}
Here,
\begin{equation}
\label{kron-delta}
\delta^{i_1 ... i_n}_{j_1 ... j_n} \equiv n! \delta^{i_1 ... i_n}_{[j_1 ... j_n]}
\end{equation}
is the generalized Kronecker delta\footnote{
In particular, $\delta^{i_1 i_2}_{j_1 j_2} = \delta^{i_1}_{j_1}\delta^{ i_2}_{j_2} - \delta^{i_1}_{j_2}\delta^{ i_2}_{j_1} $; and
$\delta^{i_1 i_2 i_3}_{j_1 j_2 j_3} = \delta^{i_1}_{j_1}\delta^{ i_2}_{j_2}\delta^{ i_3}_{j_3} - \delta^{i_1}_{j_1}\delta^{ i_2}_{j_3}\delta^{ i_3}_{j_2} + \delta^{i_1}_{j_2}\delta^{ i_2}_{j_3}\delta^{ i_3}_{j_1} - \delta^{i_1}_{j_2}\delta^{ i_2}_{j_1}\delta^{ i_3}_{j_3} + \delta^{i_1}_{j_3}\delta^{ i_2}_{j_1}\delta^{ i_3}_{j_2} - \delta^{i_1}_{j_3}\delta^{ i_2}_{j_2}\delta^{ i_3}_{j_1}$.
}; 
\begin{equation}
R_{\alpha \beta \gamma}{}^{ \nu} \equiv \partial_{\beta}\Gamma_{\gamma \alpha}^{\nu} - \partial_{\gamma}\Gamma_{\beta \alpha}^{\nu} + \Gamma_{\beta \lambda}^{\nu} \Gamma_{\gamma \alpha}^{\lambda} - \Gamma_{\gamma \lambda}^{\nu} \Gamma_{\beta \alpha}^{\lambda}
\end{equation}
is the Riemann tensor;  and 
\begin{equation}
R_{\alpha \beta}{}^{\mu \nu} \equiv g^{\mu \gamma} R_{\alpha \beta \gamma}{}^{\nu}
\,.
\end{equation}
The Ricci tensor is given by
\begin{equation}
R_{\mu \nu} \equiv R_{\mu\lambda\nu}{}^{\lambda}\,;
\end{equation}
and we re-express the Einstein tensor in terms of the Riemann tensor as
\begin{equation}
G_{\nu}^{\mu} = -\frac14 \delta^{\mu \alpha_1 \alpha_2}_{\nu  \beta_1 \beta_2}  R_{\alpha_1 \alpha_2}{}^{\beta_1 \beta_2}
\,.
\end{equation}
In addition, we write two-derivative expressions as follows
\begin{eqnarray}
\Box \phi \nabla^{\mu}\phi \nabla_{\nu}\phi &=& \delta^{\mu}_{ \beta_1}\delta^{ \alpha_1}_{  \nu}\delta^{\alpha_2}_{ \beta_2}  \nabla_{\alpha_1}\phi \nabla^{\beta_1}\phi   \nabla_{\alpha_2}\nabla^{\beta_2}\phi 
\,;\\
\delta^{\mu}_{\nu} \nabla_{\lambda}\phi \nabla^{\lambda} X &=& - \delta^{\mu}_{ \nu}\delta^{ \alpha_1}_{  \beta_2}\delta^{\alpha_2}_{ \beta_1}   \nabla_{\alpha_1}\phi \nabla^{\beta_1}\phi   \nabla_{\alpha_2}\nabla^{\beta_2}\phi 
\,;\\
\nabla^{\mu}\phi \nabla_{\nu} X &=& - \delta^{\mu}_{ \beta_1}\delta^{ \alpha_1}_{  \beta_2}\delta^{\alpha_2}_{ \nu}   \nabla_{\alpha_1}\phi \nabla^{\beta_1}\phi   \nabla_{\alpha_2}\nabla^{\beta_2}\phi 
\,;\\
\nabla_{\nu}\phi \nabla^{\mu} X &=& - \delta^{\mu}_{ \beta_2}\delta^{ \alpha_1}_{  \nu}\delta^{\alpha_2}_{ \beta_1}   \nabla_{\alpha_1}\phi \nabla^{\beta_1}\phi   \nabla_{\alpha_2}\nabla^{\beta_2}\phi 
\,,
\end{eqnarray}
such that 
\begin{eqnarray}
\label{phi-eq-app}
\Box \phi \nabla^{\mu}\phi \nabla_{\nu}\phi - \delta^{\mu}_{\nu} \nabla_{\lambda}\phi \nabla^{\lambda} X + \nabla^{\mu}\phi \nabla_{\nu} X + \nabla_{\nu}\phi \nabla^{\mu} X &=& 
- \delta^{\mu \alpha_1 \alpha_2}_{ \nu \beta_1 \beta_2}  \nabla_{\alpha_1}\phi \nabla^{\beta_1}\phi   \nabla_{\alpha_2}\nabla^{\beta_2}\phi \qquad \\ 
&-& 2X \delta^{\mu \alpha}_{ \nu \beta} \nabla_{\alpha}\nabla^{\beta}\phi
\nonumber\,.
\end{eqnarray}

Taking the trace of the Einstein equations, 
\begin{equation}
R = - T_{\alpha}^{\alpha} \,,
\end{equation}
we find
\begin{eqnarray}
\label{trace}
R &=& 
6\Big( - b(\phi)X + \frac12 G_{4,\phi}\Big) \Box \phi
- 2 b(\phi) \delta^{ \alpha_1 \alpha_2}_{  \beta_1 \beta_2}  \Big(\nabla_{\alpha_1}\phi \nabla^{\beta_1}\phi \Big)  \nabla_{\alpha_2}\nabla^{\beta_2}\phi 
\quad\\
&-& 4G_2(X,\phi) + 2G_{2,X}X - 6G_{4,\phi\phi}X + 4 b_{, \phi}(\phi) X ^2   - G_4(\phi) R
 \,;
 \nonumber
\end{eqnarray}
and the trace-reversed Einstein equations 
\begin{equation}
R_{\mu\nu} = T_{\mu\nu} - \frac12g_{\mu\nu}T^{\alpha}_{\alpha}
\end{equation}
take the form
\begin{eqnarray}
\label{trace-rev-EEq}
R_{ \mu \nu}  & = & 
\Big(-G_2(X,\phi) + G_{2,X}X  - G_{4,\phi\phi}(\phi)X   \Big) g _{\mu\nu} 
\\
&+& \Big( G_{2,X}(X,\phi)  - 2 b_{,\phi}(\phi) X + G_{4,\phi\phi}(\phi) \Big)\nabla _{\mu} \phi\, \nabla _{\nu} \phi 
\nonumber\\
&+&  \Big( - b(\phi)X + \frac12 G_{4,\phi}\Big)\Big( 2\nabla _{\mu} \nabla_{\nu} \phi  +   g _{\mu\nu} \Box\phi \Big)
\nonumber\\
 &-&  b( \phi) \Big(\nabla _{ \mu} \phi\, \nabla _{ \nu} X +  \nabla _{ \nu} \phi\, \nabla _{ \mu} X + \Box \phi \nabla _{\mu} \phi \nabla _{\nu} \phi -  2X  \nabla _{\mu} \nabla_{\nu} \phi \Big)  
 - G_4(\phi) R_{ \mu \nu}
 \,;
 \nonumber
\end{eqnarray}
or equivalently,
\begin{eqnarray}
\Big(1+ G_4(\phi) \Big)R^{ \mu}_{\nu}  & = & 
\Big(-G_2(X,\phi) + G_{2,X}X - G_{4,\phi\phi}(\phi)X    \Big) \delta^{\mu}_{\nu} 
\\
&+& \Big( G_{2,X}(X,\phi)  - 2 b_{,\phi}(\phi) X + G_{4,\phi\phi}(\phi) \Big)\nabla ^{\mu} \phi\, \nabla _{\nu} \phi 
\nonumber\\
&+&  \Big( - b(\phi)X + \frac12 G_{4,\phi}\Big) \Big(  3\delta^{ \mu}_{\nu} \Box \phi
- 2\delta^{ \mu \alpha}_{\nu \beta} \nabla _{ \alpha}\nabla^{ \beta} \phi \Big)
\nonumber
\\
&+&  b(\phi) \Big( \delta^{\mu \alpha_1 \alpha_2}_{ \nu \beta_1 \beta_2} - \delta^{ \mu}_{\nu} \delta^{ \alpha_1 \alpha_2}_{  \beta_1 \beta_2}  \Big)\nabla_{\alpha_1}\phi \nabla^{\beta_1}\phi \,  \nabla_{\alpha_2}\nabla^{\beta_2}\phi 
 \,.
 \nonumber
\end{eqnarray}

In any generalized harmonic gauge,
\begin{equation}
\label{C-mu}
\Box x^{\mu} - J^{\mu} \equiv0\,,
\end{equation}
the Ricci tensor takes the form
\begin{eqnarray}
\label{def-ricci}
&& R_{\mu\nu}  = -\frac12 g^{\gamma \sigma} \partial_{\gamma} \partial_{\sigma} g_{\mu \nu}
- \partial_{(\mu}J_{\nu)} + \Gamma^{\gamma}_{\mu\nu}J_{\gamma}
- \frac12 \partial_{\nu} g^{\alpha \beta}\partial_{\alpha} g_{\beta \mu}
- \frac12 \partial_{\mu} g^{\alpha \beta}\partial_{\alpha} g_{\beta \nu}
-\Gamma^{\gamma}_{\alpha\mu}\Gamma^{\alpha}_{\gamma\nu}
\,;\quad\;\quad
\end{eqnarray}
and second covariant derivatives of the scalar field are given by
\begin{eqnarray}
%
&& \Box \phi  = 
g^{\alpha \beta} \phi_{,\beta\alpha} + J^{\alpha}\phi_{,\alpha}
\;;\\
\label{cov-der-har}
&&\nabla_{\mu}\nabla_{\nu}\phi = \phi_{,\nu\mu} - \Gamma_{\mu \nu}^{\sigma}\phi_{,\sigma}
\,.
\end{eqnarray}
Substituting  Eqs.~(\ref{def-ricci}-\ref{cov-der-har}) into Eqs.~\eqref{trace-rev-EEq} and~\eqref{sf-cov-1app}  yields the trace-reversed Einstein and scalar field equations in generalized harmonic formulation as given in~Eqs.~\eqref{trace-rev-EEq-J} and~\eqref{scalar-field-eq-J-3}; except that in Eq.~\eqref{scalar-field-eq-J-3} we did not substitute the harmonic expressions for the curvature terms $R_{\mu\nu}$ and $R$ in Eq.~~\eqref{scalar-field-eq-J-3}. This is because we want to make manifest that,
if the coupling to the 4-Ricci scalar is positive definite for all values of the field $\phi$, {\it i.e.}, $1+G_4(\phi)>0$,  we can eliminate $R$ using Eq.~\eqref{trace} and $R_{\mu\nu}$ using the trace-reversed Einstein equations~\eqref{trace-rev-EEq} such that the scalar field equation entails only second derivatives of the field but not of the metric. 

\section{Linearized Einstein equations~(\ref{trace-rev-EEq-J-lin-00}-\ref{trace-rev-EEq-J-lin-ij}) in  harmonic formulation}
\label{sec:appB}

In this Appendix, we provide all linearized curvature terms in harmonic formulation used in the perturbed Einstein equations~~(\ref{trace-rev-EEq-J-lin-00}-\ref{trace-rev-EEq-J-lin-ij}).

As above, bar denotes the unperturbed (background) metric,
\begin{equation}
\label{pert-met-app}
\bar{g}_{\mu\nu} = \left(\begin{array}{cc} \bar{g}_{00} &  0 \\0 &  \bar{g}_{ij}\end{array}\right)\,,\quad  \bar{g}_{\mu\nu, i} =0 \quad {\rm for\; all}\; i=1,2,3\,;
\end{equation}
 and 
 \begin{equation}
 h_{\mu\nu} \equiv g_{\mu\nu} - \bar{g}_{\mu\nu}
 \end{equation}
  is the linear perturbation to $\bar{g}_{\mu\nu}$.
Notice that on a Minkowski background ($\bar{g}_{\mu\nu}=\eta_{\mu\nu}$), we recover $\delta R_{\mu\nu}  = -\frac12 \Box  h_{\mu \nu}- \delta J_{(\mu, \nu)}$; in particular, $\delta R_{\mu\nu} = -\frac12 \Box h_{\mu\nu}$ in harmonic gauge ($J_{\mu}\equiv 0$).

The inverse of the linearized metric perturbation is given by
\begin{equation}
 h^{\mu\nu} = g^{\mu\nu} - \bar{g}^{\mu\nu} = -  \bar{g}^{\mu\rho}\bar{g}^{\nu\sigma}h_{\rho\sigma}\,;
\end{equation}
{\it i.e.}, 
\begin{equation}
h^{00} = - \bar{g}^{00}\bar{g}^{00}h_{00} \,,\quad h^{0i} = - \bar{g}^{ij}\bar{g}^{00}h_{0ij}\,,\quad h^{ij}= - \bar{g}^{im}\bar{g}^{jn}h_{mn}
\,.
\end{equation}
Linearizing the connection terms 
\begin{equation}
\Gamma_{\nu \lambda}^{\mu} \equiv \frac12 g^{\mu\rho}\left( g_{\rho \nu, \lambda} + g_{\rho \lambda, \nu} - g_{\lambda \nu, \rho}  \right) 
\end{equation}
yields
\begin{equation}
\delta\Gamma_{\nu \lambda}^{\mu} = \frac12 \bar{g}^{\mu\rho}\left( h_{\rho \nu, \lambda} + h_{\rho \lambda, \nu} - h_{\lambda \nu, \rho} - 2 h_{\rho\sigma} \bar{\Gamma}_{\nu \lambda}^{\sigma} \right)
\,,
\end{equation}
{\it i.e.},
\begin{eqnarray}
\label{conn-000-h}
\delta\Gamma_{00}^{0} &=& \frac12 \bar{g}^{00}\left( \dot{h}_{0 0}  - 2 h_{00} \bar{\Gamma}_{00}^{0} \right)
\,,\\
\delta\Gamma_{00}^{i} &=& \frac12 \bar{g}^{ij}\left( 2\dot{h}_{j 0}  - h_{00, j} - 2 h_{j 0} \bar{\Gamma}_{00}^{0} \right)
\,,\\
\delta\Gamma_{0i}^{0} &=& \frac12 \bar{g}^{00}\left( h_{0 0, i}  - 2 h_{0k} \bar{\Gamma}_{0i}^{k} \right)
\,,\\
\delta\Gamma_{0k}^{l} &=& \frac12 \bar{g}^{ll}\left( \dot{h}_{kl} + h_{l 0, k}  - h_{0k, l} - 2 h_{lm} \bar{\Gamma}_{0k}^{m} \right)
\,,\\
\delta\Gamma_{kl}^0 &=& \frac12 \bar{g}^{00}\left( h_{0 k, l} + h_{0 l, k} - \dot{h}_{kl} - 2 h_{00} \bar{\Gamma}_{kl}^0 \right)
\,,\\
\label{conn-ikl-h}
\delta\Gamma_{ik}^{l} &=& \frac12 \bar{g}^{lm}\left( h_{m i, k} + h_{m k, i} - h_{ik, m} - 2 h_{m0} \bar{\Gamma}_{ik}^{0} \right)
\,.
\end{eqnarray}
Here, the background expressions $\bar{\Gamma}_{\nu\lambda}^{\mu}$ are given by
\begin{equation}
\bar{\Gamma}_{00}^{0} = \frac12 \bar{g}^{00}\dot{\bar{g}}_{00}  \,, 
\quad \bar{\Gamma}_{k0}^{l} = \frac12 \bar{g}^{lm}\dot{\bar{g}}_{mk} \,, 
\quad \bar{\Gamma}_{ij}^{0} = -\frac12 \bar{g}^{00}\dot{\bar{g}}_{ij}  \,, 
\quad \bar{\Gamma}_{00}^{i} = \bar{\Gamma}_{i0}^{0} =  \bar{\Gamma}_{ij}^{k} = 0
\,.
\end{equation}

Substituting into Eq.~\eqref{def-ricci}, the homogeneous part of the Ricci tensor is given by
\begin{eqnarray}
&& \bar{R}_{\mu\nu}  = -\frac12 \bar{g}^{00} \ddot{\bar{g}}_{\mu \nu}
- \partial_{(\mu}\bar{J}_{\nu)} + \bar{\Gamma}^0_{\mu\nu}\bar{J}_0
- \frac12 \bar{g}^{00}{}_{, \nu} \dot{\bar{g}}_{0 \mu}
- \frac12 \bar{g}^{0 0}{}_{, \mu}\dot{\bar{g}}_{0 \nu}
- \bar{\Gamma}^{\gamma}_{\alpha\mu}\bar{\Gamma}^{\alpha}_{\gamma\nu}
\,;\quad
\end{eqnarray}
{\it i.e.},
\begin{eqnarray}
 \bar{R}_{00}  &=& -\frac12 g^{00} \ddot{\bar{g}}_{00}
- \dot{\bar{J}}_0 + \bar{\Gamma}^0_{00}\bar{J}_0
- \dot{\bar{g}}^{00} \dot{\bar{g}}_{0 0}
- \bar{\Gamma}^0_{0 0}\bar{\Gamma}^0_{0 0}
- \bar{\Gamma}^l_{k 0}\bar{\Gamma}^k_{l 0}
\,;\quad\\
 \bar{R}_{0i}  &=&  \bar{R}_{i0} = 0\,;\\
 \bar{R}_{ij}  &=& -\frac12 g^{00} \ddot{\bar{g}}_{ij}
+ \bar{\Gamma}^0_{ij}\bar{J}_0
- \bar{\Gamma}^k_{0 i}\bar{\Gamma}^0_{k j}
- \bar{\Gamma}^{0}_{k i}\bar{\Gamma}^k_{0 j}
\,.
\end{eqnarray}
and, finally, 
\begin{eqnarray}
\bar{R}  &=&  \bar{g}^{00} \bar{R}_{00} + \bar{g}^{ij} \bar{R}_{ij} \\
&=& \bar{g}^{00} \left(  -\frac12 \bar{g}^{00} \ddot{\bar{g}}_{00}
- \dot{\bar{J}}_0 + \bar{\Gamma}^0_{00}\bar{J}_0
- \dot{\bar{g}}^{00} \dot{\bar{g}}_{0 0}
- \bar{\Gamma}^0_{0 0}\bar{\Gamma}^0_{0 0}
- \bar{\Gamma}^l_{k 0}\bar{\Gamma}^k_{l 0}\right)
\nonumber\\
  &+&  \bar{g}^{ij} \left( -\frac12 g^{00} \ddot{\bar{g}}_{ij}
+ \bar{\Gamma}^0_{ij}\bar{J}_0
- \bar{\Gamma}^k_{0 i}\bar{\Gamma}^0_{k j}
- \bar{\Gamma}^{0}_{k i}\bar{\Gamma}^k_{0 j} \right)
\,.
\end{eqnarray}
For example, in harmonic gauge ($J_{\mu}\equiv 0$), $ \bar{R}_{00} = 3 \left( - \dot{H} + 2H^2\right); \bar{R}_{ij} = a^{-4} \dot{H}\delta_{ij}$; and $ \bar{R} = 6 \, a^{-6} \left( \dot{H} - H^2\right)$.

The linearized Ricci tensor takes the form
\begin{eqnarray}
\label{lin-Ricci}
\delta R_{\mu\nu}  &=& -\frac12 \bar{g}^{\alpha \beta}  h_{\mu \nu}{}_{,\beta\alpha}
- \delta J_{(\mu, \nu)}
-\frac12 h^{\alpha \beta}  \bar{g}_{\mu \nu}{}_{,\beta\alpha}
 + \bar{\Gamma}^{\alpha}_{\mu\nu}\delta J_{\alpha}
  + \delta\Gamma^{\alpha}_{\mu\nu}\bar{J}_{\alpha}
  \\
&-&\bar{\Gamma}^{\beta}_{\alpha\mu}\delta\Gamma^{\alpha}_{\beta\nu}
-\delta\Gamma^{\beta}_{\alpha\mu}\bar{\Gamma}^{\alpha}_{\beta\nu}
- \frac12 \bar{g}^{\alpha \beta}{}_{,\nu} h_{\beta \mu}{}_{,\alpha}
- \frac12 h^{\alpha \beta}{}_{,\nu} \bar{g}_{\beta \mu}{}_{,\alpha}
- \frac12 \bar{g}^{\alpha \beta}{}_{, \mu} h_{\beta \nu, \alpha}
- \frac12 h^{\alpha \beta}{}_{,\mu} \bar{g}_{\beta \nu, \alpha}
\,;\nonumber\quad
\end{eqnarray}
such that the components of the linearized Ricci tensor are given by
\begin{eqnarray}
\label{lin-Ricci-00}
\delta R_{00}  &=&   -\frac12 \left( \bar{g}^{00}  \ddot{h}_{00} + \bar{g}^{kl}  h_{00}{}_{,lk} \right) 
- \delta J_{0, 0}
+ \frac12 \bar{g}^{00}\dot{\bar{g}}_{00} \delta J_0
+ \delta\Gamma^{0}_{00}\bar{J}_{0}
  \\
&+& \left( \bar{g}^{00}\bar{g}^{00}  \dot{\bar{g}}_{0 0} -\dot{\bar{g}}^{00} \right)\dot{h}_{00}
- \dot{\bar{g}}^{kl} h_{l 0, k}
+ \left( 2 \dot{\bar{g}}^{00} \dot{\bar{g}}_{0 0} + \frac12 \bar{g}^{00} \ddot{\bar{g}}_{00} \right) \bar{g}^{00} h_{00}
 \nonumber\\
&-& 2\bar{\Gamma}^0_{0 0}\delta\Gamma^0_{0 0}
-2\bar{\Gamma}^{k}_{l 0}\delta\Gamma^{l}_{k 0}
\,;\nonumber
\\
\label{lin-Ricci-0i}
\delta R_{0i}  &=& -\frac12\left( \bar{g}^{00}  \ddot{h}_{0i} + \bar{g}^{kl}  h_{0i,lk} \right) 
- \delta J_{(0, i)}
 + \frac12 \bar{g}^{m n}\dot{\bar{g}}_{n i}\delta J_{m}
  + \delta\Gamma^{0}_{0i}\bar{J}_{0}
  \\
&+& \frac12 \bar{g}^{00}\bar{g}^{00} \dot{\bar{g}}_{0 0} h_{00,i}
- \frac12  \dot{\bar{g}}^{00}  \dot{h}_{0 i}
+ \frac12 \bar{g}^{00}\bar{g}^{mn} \dot{\bar{g}}_{n i} \dot{h}_{0 m}
- \frac12 \dot{\bar{g}}^{mn} h_{n i, m}
\,;
\nonumber\\
\label{lin-Ricci-ij}
\delta R_{ij}  &=&-\frac12 \left( \bar{g}^{00}  \ddot{h}_{ij} + \bar{g}^{kl}  h_{ij,lk} \right)   - \delta J_{(i, j)}
-\frac12 \bar{g}^{00}\dot{\bar{g}}_{ij} \delta J_{0}
+ \delta\Gamma^{0}_{ij}\bar{J}_{0}
  \\
&-& \frac12 (-\bar{g}^{00}) \bar{g}^{kl} \dot{\bar{g}}_{ki} h_{0l,j} 
- \frac12 (-\bar{g}^{00}) \bar{g}^{kl}\dot{\bar{g}}_{k j}h_{0k,i} 
+ \frac12  \bar{g}^{00} \bar{g}^{00}\ddot{\bar{g}}_{i j} h_{00}  
\nonumber \\
&-&\bar{\Gamma}^k_{0 i}\delta\Gamma^{0}_{k j}
-\bar{\Gamma}^0_{k i}\delta\Gamma^k_{0 j}
- \bar{\Gamma}^0_{k j}\delta\Gamma^{k}_{0 i}
- \bar{\Gamma}^k_{0 j} \delta\Gamma^0_{k i}
\,;\nonumber
\end{eqnarray}
and the linearized Ricci scalar is given by
\begin{eqnarray}
\label{lin-Ricci-sc}
\delta R  &=& \bar{g}^{00}\delta R_{00} + \bar{g}^{kl}\delta R_{kl} - \bar{g}^{00}\bar{g}^{00}h_{00}\bar{R}_{00} - \bar{g}^{km}\bar{g}^{ln} h_{mn}\bar{R}_{kl}
\,.
\end{eqnarray}

Finally, evaluating the covariant trace equation as given in Eq.~\eqref{trace} for the background and the linearized metric yields in generalized harmonic gauge
\begin{eqnarray}
\label{trace-bgr}
 \Big(1+ \bar{G}_4(\phi) \Big)   \bar{R} &=& 
- 4 \,\bar{G}_2(X,\phi) + \Big( \bar{G}_{2,X} - 3 G_{4,\phi\phi}\Big) (-\bar{g}^{00}) \dot{\phi}^2 
+ \bar{b}_{,\phi}(\phi)\bar{g}^{00}\bar{g}^{00}\dot{\phi}^4 
\\
&+& 3 \Big(  \bar{b}(\phi) \bar{g}^{00} \dot{\phi}^2 + \bar{G}_{4,\phi} \Big) \Big(\bar{g}^{00}\ddot{\phi} + \bar{g}^{00}\bar{J}_0\dot{\phi}\Big) 
- 2 \bar{b}(\phi) \bar{g}^{00}\bar{g}^{00} \dot{\phi}^3 \Big( \bar{J}_0 + \bar{\Gamma}^0_{00} \Big)
 \,;
 \nonumber
\\
\label{trace-lin}
\Big(1+ \bar{G}_4(\phi) \Big)   \delta R &=& 
3 \Big( \bar{b}(\phi) \bar{g}^{00}\dot{\phi}^2 + \bar{G}_{4,\phi} \Big) \left( \bar{g}^{\alpha\beta} \pi,_{\beta\alpha} - \bar{g}^{00}\bar{g}^{00} \ddot{\phi} h_{00} + \bar{g}^{00} \bar{J}_0 \dot{\pi} + \dot{\phi}\delta J^0 \right)
\\
&-& 2 \bar{b}(\phi)\bar{g}^{00}\dot{\phi}^2 \left( \bar{g}^{kl} \pi,_{lk}  + \dot{\phi} \Big(\delta J^0 + \bar{g}^{00}\delta \Gamma^0_{00} \Big) \right)
\nonumber
\\
&+& \left( -\bar{G}_{2,X}  - \frac12 \Big( \bar{G}_{2,XX}  + 4 \bar{b}_{,\phi} \Big) \bar{g}^{00}\dot{\phi}^2 - 3G_{4,\phi\phi}  \right)\Big( 2(-\bar{g}^{00}) \dot{\phi}\dot{\pi} + \bar{g}^{00}\bar{g}^{00} \dot{\phi}^2 h_{00} \Big) 
\nonumber
\\
&-&  \bar{b}(\phi)  \bar{g}^{00}\bar{g}^{00}  \Big( 3 \ddot{\phi} +  \big(\bar{J}_0  - 2 \bar{\Gamma}^0_{00}\big) \dot{\phi} \Big) \dot{\phi}^2 \, \bar{g}^{00} h_{00} 
+  6 \bar{b}(\phi) \bar{g}^{00} \bar{g}^{00} \Big(  \ddot{\phi} - \bar{\Gamma}^0_{00}\dot{\phi} \Big)  \dot{\phi}\dot{\pi}  
\nonumber
\\
&+&  \left( - 4 \bar{G}_{2,\phi} + \Big( \bar{G}_{2,X\phi} - 3 \bar{G}_{4,\phi\phi\phi}\Big)(-\bar{g}^{00})\dot{\phi}^2 +  \bar{b}_{,\phi\phi} \bar{g}^{00}\bar{g}^{00}\dot{\phi}^4   \right) \pi 
\nonumber
\\
&+&  \bar{b}_{,\phi}\bar{g}^{00}\bar{g}^{00}\dot{\phi}^2  \left( 3\ddot{\phi} + \bar{J}_0 \dot{\phi} - 2\bar{\Gamma}^0_{00}\dot{\phi} \right) \pi
- 3 \bar{G}_{4,\phi\phi} \left( \ddot{\phi} + \bar{J}_0 \dot{\phi} \right) (-\bar{g}^{00})\pi
-  \bar{G}_{4,\phi} \bar{R}\, \pi 
 \,.
 \nonumber
\end{eqnarray}


For completeness, we also derive the linearized expressions for the harmonic source functions in terms of the linearized metric:

Substituting $\Box \equiv (1/\sqrt{-g}) \partial_{\alpha}(\sqrt{-g}g^{\alpha \beta} \partial_{\beta})$,
the harmonic gauge condition in Eq.~\eqref{C-mu} can be re-written as
\begin{equation}
\label{harmonic-cond}
J_{\mu} = \ln(\sqrt{-g})_{,\mu} - g^{\alpha \beta} g_{\beta \mu,\alpha} 
= \frac12 g^{\alpha \beta} g_{\alpha \beta,\mu} - g^{\alpha \beta} g_{\beta \mu,\alpha} 
\,.
\end{equation}
 Linearizing Eq.~\eqref{harmonic-cond}, the perturbed harmonic source function $\delta J^{\mu}$ is given by
 \begin{eqnarray}
\label{harmonic-cond-lin1}
\delta J_{\mu} &=&  \frac12 \bar{g}^{\alpha \beta} h_{\alpha \beta,\mu} + \frac12 h^{\alpha \beta} \bar{g}_{\alpha \beta,\mu}
 - h^{\alpha \mu} \bar{g}_{\mu \mu,\alpha} - \bar{g}^{\alpha \beta} h_{\beta \mu,\alpha} 
 \\
&=& \frac12 \bar{g}^{00} h_{00,\mu} + \frac12 \bar{g}^{kl} h_{kl,\mu} 
- \frac12 \bar{g}_{\alpha \beta,\mu} \bar{g}^{\alpha \rho}\bar{g}^{\beta \sigma}h_{\rho\sigma}
+ \bar{g}^{00}\bar{g}^{\mu \mu} \dot{\bar{g}}_{\mu \mu} h_{0 \mu}  
 - \bar{g}^{\alpha \beta} h_{\beta \mu,\alpha} 
\,;\nonumber
\end{eqnarray} 
in particular, for a homogeneous background as specified in Eq.~\eqref{pert-met-app},
\begin{eqnarray}
\delta J_0 &=& 
-\frac12 \bar{g}^{00} \dot{h}_{00} + \frac12 \bar{g}^{kl} \dot{h}_{kl}  - \bar{g}^{kl} h_{l 0,k} 
+ \frac12 \bar{g}^{00}\bar{g}^{00} \dot{\bar{g}}_{00} h_{00}  
- \frac12 \dot{\bar{g}}_{kl} \bar{g}^{k m}\bar{g}^{l n}h_{m n}
\,,\\ 
\delta J_i &=& \frac12 \bar{g}^{00} h_{00,i} 
 + \frac12 \bar{g}^{kl} h_{kl,i} 
  - \bar{g}^{00} \dot{h}_{0 i}
   - \bar{g}^{kl} h_{l i,k} 
+ \bar{g}^{00}\bar{g}^{ii} \dot{\bar{g}}_{ii} h_{0 i}  
\,.
\end{eqnarray}

%
We note that setting $\bar{g}_{00}=-1, \bar{g}_{ij} = a^2(t)\delta_{ij}$, $\bar{J}_0 = 3H; \bar{J}_i = 0$ and the components of the Ricci scalar take the form 
\begin{eqnarray}
\label{lin-Ricci-00-W}
\delta R_{00}  &=&    \frac12 \left(  \ddot{h}_{00} - a^{-2}\delta^{kl}  h_{00}{}_{,lk} \right) 
- \delta J_{0, 0}
-  \frac32 H  \dot{h}_{0 0}    
 - a^{-2} H\delta^{kl} \left( \dot{h}_{kl}  - 2h_{l 0, k}- 2 H h_{lk}  \right); \qquad
\\
\label{lin-Ricci-0i-W}
\delta R_{0i}  &=& \frac12\left( \ddot{h}_{0i} - a^{-2}\delta^{kl} h_{0i,lk} \right) 
- \delta J_{(0, i)}
 + H \delta J_{i}
  \\
&-&  2H  \dot{h}_{0 i} - \frac12   H h_{0 0, i} 
+ a^{-2} H\delta^{kl} \left( h_{l i, k} - \frac12 h_{l k, i} \right) 
+ 5H^2 h_{0 i};
\nonumber\\
\label{lin-Ricci-ij-W}
\delta R_{ij}  &=& \frac12 \left( \ddot{h}_{ij} - a^{-2}\delta^{kl} h_{ij,lk} \right)   - \delta J_{(i, j)}
+ a^2H \delta_{ij} \delta J_{0}
  \\
&-& \frac32 H \left( h_{0i,j} + h_{0j,i} \right) - \frac12 H \dot{h}_{ij} 
+ a^2 \left( \dot{H} + 3H^2\right) \delta_{i j} h_{00}  
+ 2H^2 h_{ij}  
\,.\nonumber
\end{eqnarray}
Using the same background time-slicing, the linearized harmonic source functions take the form 
 \begin{eqnarray}
\delta J_0 &=&  \frac12 \dot{h}_{00} + \frac{1}{2a^2} \delta^{kl} \dot{h}_{kl}  
- \frac{1}{a^2}\delta^{kl} h_{l 0,k} 
- \frac{1}{a^2}H\delta^{kl}h_{kl}
\,,\\ 
\delta J_i &=& -\frac12 h_{00,i} + \frac{1}{2a^2} \delta^{kl} h_{kl,i}  + \dot{h}_{0 i}
- 2H h_{0 i}  
- \frac{1}{a^2}\delta^{kl}  h_{l i,k} 
\,.
\end{eqnarray}
Now, substituting  into Eqs.~(\ref{lin-Ricci-00-W}-\ref{lin-Ricci-ij-W}), one can easily verify agreement with results in the literature that were used to derive the linearized metric in SVT decomposition. 

We stress though that, in the generalized {\it harmonic} formulation the expressions for the harmonic source functions in terms of the metric should not be substituted back into the linearized Einstein equations because the underlying idea of the harmonic decomposition is exactly to `trade' second metric derivatives for first derivatives of functions that only depend on the coordinates. For example, the analysis of linearized ${\cal L}_3$-Horndeski by Battarra et al. in Ref.~\cite{Battarra:2014tga} used harmonic coordinates in the ADM and {\it not} the harmonic decomposition of the field equations. This is obvious from the fact that the $00$- and $0i$-components take the form of constraint equations. But the introduction of harmonic coordinates without the harmonic formulation cannot be used to determine whether the theory is well-posed, and so the conclusion in Ref.~\cite{Battarra:2014tga} is not valid.

%
\section{Equivalence of formulations of the linearized theory}
\label{sec:appC}

In this section we show that re-ordering the linearized Einstein and scalar field equations does not change the principal symbol.

We consider following generic, second-order system of coupled PDEs
\begin{eqnarray}
&& \Box h_{\mu\nu} + Q^{\alpha\beta}_{\mu\nu}\pi_{,\alpha\beta} + ... = 0\,,\\
&&\Box \pi + N^{\alpha\beta}\pi_{,\alpha\beta} + M^{\mu\nu} \Box h_{\mu\nu} + ... =0\,.
\end{eqnarray}
Here, $h_{\mu\nu}$ is the linearized metric with $\mu,\nu=0,...,3$, $\pi$ the linearized scalar field, and $Q, N, M$ real coefficient matrices.

Substituting for $\Box h_{\mu\nu}$ in the second equation, the system takes the form
\begin{eqnarray}
&&\Box h_{\mu\nu} + Q^{\alpha\beta}_{\mu\nu}\pi_{,\alpha\beta} + ... =0\,,\\
&& \Box \pi + N^{\alpha\beta}\pi_{,\alpha\beta} - M^{\mu\nu} Q^{\alpha\beta}_{\mu\nu}\pi_{,\alpha\beta} + ... = 0\,.
\end{eqnarray}

Denoting the coefficient matrices of the original system by 
\begin{equation}
A(t) = \left(
\begin{array}{cc}
 A_{h}  & \; A_{h \pi}  \\
A_{\pi h} & \; A_{\pi}
\end{array}
\right), \quad B^{mn}(t) = \left(
\begin{array}{cc}
B^{mn}_{h} & \; B^{mn}_{h \pi}  \\
B^{mn}_{\pi h} & \; B^{mn}_{\pi}
\end{array}
\right),\;  
D^m(t) = \left(
\begin{array}{cc}
D^m_h & \; D^m_{h \pi}  \\
D^m_{\pi h} & \; D^m_{\pi}
\end{array}
\right)\,,
\end{equation}
where $A,B, D$ are defined as in Eq.~\eqref{lin-eq}, 
\begin{align}
\label{def-apih}
&A_{h\pi} = (Q^{00}_{tt}, Q^{00}_{tx}, ..., Q^{00}_{zz})^{\rm T}\,, \quad &A_{\pi h} = \bar{g}^{00}(M^{tt}, M^{tx}, ..., M^{zz})
\,,\\
&B^{mn}_{h\pi} = (Q^{mn}_{tt}, Q^{mn}_{tx}, ..., Q^{mn}_{zz})^{\rm T}\,, 
&B^{mn}_{\pi h} = \bar{g}^{mn}(M^{tt}, M^{tx}, ..., M^{zz})
\,,\\
\label{def-dpih}
& D^{m}_{h\pi} = (Q^{0m}_{tt}, Q^{0m}_{tx}, ..., Q^{0m}_{zz})^{\rm T}\,,  
&D^{m}_{\pi h} = \bar{g}^{0m}(M^{tt}, M^{tx}, ..., M^{zz})
\,,
\end{align}
the coefficient matrices corresponding to the new system change to
 \begin{equation}
\tilde{A}(t) = \left(
\begin{array}{cc}
 A_{h}  & \; A_{h \pi}  \\
0 & \; \tilde{A}_{\pi}
\end{array}
\right), \quad \tilde{B}^{mn}(t) = \left(
\begin{array}{cc}
B^{mn}_{h} & \; B^{mn}_{h \pi}  \\
0 & \; \tilde{B}^{mn}_{\pi}
\end{array}
\right),\;  
D^m(t) = \left(
\begin{array}{cc}
D^m_h & \; D^m _{h \pi}  \\
0 & \; \tilde{D}^m_{\pi}
\end{array}
\right)\,,
\end{equation}
where $\tilde{A}_{\pi}, \tilde{D}^{mn}_{\pi}$ and $\tilde{D}^m_{\pi}$ are real scalars given by
\begin{equation}
\tilde{A}_{\pi} = A_{\pi} - \bar{g}_{00} A_{h\pi}\cdot A_{\pi h} \,, \quad 
\tilde{B}^{mn}_{\pi} = B^{mn}_{\pi} - \bar{g}_{mn} B^{mn}_{h\pi}\cdot B^{mn}_{\pi h} \,, \quad 
\tilde{D}^{m}_{\pi} = D^m_{\pi} - \bar{g}_{0m} D^{m}_{h\pi}\cdot D^{m}_{\pi h}\,.
\end{equation}
Note that only the last line of the matrices changes since we left the trace-reversed Einstein equations unchanged. 

To make the algebraic operation (`re-ordering') manifest, we can re-express the new coefficient matrices as a result of matrix multiplications,
\begin{equation}
\tilde{A} = M_A \times A, \quad \tilde{B}^{mn} = M_B \times B^{mn}\,,  \quad \tilde{D}^{m} = M_D \times D^{m}\,, \quad M_A, M_B, M_D \in {\mathbb R}^{11\times 11},
\end{equation}
such that the principal symbol $\tilde{{\cal P}}^0$ of the new system becomes  
 \begin{equation}
\tilde{{\cal P}}^0 = |k| \left(
\begin{array}{cc}
0&  {\mathbb I}_{11}   \\
A^{-1}M_A^{-1}M_B B^{mn} \tilde{k}_m\tilde{k}_n   & \;   A^{-1}M_A^{-1}M_D D^{m}\tilde{k}_m  
\end{array}
\right)\,.
\end{equation}

From the definition of $A_{\pi h}, B^{mn}_{\pi h}, D^m_{\pi h}$ in Eqs.~(\ref{def-apih}-\ref{def-dpih}), it is easy to see that all three matrices 
 \begin{equation}
M_A \equiv \tilde{A} A^{-1} = \left(
\begin{array}{cc}
 {\mathbb I}_{10}  & \; 0  \\
- \bar{g}_{00} A_{\pi h} & \; 1
\end{array}
\right), \quad M_B \equiv \tilde{B}^{mn} (B^{mn})^{-1} = \left(
\begin{array}{cc}
 {\mathbb I}_{10}  & \; 0  \\
- \bar{g}_{mn} B^{mn}_{\pi h} & \; 1
\end{array}
\right)\,,
\end{equation}
and 
 \begin{equation}
M_D \equiv \tilde{D}^m (D^m)^{-1} = \left(
\begin{array}{cc}
 {\mathbb I}_{10}  & \; 0  \\
- \bar{g}_{0m} D^m_{\pi h} & \; 1
\end{array}
\right)\,,
\end{equation}
are equivalent, and, hence, both systems have the same principal symbol, $\tilde{{\cal P}}^0 \equiv {\cal P}^0$.


\section{Characteristic analysis on generic backgrounds} 
\label{sec:appD}

In this Appendix we present necessary conditions for the initial value problem for ${\cal L}_3$-Horndeski theories is strongly hyperbolic on an arbitrary background. Our results apply to conformally-coupled ${\cal L}_4$-Horndeski theories as well, as can be straightforwardly verified.

Keeping only terms that are second-order in derivatives, the unperturbed field equations read
\begin{eqnarray}
\label{trace-rev-EEq-J-L3-inh}
&-& \frac12 g^{\alpha \beta}g_{\mu \nu}{}_{,\beta\alpha}
 \\ 
&+&  b( \phi) \left( \left( \phi _{,\mu}  \phi_{,\nu} -\frac12 \phi _{,\alpha}  \phi_{,\beta} g ^{\alpha\beta} g _{\mu\nu} \right)   g^{\alpha \beta} \phi_{,\beta\alpha} 
-   g^{\rho\sigma}  \phi_{,\rho}  \Big( \phi_{,\mu}  \phi_{,\sigma\nu} +  \phi_{,\nu} \phi_{,\sigma\mu}  \Big)   \right)
+ ...= 0
 \,;
 \nonumber\\
 \label{trace-rev-phiEq-J-L3-inh}
 &-& G_{2,X}  g^{\alpha \beta} \phi_{,\beta\alpha} 
+  \Big( G_{2,XX} - 2b_{,\phi} \Big) g^{\alpha \beta} g^{\mu\nu} \phi_{,\alpha} \phi_{,\nu}\phi_{,\beta\mu}
\\
&+& b(\phi) \delta^{ \alpha_1 \alpha_2}_{ \beta_1 \beta_2} \nabla_{\alpha_1}\nabla^{\beta_1}\phi 
 \nabla_{\alpha_2}\nabla^{\beta_2}\phi 
- b(\phi)  g^{\mu \alpha} g^{\nu \beta} \phi_{,\alpha} \phi_{,\beta} R_{\mu\nu} 
+ ... = 0\,.
\nonumber
\end{eqnarray}
As above, $...$ denotes lower than second order terms that do not contribute to the principal symbol.

Linearizing around an arbitrary background, Eqs.(\ref{trace-rev-EEq-J-L3-inh}-\ref{trace-rev-phiEq-J-L3-inh}) become 
\begin{eqnarray}
\label{trace-rev-EEq-J-L3-inh-lin}
&& -\frac12 \bar{g}^{00} \ddot{h}_{00} - \frac12 \bar{g}^{mn}h_{00,mn} -  \bar{g}^{0m} \dot{h}_{00,m} 
 \\ 
&+&  \bar{b}( \phi)  \left( -\frac32 \bar{g}^{00} \dot{\phi}^2   - \frac12 \phi _{,m}  \phi_{,n} \bar{g}^{mn}  -  3 \dot{\phi}  \phi_{,m} \bar{g}^{0m} \right)   \ddot{\pi} 
\nonumber\\
&+&  \bar{b}( \phi) \left( \frac12 \dot{\phi}^2   -\frac12 \phi _{,m}  \phi_{,n} \bar{g}^{mn} \bar{g} _{00} -  \dot{\phi}  \phi_{,m} \bar{g}^{0m} \bar{g} _{00}\right)  \bar{g}^{mn} \pi_{,nm} 
\nonumber\\
&-&  \bar{b}( \phi)  \left( \dot{\phi}^2   + \phi _{,m}  \phi_{,n} \bar{g}^{mn} \bar{g} _{00} +2 \dot{\phi}  \phi_{,m} \bar{g}^{0m} \bar{g} _{00}\right) \bar{g}^{0m} \dot{\pi}_{,m} 
- 2\bar{b}( \phi) \bar{g}^{m n}  \phi_{,n}   \dot{\phi} \dot{\pi}_{,m}   
+ ...= 0
 \,;
 \nonumber\\
 && -\frac12 \bar{g}^{00} \ddot{h}_{0i} - \frac12 \bar{g}^{mn}h_{0i,mn} -  \bar{g}^{0m} \dot{h}_{0i,m} 
 \\ 
&-&  b( \phi) \left(  \frac12 \phi _{,\alpha}  \phi_{,\beta} \bar{g}^{\alpha\beta} \bar{g} _{0i} \bar{g}^{00} + \bar{g}^{0m}  \phi_{,m}   \phi_{,i} \right) \ddot{\pi} 
\nonumber\\
&+&  b( \phi) \left( \dot{\phi}\phi_{,i} -\frac12 \phi _{,\alpha}  \phi_{,\beta} \bar{g}^{\alpha\beta} \bar{g}_{0i} \right) \bar{g}^{mn} \pi_{,nm} 
- b( \phi)  \bar{g}^{\rho m}  \phi_{,\rho} \dot{\phi}  \pi_{,m i} 
\nonumber\\
&+&  b( \phi) \left( \dot{\phi}\phi_{,i} - \phi _{,\alpha}  \phi_{,\beta} \bar{g}^{\alpha\beta} \bar{g}_{0i} \right)  \bar{g}^{0m} \dot{\pi}_{,m} 
- b( \phi) \bar{g}^{m n}  \phi_{,n}  \phi_{,i} \dot{\pi}_{,m}  
- b( \phi)  \bar{g}^{\rho 0}  \phi_{,\rho} \dot{\phi}  \dot{\pi}_{, i} 
+ ...= 0
 \,;
 \nonumber\\
 \label{ij-gb}
 && -\frac12 \bar{g}^{00} \ddot{h}_{ij} - \frac12 \bar{g}^{mn}h_{ij,mn} -  \bar{g}^{0m} \dot{h}_{ij,m} 
 \\ 
 &+&  b( \phi)  \left( \phi _{,i}  \phi_{,j} -\frac12 \phi _{,\alpha}  \phi_{,\beta} \bar{g}^{\alpha\beta} \bar{g} _{ij} \right)   \bar{g}^{00} \ddot{\pi} 
 \nonumber\\
   &+&  b( \phi)  \left( \phi _{,i}  \phi_{,j} -\frac12 \phi _{,\alpha}  \phi_{,\beta} \bar{g}^{\alpha\beta} \bar{g} _{ij} \right)   \bar{g}^{mn} \pi_{,nm} 
 - b( \phi)  \bar{g}^{\rho m}  \phi_{,\rho}  \Big( \phi_{,i}  \pi_{,m j} +  \phi_{,j} \pi_{,m i}  \Big)  
 \nonumber\\
  &+&  b( \phi)  \left( \phi _{,i}  \phi_{,j} -\frac12 \phi _{,\alpha}  \phi_{,\beta} \bar{g}^{\alpha\beta} \bar{g} _{ij} \right)   \bar{g}^{0m} \dot{\pi}_{,m} 
-  b( \phi)  \bar{g}^{\rho0}  \phi_{,\rho}  \Big( \phi_{,i}  \dot{\pi}_{, j} +  \phi_{,j} \dot{\pi}_{, i}  \Big) 
+ ...= 0
 \,;
 \nonumber\\
 \label{s-f-gb}
&&\left( - G_{2,X}  g^{00} 
+  \Big( G_{2,XX} - 2b_{,\phi} \Big) g^{\alpha 0} g^{\nu0} \phi_{,\alpha} \phi_{,\nu} 
\right)\ddot{\pi} 
\\
&+& 2 b(\phi) \left( \bar{g}^{00} \Big( \bar{g}^{\alpha\beta}\phi_{,\beta\alpha} + \bar{J}^{\alpha}\phi_{,\alpha} \Big)  
- \bar{g}^{0\alpha}\bar{g}^{0\rho}  \Big( \phi_{,\rho\alpha} - \Gamma_{\alpha\rho}^{\lambda}\phi_{,\lambda} \Big) \right)\ddot{\pi}
  \nonumber\\
 &-& \left( 2 G_{2,X}  g^{0m}  
-  2\Big( G_{2,XX} - 2b_{,\phi} \Big) g^{\alpha 0} g^{m \nu} \phi_{,\alpha} \phi_{,\nu} \right) \dot{\pi}_{,m}
\nonumber\\
 &+& 4 b(\phi) \left( \bar{g}^{0m}\Big( \bar{g}^{\alpha\beta}\phi_{,\beta\alpha} + \bar{J}^{\alpha}\phi_{,\alpha} \Big)  -  \bar{g}^{0\alpha}\bar{g}^{m \rho}  \Big( \phi_{,\rho\alpha} - \Gamma_{\alpha\rho}^{\lambda}\phi_{,\lambda} \Big) \right)\dot{\pi}_{,m} 
 \nonumber\\
 &-& G_{2,X}  g^{mn} \pi_{,mn} 
+  2\Big( G_{2,XX} - 2b_{,\phi} \Big) g^{\alpha n} g^{m\nu} \phi_{,\alpha} \phi_{,\nu}\pi_{,mn}
\nonumber\\
 &+& 2 b(\phi) \left( \bar{g}^{mn} \Big( \bar{g}^{\alpha\beta}\phi_{,\beta\alpha} + \bar{J}^{\alpha}\phi_{,\alpha} \Big) 
-  \bar{g}^{m \alpha}\bar{g}^{n\rho}  \Big( \phi_{,\rho\alpha} - \Gamma_{\alpha\rho}^{\lambda} \phi_{,\lambda} \Big)
\right) \pi_{,mn}
 \nonumber\\
&-& b(\phi)  \bar{g}^{\mu \alpha} \bar{g}^{\nu \beta} \phi_{,\alpha} \phi_{,\beta} \delta R_{\mu\nu} 
+ ... = 0\,.
\nonumber
\end{eqnarray}

Again, re-ordering the system by substituting the right hand side of the trace-reversed Einstein equations~(\ref{trace-rev-EEq-J-L3-inh-lin}-\ref{ij-gb}) for $\delta R_{\mu\nu}$ in the scalar field equation~\eqref{s-f-gb}, the three matrices $A, B^{mn},$ and $D^m$ introduced in Eq.~\eqref{lin-eq} take the same form as in the case of homogeneous background, with $A_h, B_h^{mn}, D_h^{m}$ being diagonal $10\times 10$ matrices as given in Eq.~\eqref{coeff-matrices}. 

\subsection{Necessary conditions for mode stability}

In this subsection, we present all eigenvalues and eigenvectors of the principal symbol corresponding to ${\cal L}_3$-Horndeski theories as given in Eqs.~(\ref{trace-rev-EEq-J-L3-inh-lin}-\ref{ij-gb}, \ref{s-f-gb}) and explicitly list necessary conditions required for weak and strong hyperbolicity of the initial value problem as introduced above in Sec.~\ref{sec:harmonic}. 

\subsubsection{Weak hyperbolicity}

For the initial value problem to be weakly hyperbolic, all eigenvalues of the principal symbol ${\cal P}^0$ must be real and finite.
The characteristic polynomial of ${\cal P}^0$ is given by
\begin{equation}
\label{char-gb}
\chi(\lambda) =  \left( \Big( \lambda - (-\bar{g}_{00}) \bar{g}^{0m}\tilde{k}_m \Big) \lambda - (-\bar{g}_{00}) \bar{g}^{mn}\tilde{k}_m\tilde{k}_n  \right)^{10}  \left( \Big(\lambda - \xi_D \Big) \lambda - \xi_B  \right)
\,,
\end{equation}
where the quantities $\xi_B$ and $\xi_D$ are defined as
\begin{eqnarray}
\xi_B &=& A_{\pi}^{-1}B_{\pi}^{mn}\tilde{k}_m \tilde{k}_n
\,,\\
\xi_D & = & A_{\pi}^{-1} D_{\pi}^m \tilde{k}_m\,;
\end{eqnarray}
with $A_{\pi}, B_{\pi}^{mn}, D_{\pi}^m$ being the coefficients of the second-order terms in the scalar field equation~\eqref{s-f-gb},
\begin{eqnarray}
A_{\pi} &=& (- \bar{g}^{00})  \bar{G}_{2,X}  +  \Big( \bar{G}_{2,XX} - 2\bar{b}_{,\phi} \Big) \bar{g}^{\alpha 0} \bar{g}^{\beta0} \phi_{,\alpha} \phi_{,\beta} \\
&+& 2 \bar{b}(\phi) \left( \bar{g}^{00} \Big( \bar{g}^{\alpha\beta}\phi_{,\beta\alpha} + \bar{J}^{\alpha}\phi_{,\alpha} \Big)  
- \bar{g}^{0\alpha}\bar{g}^{0\rho}  \Big( \phi_{,\rho\alpha} - \bar{\Gamma}_{\alpha\rho}^{\lambda}\phi_{,\lambda} \Big) \right) 
\nonumber\\
&+& \bar{b}(\phi)^2 \left( \bar{g}^{00}  \bar{g}^{\mu \alpha} \bar{g}^{\nu \beta} \phi_{,\alpha} \phi_{,\beta}  \left( \phi _{,\mu}  \phi_{,\nu} -\frac12 \phi _{,\alpha}  \phi_{,\beta} \bar{g}^{\alpha\beta} \bar{g}_{\mu\nu} \right)      
- 2 (\bar{g}^{0\mu}  \phi_{,\mu})^2  \bar{g}^{\alpha\beta}  \phi_{,\alpha} \phi_{,\beta}  \right)
\,,\nonumber \\
B_{\pi}^{mn} & = & \bar{G}_{2,X}  \bar{g}^{mn} 
-  2\Big( \bar{G}_{2,XX} - 2\bar{b}_{,\phi} \Big) \bar{g}^{m \alpha } \bar{g}^{n \beta} \phi_{,\alpha} \phi_{,\beta}
\\
 &-& 2 \bar{b}(\phi) \left( \bar{g}^{mn} \Big( \bar{g}^{\alpha\beta}\phi_{,\beta\alpha} + \bar{J}^{\alpha}\phi_{,\alpha} \Big) 
-  \bar{g}^{m \alpha}\bar{g}^{n\beta}  \Big( \phi_{,\beta\alpha} - \bar{\Gamma}_{\alpha\beta}^{\lambda} \phi_{,\lambda} \Big)
\right) 
 \nonumber\\
&-& \bar{b}(\phi)^2  \left( \bar{g}^{\mu \alpha} \bar{g}^{\nu \beta} \phi_{,\alpha} \phi_{,\beta}  \left( \phi _{,\mu}  \phi_{,\nu} -\frac12 \phi _{,\alpha}  \phi_{,\beta} \bar{g}^{\alpha\beta} \bar{g} _{\mu\nu} \right)  \bar{g}^{mn}  
-  2   \bar{g}^{\mu \nu} \phi_{,\mu} \phi_{,\nu} \bar{g}^{n \alpha }  \bar{g}^{m \beta}   \phi_{,\alpha} \phi_{,\beta}     \right)
\,,\nonumber \\
D_{\pi}^m & = & 2 \left( \bar{G}_{2,X}  \bar{g}^{0m}  
-  \Big( \bar{G}_{2,XX} - 2\bar{b}_{,\phi} \Big) g^{\alpha 0} \bar{g}^{\beta m} \phi_{,\alpha} \phi_{,\beta} \right) 
\\
 &-& 4 \bar{b}(\phi) \left( \bar{g}^{0m}\Big( \bar{g}^{\alpha\beta}\phi_{,\beta\alpha} + \bar{J}^{\alpha}\phi_{,\alpha} \Big)  -  \bar{g}^{0\alpha}\bar{g}^{m \rho}  \Big( \phi_{,\rho\alpha} - \bar{\Gamma}_{\alpha\rho}^{\lambda}\phi_{,\lambda} \Big) \right)
 \nonumber\\
 &-& \bar{b}(\phi)^2  \left(  \bar{g}^{\mu \alpha} \bar{g}^{\nu \beta} \phi_{,\alpha} \phi_{,\beta}  \left( \phi _{,\mu}  \phi_{,\nu} -\frac12 \phi _{,\alpha}  \phi_{,\beta} \bar{g}^{\alpha\beta} \bar{g}_{\mu\nu} \right)   \bar{g}^{0m} 
-   2  \bar{g}^{\mu \nu} \phi_{,\mu}  \phi_{,\nu} \bar{g}^{0\alpha}  \phi_{,\alpha}  \bar{g}^{m \beta}  \phi_{,\beta}       \right)
\,.\nonumber
\end{eqnarray}
It is clear from Eq.~\eqref{char-gb} that ${\cal P}^0$ has four distinct eigenvalues, namely
\begin{eqnarray}
\lambda^{\pm} &=& \frac12(-\bar{g}_{00})  \left( \bar{g}^{0m}\tilde{k}_m \pm \sqrt{ (\bar{g}^{0m}\tilde{k}_m)^2 + 4 (-\bar{g}^{00}) \bar{g}^{mn}\tilde{k}_m \tilde{k}_n }\, \right) \,,\quad \\
c_S^{\pm} &=& \frac12 \Big( \xi_D \pm \sqrt{4 \xi_B + \xi_D^2}\Big) \,.
\end{eqnarray}
The eigenvalues $\lambda_{\pm}$ are inherited from Einstein gravity and are all manifestly real. The eigenvalues $c_S^{\pm}$ are due to the Horndeski scalar field and can be interpreted as the `characteristic speeds' associated with the linearized scalar field $\pi$. These eigenvalues are  real if  and only if 
\begin{equation}
4 \xi_B + \xi_D^2 \geq 0\,.
\end{equation}
In addition, for $c_S^{\pm}$ to be finite, we must require $A_{\pi} \neq 0$, which exactly coincides with the invertibility condition of the kinetic matrix $A$. Note that in the homogeneous case $c_S^{\pm} \equiv \pm \sqrt{\xi_B}$ and $\xi_B$ is equivalent to the quantity often defined as $c_S^2$ in the cosmology literature. But this notation is unfortunate since it does not account for the fact that $\xi_B$ can be negative and hence we do not adapt it in our analysis. In particular, what is called `stability analysis' in the cosmology literature is in reality only a test of weak hyperbolicity but {\it not} a test of stability against mode fluctuations.

\subsubsection{Strong hyperbolicity}

For the initial value problem to be strongly hyperbolic, there must be a complete set of eigenvectors and the eigenvectors have to be finite and depend smoothly on the initial data. These three criteria are necessary such that, for any initial data, we can find an energy estimate that bounds the solution from above. That means, the mode fluctuations are under perturbative control (`mode stability').

The first two eigenvalues $\lambda^{\pm}$ each have ten corresponding eigenvectors ${\bf l}^{\pm}_i$ $(1\leq i\leq10)$ that each take the form
\begin{equation}
{\bf l}^{\pm}_i = p^{\pm} \delta_i^j {\bf e}_j  + \delta_i^j {\bf e}_{j+11}\,, \quad 1\leq j \leq 10\,,
\end{equation}
where ${\bf e}_n$ is the $n$th column of the ($22\times22$) identity matrix and 
\begin{equation}
p^{\pm} = - \frac12 \frac{g^{0m}\tilde{k}_m \mp \sqrt{
 (\bar{g}^{0m}\tilde{k}_m)^2 + 4 (-\bar{g}^{00}) \bar{g}^{mn}\tilde{k}_m\tilde{k}_n } }{\bar{g}^{mn}\tilde{k}_m\tilde{k}_n}
,
\end{equation}
{\it i.e.}, each eigenvector ${\bf l}^{\pm}_i$ has exactly two non-zero entries, namely the $i$th  and the $(i+11)$th entries, the former being equal to $p^{\pm}$ and latter being equal to one. Note that the $11$th and $22$nd entries are zero for all ${\bf l}_i^{\pm}$.
Obviously, all the twenty eigenvectors are linearly independent and finite as they should because these eigenvectors describe the characteristic structure of Einstein gravity.

The eigenvectors corresponding to the remaining two eigenvalues $c_S^{\pm}$ take the form
\begin{equation}
{\bf s}^{\pm} = \left(v_{tt}^{\pm}, ..., v_{zz}^{\pm}, - c_S^{\pm}/\xi_B , w_{tt}^{\pm}, ..., w_{zz}^{\pm},1 \right)\,,
\end{equation}
where
\begin{eqnarray}
\label{v-pm-inh}
v_{\mu\nu}^{\pm} &=& \frac{   c_S^- A_{h \pi}^{\mu\nu} + (c_S^+ / \xi_B)  \, B_{h \pi}^{\mu\nu}
- D_{h \pi}^{\mu\nu}}{ \bar{g}^{00}  \xi_B
+  \left( \bar{g}^{00}\xi_D + \bar{g}^{0m}\tilde{k}_m  \right) c_S^{\pm}
+  \bar{g}^{mn} \tilde{k}_m \tilde{k}_n
   }\,,\quad \\
   \label{w-pm-inh}
w_{\mu\nu}^{\pm} &=& \frac{  \xi_B A_{h \pi}^{\mu\nu}  - B_{h \pi}^{\mu\nu} +  c_S^{\pm} \left( \xi_D A_{h \pi}^{\mu\nu}  - D_{h \pi}^{\mu\nu} \right)
}{
  \bar{g}^{00}  \xi_B
 +  \left( \bar{g}^{00}\xi_D + \bar{g}^{0m}\tilde{k}_m  \right) c_S^{\pm} +  \bar{g}^{mn} \tilde{k}_m \tilde{k}_n  
   }\,;\quad 
\end{eqnarray}
and the coefficients $A_{h \pi}^{\mu\nu}, B_{h \pi}^{\mu\nu}, D_{h \pi}^{\mu\nu}$ can be read off from the perturbed Einstein equations~(\ref{trace-rev-EEq-J-L3-inh-lin}-\ref{ij-gb}) and take the following form,
\begin{eqnarray}
A_{h \pi}^{tt} &=&  \bar{b}( \phi)  \left( \frac32 (-\bar{g}^{00}) \dot{\phi}^2   - \frac12 \bar{g}^{mn} \phi _{,m}  \phi_{,n}   -  3 \dot{\phi} \bar{g}^{0m}  \phi_{,m}  \right),
\\
A_{h \pi}^{ti} &=&  \bar{b}( \phi) \left(  \frac12 (-\bar{g}^{00})\bar{g} _{0i} \bar{g}^{\alpha\beta}  \phi _{,\alpha}  \phi_{,\beta}   - \bar{g}^{0m}  \phi_{,m}   \phi_{,i} \right),
\\
A_{h \pi}^{ij} &=& \bar{b}( \phi)  (-\bar{g}^{00}) \left(\frac12  \bar{g} _{ij} \bar{g}^{\alpha\beta} \phi _{,\alpha}  \phi_{,\beta} - \phi _{,i}  \phi_{,j}  \right)   \,,
\\ 
B_{h \pi}^{tt} &=& - \frac12 \bar{b}( \phi) \left(  \dot{\phi}^2   - \bar{g} _{00} \bar{g}^{mn} \phi _{,m}  \phi_{,n}   - 2 \bar{g} _{00} \bar{g}^{0m} \dot{\phi}  \phi_{,m}\right)  \bar{g}^{mn}\tilde{k}_m\tilde{k}_n \,,
\\
B_{h \pi}^{ti} &=& - \bar{b}( \phi) \left( \dot{\phi}\phi_{,i} -\frac12 \bar{g}_{0i} \bar{g}^{\alpha\beta}  \phi _{,\alpha}  \phi_{,\beta}  \right) \bar{g}^{mn} \tilde{k}_m\tilde{k}_n
+ \bar{b}( \phi)  \dot{\phi}\,  \bar{g}^{\rho m}  \phi_{,\rho} \tilde{k}_m \tilde{k}_i \,,
\\
B_{h \pi}^{ij} &=& - \bar{b}( \phi)\left( \phi _{,i}  \phi_{,j} - \frac12 \bar{g} _{ij} \bar{g}^{\alpha\beta}  \phi _{,\alpha}  \phi_{,\beta}   \right)   \bar{g}^{mn} \tilde{k}_n \tilde{k}_m 
+ \bar{b}( \phi)  \bar{g}^{\rho m} \phi_{,\rho} \tilde{k}_m  \Big( \phi_{,i}  \tilde{k}_ j +  \phi_{,j}  \tilde{k}_i  \Big),\qquad
\\
D_{h \pi}^{tt}  &=&  - \bar{b}( \phi)  \left( \dot{\phi}^2   + \bar{g} _{00} \bar{g}^{mn} \phi _{,m}  \phi_{,n}   
+2 \bar{g} _{00} \dot{\phi}  \bar{g}^{0m} \phi_{,m}  \right) \bar{g}^{0m} \tilde{k}_m 
+ 2\bar{b}( \phi)  \dot{\phi} \bar{g}^{m n}  \phi_{,n} \tilde{k}_m   ,
\\
D_{h \pi}^{ti}  &=&  - \bar{b}( \phi) \left( \dot{\phi}\phi_{,i} - \bar{g}_{0i} \bar{g}^{\alpha\beta} \phi _{,\alpha}  \phi_{,\beta}  \right)  \bar{g}^{0m} \tilde{k}_m
+ \bar{b}( \phi) \phi_{,i} \bar{g}^{nm}  \phi_{,n} \tilde{k}_m  
+ \bar{b}( \phi)  \dot{\phi} \bar{g}^{\rho 0}  \phi_{,\rho}  \tilde{k}_i ,
\\
D_{h \pi}^{ij} &=& - \bar{b}( \phi)  \left( \phi _{,i}  \phi_{,j} -\frac12 \bar{g} _{ij}  \bar{g}^{\alpha\beta} \phi _{,\alpha}  \phi_{,\beta}  \right)   \bar{g}^{0m} \tilde{k}_m
+\bar{b}( \phi)  \bar{g}^{0\rho}  \phi_{,\rho}  \Big( \phi_{,i}  \tilde{k}_j +  \phi_{,j} \tilde{k}_i \Big) \,.
\end{eqnarray}
It is immediately apparent that both $\pi$-eigenvectors ${\bf s}^{\pm}$ are linearly independent and they are linearly independent of the another twenty eigenvectors. But, in order for ${\bf s}^{\pm}$ to be bounded from above, we have to require that (i) neither denominator vanishes at any point in time and  (ii)  no numerator blows up. By direct inspection of the expressions, it is straightforward to see that condition (ii) is equivalent to requiring $\xi_B\neq 0$.

Comparing Eqs.~(\ref{v-pm-inh}-\ref{w-pm-inh}) for an inhomogeneous background to Eqs.~(\ref{v-pm-h}-\ref{w-pm-h}) for a homogeneous background, we see that the difference is the middle term in the denominator.  In the homogeneous case, it is straightforward to construct backgrounds where the denominator  is substantially different from zero.  In these cases, there is a finite range of background inhomogeneity that can be added so it leaves the middle term small enough for the denominator to remain non-zero.  This shows that the homogeneous solution is not on a `knife-edge' of mode instability.  Furthermore, in setting a numerical code, the denominator can be used as a diagnostic to test whether the simulation is approaching a mode instability. 

Our analysis yields a similar overall conclusion as the finding in Ref.~\cite{Papallo:2017qvl}, where Papallo and Reall concluded strong hyperbolicity on generic `weak-field backgrounds' requires a specific `deformation' of the theory that involves a particular gauge condition.
In their proposed gauge, the coordinates are sourced by first derivatives of the linearized scalar-field, {\it i.e.},
\begin{equation}
\label{harvey-gauge}
G_{\mu}{}^{\nu \alpha\beta} \nabla_{\nu} h_{\alpha\beta} = {\cal H}_{\mu}{}^{\nu} \nabla_{\nu} \pi
\,,
\end{equation}
where both tensors $G^{\mu\nu \alpha\beta} = (1/2)(\bar{g}^{\mu\alpha}\bar{g}^{\nu\beta} + \bar{g}^{\mu\beta}\bar{g}^{\nu\alpha} - \bar{g}^{\mu\nu}\bar{g}^{\alpha\beta} )$ and ${\cal H}_{\mu}{}^{\nu}$ depend only on background quantities.
Substituting this gauge condition into Eqs.~(\ref{trace-rev-EEq-J-L3-inh-lin}--\ref{ij-gb}) via
\begin{equation}
\label{harvey-gauge-cov}
\delta J_{\mu} = {\cal H}_{\mu}{}^{\alpha} \nabla_{\alpha} \pi
\,,
\end{equation}
we obtain
\begin{eqnarray}
\label{trace-rev-EEq-J-harvey-gauge-lin}
&-& \frac12 \bar{g}^{\alpha \beta}h_{\mu \nu}{}_{,\beta\alpha}
+   b( \phi)  \left( \phi _{,\mu}  \phi_{,\nu} -\frac12 \phi _{,\alpha}  \phi_{,\beta} \bar{g} ^{\alpha\beta} \bar{g}_{\mu\nu} \right)  \bar{g}^{\alpha \beta} \pi_{,\beta\alpha} 
\qquad\\
&-& 
 \left( \phi_{,\mu} b( \phi) \bar{g}^{\rho\sigma}  \phi_{,\rho}   
 + \frac12 {\cal H}_{\mu}{}^{\sigma}  \right) \pi_{,\sigma \nu}
-  \left(  b( \phi) \bar{g}^{\rho\sigma}  \phi_{,\rho}  \phi_{,\nu} 
+ \frac12 {\cal H}_{\nu}{}^{\sigma} \right)\pi_{,\sigma\mu} 
+ ...
= 0
 \,.
 \nonumber
\end{eqnarray}
Manifestly, there is a single functional form for ${\cal H}_{\mu}{}^{\alpha}$ that removes all terms in the second line of Eq.~\eqref{trace-rev-EEq-J-harvey-gauge-lin}, such that the Einstein equations take the form 
\begin{equation}
\label{g-h-form}
\Box h_{\mu\nu} +  b( \phi)  \left( \phi _{,\mu}  \phi_{,\nu} -\frac12 \phi _{,\alpha}  \phi_{,\beta} \bar{g} ^{\alpha\beta} \bar{g}_{\mu\nu} \right) \Box \pi + ... = 0.
\end{equation} 
This is exactly the form for ${\cal H}_{\mu}{}^{\alpha}$ that Papallo and Reall proposed. 

In terms of our analysis, it is straightforward to understand why Papallo and Reall were forced to make this gauge choice: unless the off-diagonal terms $A_{h\pi}, B^{mn}_{h\pi}, D^{m}_{h\pi}$ in the coefficient matrices $A, B^{mn}, D^{m}$ as defined in Eq.~\eqref{coeff-matrices} are removed, there will always be a background such that the two $\pi$-eigenvectors as given in Eqs.~(\ref{v-pm-inh}-\ref{w-pm-inh}) blow up. While weak hyperbolicity is maintained, strong hyperbolicity is broken. Since they were interested in linear well-posedness on {\it generic} backgrounds, Papallo and Reall had to eliminate the off-diagonal terms. In other words, to show local well-posedness on generic backgrounds, they deformed the theory in a way that ensures conformal (or disformal) equivalence to Einstein gravity.  We are interested in well-posedness on and around certain homogenous cosmological backgrounds in which case, apparently, we are not forced to their gauge choice but can utilize any generalized harmonic source function.

It is well-known that ${\cal L}_3$-Horndeski theories are neither conformally nor disformally equivalent to Einstein gravity. For this reason, it is not surprising that, for generic backgrounds, the deformed gauge condition in Eq.~\eqref{harvey-gauge-cov} does not have a covariant lift in the case of ${\cal L}_3$-Horndeski theories while it does in the case of Brans-Dicke gravity.


\section{Scalar-Vector-Tensor Decomposition of the linearized metric in generalized harmonic gauge} 
\label{sec:appE}

For an FRW background (${\rm d}s^2 = \bar{g}_{00}{\rm d}t^2 + g_{ij}{\rm d}x^i{\rm d}x^j$ where $g_{ij}=a^2(t)\delta_{ij}$), up to linear order, we can decompose the metric as
\begin{eqnarray}
h_{00} &=& 2\,\bar{g}_{00} \alpha 
\,,\\
h_{0i} &=& \sqrt{-\bar{g}_{00}}\, a(t) \big(\beta_{,i} + B_i \big) 
\,,\\
h_{ij} &=&  2\,a^2(t)  \Big( -\psi \delta_{ij} + \epsilon_{,ij} + 2S_{(i,j)} + u_{ij} \Big) \,,
\end{eqnarray}
where
\begin{equation}
\label{SVT-constraints}
\partial^i B_i = 0\,; \quad \partial^i S_i = 0\,; \quad u_{ij} = u_{ji}\,; \quad \partial^i u_{ij} = 0\,; \quad u^i_i = 0\,.
\end{equation}
Here, $\alpha, \beta, \psi$ and $\epsilon$ are the scalar components; $B_i$ and $S_i$ are the vector components; and $u_{ij}$ are the tensor components of $h_{\mu \nu}$.

Substituting into Eqs.~(\ref{conn-000-h}-\ref{conn-ikl-h}), the connection coefficients take the form
\begin{eqnarray}
\delta\Gamma_{00}^{0} &=&   \dot{\alpha} 
\,,\\
\delta\Gamma_{00}^{i} &=&   a^{-2}  \left(  -  \bar{g}_{00} \alpha +  \sqrt{-\bar{g}_{00}}\, a \left(\dot{\beta} + H\beta \right) \right)_{, i}
+ a^{-1} \sqrt{-\bar{g}_{00}} \left( \dot{B}_i + HB_i \right)
\,,\\
\delta\Gamma_{0i}^{0} &=&  \left(  \alpha +   \sqrt{-\bar{g}^{00}}\, a H \beta \right)_{, i}  +   \sqrt{-\bar{g}^{00}}\, a H B_i 
\,,\\
\delta\Gamma_{0k}^{l} &=&  \left( -\dot{\psi} \delta_{kl} + \dot{\epsilon}_{,kl} + 2\dot{S}_{(k,l)} + \dot{u}_{kl} \right) 
+ \frac12 \ \sqrt{-\bar{g}^{00}} \,a^{-1} \left(  B_{l , k}  - B_{k, l}  \right)
\,,\\
\delta\Gamma_{kl}^0 &=& \frac12 (-\bar{g}^{00})\left( \dot{h}_{kl} - h_{0 k, l} - h_{0 l, k} \right) 
- 2\, (-  \bar{g}^{00}) a^2(t)H(t) \, \alpha\, \delta_{kl}
\,,\\
\delta\Gamma_{ik}^{l} &=& \frac12 a^{-2}\left( h_{l i, k} + h_{l k, i} - h_{ik, l}\right) 
-  \sqrt{-\bar{g}^{00}}\, a(t) H\big(\beta_{,l} + B_l \big) \delta_{ik}
\,.
\end{eqnarray}
and, substituting into Eq.~\eqref{harmonic-cond-lin1}, the linearized harmonic gauge condition takes the form
\begin{eqnarray}
\label{SVT-J0}
\delta J_0 &=& 
-  \dot{\alpha}  - 3 \dot{\psi} +  \delta^{kl} \left(\dot{\epsilon} - a^{-1} \sqrt{-\bar{g}_{00}}\,\beta \right)_{,lk} 
\,,\\ 
\label{SVT-Ji}
\delta J_i &=&  \left(\alpha - \psi - \delta^{kl}\epsilon_{,lk}  \right)_{,i} 
  -   a\sqrt{-\bar{g}^{00}}\, \left(\dot{\beta}
-  \left(H  - \frac12 \dot{\bar{g}}_{00} \bar{g}^{00} \right) \beta  \right)_{,i} 
 \\ 
 &-&   a\sqrt{-\bar{g}^{00}}\, \left( \dot{B}_i 
-  \left(H  - \frac12 \dot{\bar{g}}_{00} \bar{g}^{00} \right)  B_i  \right)
- 2  \delta^{kl} S_{i,lk}
\,. \nonumber
\end{eqnarray}
Note that due to the constraints in Eq.~\eqref{SVT-constraints}, tensors do not contribute to the harmonic source functions (or, equivalently, tensors satisfy the harmonic gauge condition $\delta J_{\mu}\equiv 0$).

Finally, the linearized Einstein equations in SVT decomposition are given by 
\begin{equation}
\label{lin-Eeq.-gen-harm}
\delta R_{\mu\nu} = \delta T_{\mu\nu} - \frac12h_{\mu\nu}\bar{T}^{\alpha}_{\alpha} - \frac12\bar{g}_{\mu\nu}\delta{T}^{\alpha}_{\alpha}
\,,
\end{equation}
where the components of the linearized Ricci tensor are given by
\begin{eqnarray}
\label{lin-Ricci-00}
\delta R_{00}  &=&   -  \ddot{\alpha} +  a^{-2} (-\bar{g}_{00})\delta^{mn}  \alpha_{,nm} 
+  \Big( \bar{g}^{00}  \dot{\bar{g}}_{00} + \bar{J}_{0}\Big)  \dot{\alpha} 
- \delta \dot{J}_{0} + \frac12 \bar{g}^{00}\dot{\bar{g}}_{00} \delta J_0
  \\
&+&   6H \dot{\psi}  
-  2 H \delta^{mn} \left(   \dot{\epsilon} -  a^{-1} \sqrt{-\bar{g}_{00}} \, \beta \right)_{,nm}
\,;\nonumber
\\
\label{lin-Ricci-0i}
\delta R_{0i}  &=& 
\sqrt{-\bar{g}_{00}}\,a(t) \left( \frac12 (-\bar{g}^{00})  \ddot{\beta} 
- \frac12 a^{-2}\delta^{mn} \beta _{,nm} -    H(t) \dot{\beta} \right)_{,i} 
\\
&+ & \frac12 \, \sqrt{-\bar{g}^{00}}\,a(t) \left(   \dot{H} + \big(H +  2 \bar{J}_{0} \big) H  
+ \frac34 \dot{\bar{g}}_{00}  \dot{\bar{g}}^{00}  
+  \frac12 \bar{g}^{00}  \ddot{\bar{g}}_{00} 
\right)  \beta_{,i}   
\nonumber\\
&+ & \left( \bar{J}_{0} - 2H  + \frac12 \bar{g}^{00}  \dot{\bar{g}}_{00} \right) \alpha_{, i}      
+ H \Big( \psi +  \delta^{mn} \epsilon_{,n m} \Big)_{,i} 
 - \delta J_{(0, i)} + H \delta J_i
\nonumber\\
&+& \sqrt{-\bar{g}_{00}}\,a(t) \left( \frac12 (-\bar{g}^{00})  \ddot{B}_i  - \frac12 a^{-2}\delta^{mn}   B_{i,nm} -    H(t)   \dot{B}_i   \right) 
\nonumber\\
&+ & \frac12 \, \sqrt{-\bar{g}^{00}}\,a(t) \left(   \dot{H} + \big(H +  2 \bar{J}_{0} \big) H  
+ \frac34 \dot{\bar{g}}_{00}  \dot{\bar{g}}^{00}  
+  \frac12 \bar{g}^{00}  \ddot{\bar{g}}_{00} \right)  B_i 
+ 2 H\delta^{mn} S_{i,n m} 
\,;
\nonumber\\
\label{lin-Ricci-ij}
\delta R_{ij}  &=& 
\delta_{ij}  (-\bar{g}^{00} ) a^2(t) \Big( -\ddot{\psi} + a^{-2} \delta^{kl} \psi_{,lk}    - \bar{J}_{0}  \dot{\psi}  
- 2  \Big( \dot{H}  + \bar{J}_{0} H \Big) \alpha +  H \delta J_{0} \Big) 
\\
&+& (-\bar{g}^{00} ) a^2(t) \Big( \ddot{\epsilon} - a^{-2} \delta^{kl} \epsilon_{,lk}  
+ \bar{J}_{0}  \dot{\epsilon} 
-  \sqrt{-\bar{g}_{00}}\, a^{-1}(t)  \bar{J}_{0} \beta \Big)_{,ji} 
\nonumber \\
&+& 2\,(-\bar{g}^{00} ) a^2(t) \left(   \ddot{S}_{(i,j)}  - \delta^{kl} S_{(i,j),lk} 
+ \bar{J}_{0} \dot{S}_{(i,j)} -  \frac12 \sqrt{-\bar{g}_{00}}\, a^{-1}(t)  \bar{J}_{0} B_{(i,j)}  \right)
- \delta J_{(i, j)}
\nonumber \\
&+& (-\bar{g}^{00} ) a^2(t) \Big( \ddot{u}_{ij} - \delta^{kl}u_{ij ,lk} + \bar{J}_{0} \dot{u}_{ij}  \Big)   
\nonumber\\
&+&  2 (-\bar{g}^{00} )\,  a^2(t) \Big(  \dot{H} +\bar{J}_{0}H   \Big)  \Big( -\psi \delta_{ij} + \epsilon_{,ji} + 2 S_{(i,j)} + u_{ij}  \Big)
\,;\nonumber
\\
\label{lin-Ricci-svt}
\delta R  &=&  (- \bar{g}^{00}) \ddot{\alpha} -  a^{-2} \delta^{mn}  \alpha_{,nm} 
- \Big( \dot{\bar{g}}^{00} - \bar{g}^{00} \bar{J}_{0}\Big)  \dot{\alpha} 
  \\
&-& (- \bar{g}^{00}) \Big( 6H^2 + 6  \dot{H}  + 2 \dot{\bar{J}}_0 + 6 H\bar{J}_{0}  + \bar{g}_{00} \dot{\bar{g}}^{00}  \bar{J}_0 
- \bar{g}_{00}\ddot{\bar{g}}^{00}  - \frac12 \dot{\bar{g}}^{00} \dot{\bar{g}}_{00}  \Big) \alpha 
\nonumber\\
&+&  3 (-\bar{g}^{00} )  \left( -\ddot{\psi} + a^{-2} \delta^{kl} \psi_{,lk} - \Big(2H + \bar{J}_{0}\Big)  \dot{\psi}  \right) 
  \nonumber\\
&+&  (-\bar{g}^{00} ) \delta^{mn} \left( \ddot{\epsilon} - a^{-2} \delta^{kl} \epsilon_{,lk}  
+ \Big(2 H + \bar{J}_{0} \Big)\Big( \dot{\epsilon} -   a^{-1} \sqrt{-\bar{g}_{00}} \, \beta \Big)
 \right)_{,nm}  
 \nonumber\\
&+& (- \bar{g}^{00} ) \delta \dot{J}_{0} 
- \frac12 \Big( \dot{\bar{g}}^{00} - 6(-\bar{g}^{00} ) H \Big) \delta J_{0}  
- a^{-2} \delta^{ij} \delta J_{(i, j)}
\,.\nonumber
\end{eqnarray}
For consistency, we checked that substituting Eqs.~(\ref{SVT-J0}-\ref{SVT-Ji}) for the harmonic source functions, Eqs.~(\ref{lin-Ricci-00}-\ref{lin-Ricci-ij}) yield the known results as presented in Sec.~\ref{sec:review}.

\bibliographystyle{plain}
\bibliography{long-paper}

\end{document}